\pdfoutput=1

\documentclass[11pt,twoside,a4paper,cmspaper,final,collab]{cms-tdr}
\def\svnVersion{242671}\def\cmsCernNo{2013-108}\def\cmsCernDate{\today}\def\cmsMessage{Published in Physical Review D as \href{http://dx.doi.org/10.1103/PhysRevD.89.092005}{\doi{10.1103/PhysRevD.89.092005.}}}

\begin{document}\cmsNoteHeader{EWK-11-009}

\hyphenation{had-ron-i-za-tion}
\hyphenation{cal-or-i-me-ter}
\hyphenation{de-vices}

\RCS$Revision: 230632 $
\RCS$HeadURL: svn+ssh://svn.cern.ch/reps/tdr2/papers/EWK-11-009/trunk/EWK-11-009.tex $
\RCS$Id: EWK-11-009.tex 230632 2014-03-07 20:06:12Z alverson $

\providecommand{\rlumi}{\ensuremath{\mathrm{L}}\xspace} 
\newcommand{\DE}{$\Delta\eta_\text{in}$}
\newcommand{\DP}{$\Delta\phi_\text{in}$}
\newcommand{\SEE}{$\sigma_{\eta\eta}$}
\newcommand{\SEP}{$\sigma_{\eta\phi}$}
\newcommand{\SPP}{$\sigma_{\phi\phi}$}
\newcommand{\SXY}{$\sigma_{XY}$}
\newcommand{\IECAL}    {I_{\textrm{ECAL}}}%
\newcommand{\IHCAL}    {I_{\textrm{HCAL}}}%
\newcommand{\ITRK}     {I_{\textrm{trk}}}%
\newcommand{\IRelComb} {I^{\textrm{rel}}_{\textrm{comb}}}%
\newcommand{\vb}{\ensuremath{\cmsSymbolFace{V}}\xspace}
\providecommand{\MT}{\ensuremath{M_\mathrm{T}}\xspace}
\newlength\cmsFigWidth
\ifthenelse{\boolean{cms@external}}{\setlength\cmsFigWidth{0.95\columnwidth}}{\setlength\cmsFigWidth{0.85\textwidth}}
\ifthenelse{\boolean{cms@external}}{\providecommand{\cmsLeft}{top}}{\providecommand{\cmsLeft}{left}}
\ifthenelse{\boolean{cms@external}}{\providecommand{\cmsRight}{bottom}}{\providecommand{\cmsRight}{right}}

\newcommand{\sihih}{\ensuremath{\sigma_{\eta\eta} } }%
\newcommand{\nsig} {\ensuremath{N_\text{sig}}}
\ifthenelse{\boolean{cms@external}}{\providecommand{\CL}{\ensuremath{\mathrm{C.L.}}\xspace}}{\providecommand{\CL}{\ensuremath{\mathrm{CL}}\xspace}}

\cmsNoteHeader{EWK-11-009} 
\title{\texorpdfstring{Measurement of the $\PW\gamma$ and $\cPZ\gamma$ inclusive cross sections in $\Pp\Pp$ collisions at $\sqrt{s} = 7\TeV$ and limits on anomalous triple gauge boson couplings}{Measurement of the W gamma and Z gamma inclusive cross sections in pp collisions at sqrt(s) = 7 TeV and limits on anomalous triple gauge boson couplings}}

\date{\today}

\abstract{
Measurements of $\PW\gamma$ and $\cPZ\gamma$ production in proton-proton
collisions at $\sqrt{s} = 7\TeV$ are used to extract limits on anomalous triple gauge couplings.
The results are based on data recorded by the
CMS experiment at the LHC that correspond to an integrated luminosity of
5.0\fbinv. The cross sections are measured for photon transverse momenta
$\pt^\gamma>15\GeV$, and for separations between photons
and final-state charged leptons in the pseudorapidity-azimuthal plane of
$\Delta R(\ell, \gamma) > 0.7$ in $\ell\nu\gamma$ and $\ell\ell\gamma$ final
states, where $\ell$ refers either to an electron or a muon.
A dilepton invariant mass requirement of $m_{\ell\ell} > 50\GeV$ is imposed for
the $\cPZ\gamma$ process. No deviations are observed relative to
predictions from the standard model, and limits are set on anomalous $\PW\PW\gamma$,
$\cPZ\cPZ\gamma$, and $\cPZ\gamma\gamma$ triple gauge couplings.
}

\hypersetup{%
pdfauthor={CMS Collaboration},%
pdftitle={Measurement of the W gamma and Z gamma inclusive cross sections in pp collisions at sqrt(s) = 7 TeV and limits on anomalous triple gauge boson couplings},%
pdfsubject={CMS},%
pdfkeywords={CMS, physics, vector boson production}}

\maketitle 

\section{Introduction}
The standard model (SM) has been enormously successful in describing the
electroweak (EW) and strong interactions. However, important
questions remain unanswered regarding possible extensions of the SM that incorporate
new interactions and new particles.
The self-interactions of the electroweak gauge bosons comprise an important and
sensitive probe of the SM, as their form and strength are determined by the
underlying $SU(2)\times U(1)$ gauge symmetry. A precise measurement of the production
of pairs of EW bosons (``diboson" events) provides direct information on the
triple gauge couplings (TGCs), and any deviation of these couplings from their
SM values would be indicative of new physics.
Even if the new phenomena involve the presence of objects that can
only be produced at large energy scales, \ie, beyond the
reach of the Large Hadron Collider (LHC), they can nevertheless induce changes in the TGCs.
In addition, since diboson processes represent the primary background to the SM Higgs production,
their precise measurement is important for an
accurate evaluation of Higgs boson production at the LHC, particularly
in association with gauge bosons.

Aside from $\gamma\gamma$ production, the EW $\PW\gamma$ and $\cPZ\gamma$
production processes at hadron colliders provide the largest and cleanest yields,
as backgrounds to $\PW\gamma$ and $\cPZ\gamma$ production can be significantly
suppressed through the identification of the massive \PW\ and \cPZ\ vector bosons
via their leptonic decay modes.
Measurements from LEP~\cite{lep, l3_1, l3_2, opal}, the
Tevatron~\cite{tevatron_wg1, tevatron_wg2, tevatron_zg0, tevatron_zg1, tevatron_zg2},
and from initial analyses at the LHC~\cite{cms_ww, cms_vgamma, atlas_vgamma_5fb} have
already explored some of the parameter space of anomalous TGCs (ATGCs) in $\PW\gamma$
and $\cPZ\gamma$ processes.

We describe an analysis of inclusive $\PW\gamma$ and $\cPZ\gamma$
events, collectively referred to as ``$\vb\gamma$'' production,
based on the leptonic decays $\PW\to \Pe\nu$, $\PW\to \mu\nu$,
$\cPZ\to \Pe\Pe$, and $\cPZ\to \mu\mu$, observed in $\Pp\Pp$ collisions at
a center-of-mass energy of 7\TeV. The data, corresponding to
an integrated luminosity $\rlumi = 5.0\fbinv$, were collected in 2011 with the Compact Muon
Solenoid (CMS) detector at the LHC. The previous results from $\Pp\Pp$ collisions at $\sqrt{s} = 7\TeV$ at the LHC
were limited by the statistics of the data samples, and this analysis achieves a significant
improvement in precision.

\begin{figure}[htb]
\begin{center}
  \includegraphics[width=\cmsFigWidth]{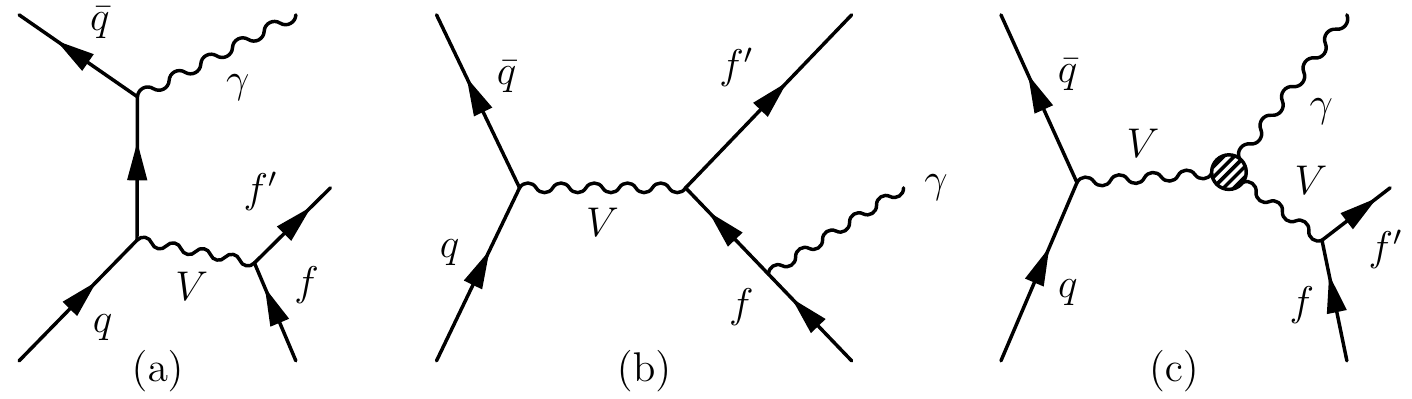}
\caption{The three lowest order diagrams for $\vb\gamma$ production, with \vb corresponding to both virtual and on-shell $\gamma$, \PW, and \cPZ\ bosons.
The three diagrams reflect contributions from (a)~initial-state and
(b)~final-state radiation and (c)~TGC. The TGC diagram does not
contribute at the lowest order to SM $\cPZ\gamma$ production since photons
do not couple to particles without electric charge.}
\label{fig:feynman_vg}
\end{center}
\end{figure}

$\vb\gamma$ production can be represented by the Feynman diagrams of Fig.~\ref{fig:feynman_vg}.
Three processes contribute: (a)~initial-state radiation, where a photon
is radiated by one of the incoming virtual partons, (b)~final-state radiation,
where a photon is radiated by one of the charged leptons from \vb decay,
and (c)~TGC at the $\PW\PW\gamma$ vertex in $\PW\gamma$ production, and
the $\cPZ\cPZ\gamma$ and $\cPZ\gamma\gamma$ vertices in $\cPZ\gamma$ production.
In the SM, contributions from the TGC process are expected only for
$\PW\gamma$ production, because neutral TGCs are forbidden at tree
level~\cite{baurWg, bho1998}.

This paper is organized as follows. Brief descriptions of the CMS detector
and Monte Carlo (MC) simulations are given in Section~\ref{sec:CMSDetector}. Selection
criteria used to identify the final states are given in
Section~\ref{sec:ObjectSelection}. Dominant backgrounds to $\vb\gamma$ production
are described in Section~\ref{sec:dataDrivenBackgrounds}, along with methods used to
estimate background contributions. Measurements of cross sections and limits on
ATGCs are given, respectively, in Section~\ref{sec:Results} and \ref{sec:ATGC}, and
the results are summarized in Section~\ref{sec:Summary}.

\section{CMS detector and Monte Carlo simulation}
\label{sec:CMSDetector}
The central feature of the CMS apparatus is a superconducting
solenoid, which is 13\unit{m} long and 6\unit{m} in diameter, and provides
an axial magnetic field of 3.8\unit{T}. The bore of the solenoid is instrumented
with detectors that provide excellent performance for
reconstructing hadrons, electrons, muons, and photons.
Charged particle trajectories are measured with silicon pixel and strip trackers
that cover all azimuthal angles $0 < \phi < 2\pi$ and pseudorapidities
$\abs{\eta} < 2.5$, where $\eta$ is defined as $-\ln[\tan(\theta/2)]$,
with $\theta$ being the polar angle of the trajectory of the particle
relative to the counterclockwise-beam direction.
A lead tungstate crystal electromagnetic calorimeter (ECAL)
and a brass/scintillator hadron calorimeter (HCAL) surround the
tracking volume.
Muons are identified and measured in gas-ionization detectors embedded in
the steel flux return yoke outside of the solenoid. The detector coverage is nearly
hermetic, providing thereby accurate measurements of the imbalance
in momentum in the plane transverse to the beam direction.
A two-tier trigger system selects the most interesting $\Pp\Pp$ collisions
for use in analyses. A more detailed description of the CMS detector can be
found in Ref.~\cite{CMS:2008zzk}.

The main background to $\PW\gamma$ and $\cPZ\gamma$ production arises from
\PW+jets and \cPZ+jets events, respectively,
in which one of the jets is misidentified as a photon.
To minimize systematic uncertainties associated with the modeling
of parton fragmentation through MC simulation, this
background is estimated from multijet events in data, as described in
Section~\ref{sec:dataDrivenBackgrounds}.
The background contributions from other processes, such as $\ttbar$, $\gamma$+jets, and
multijet production, are relatively small, and are estimated using MC simulation.

The MC samples for the signal processes $\PW\gamma + n\;\text{jets}$ and
$\cPZ\gamma+n\;\text{jets}$, where $n<3$, are generated with
\MADGRAPH~v5.1.4.2~\cite{MadGraph} and interfaced to
\PYTHIA~v6.424~\cite{Pythia} for parton showering and hadronization.
The kinematic distributions for these processes are
cross-checked with expectations from \SHERPA~v1.2.2~\cite{sherpa}, and the
predictions from the two programs are found to agree.
The signal samples are normalized to the predictions of
next-to-leading-order (NLO) quantum chromodynamics
from the \MCFM~v6.5 generator~\cite{mcfm1,mcfm2} using the
CTEQ6.6 NLO parton distribution functions (PDF)~\cite{cteq66}.

Backgrounds from $\ttbar$, \PW+jets, \cPZ+jets, \PW\PW, and
$\gamma\gamma$ events are also simulated with the \MADGRAPH program
interfaced with \PYTHIA. Multijet, $\gamma$+jets, and \PW\cPZ\ and \cPZ\cPZ\ diboson
events are generated using the stand-alone \PYTHIA MC program,
and have negligible impact on the analysis.
All these MC event samples, generated using the CTEQ6L1 leading-order (LO)
PDF~\cite{CTEQ6}, are passed through a detailed simulation of the CMS detector based
on~\GEANTfour~\cite{geant4}, and
reconstructed with the same software that is used for data.

\section{Selection of candidate events}
\label{sec:ObjectSelection}
The requirements for selecting isolated muons follow closely the
standard CMS muon identification criteria~\cite{muperf}.
However, electron and photon selection criteria are optimized specifically
for this analysis, and are described in greater detail in the following subsections,
as are the reconstruction of transverse momentum imbalance or the ``missing''
transverse momentum ($\ETslash$),
all trigger requirements, and the selections used to enhance the purity of signal.

The presence of pileup from additional overlapping interactions is taken
into account in the analysis, and cross-checked by studying the effectiveness of
the selection criteria, separately, for small and large pileup rates in data.
There are on average 5.8 overlapping interactions per collision for
low-pileup data, and 9.6 interactions for high-pileup data, which correspond, respectively,
to integrated luminosities of $\rlumi\approx 2.2\fbinv$
(referred to subsequently as Run 2011A) and to $\rlumi\approx 2.7\fbinv$
(referred to as Run 2011B).

\subsection{Electron identification and selection}
\label{sec:eid}

Electrons are identified as ``superclusters'' (SC) of energy deposition~\cite{photon_prod}
in the ECAL fiducial volume that are matched to tracks from the silicon tracker.
Tracks are reconstructed using a Gaussian-sum filter algorithm
that takes into account possible energy loss due to bremsstrahlung in the tracker.
The SC are required to be located within the acceptance of the tracker ($\abs{\eta} < 2.5$).
Standard electron reconstruction in the transition regions between the central barrel (EB) and the
endcap (EE) sections of the ECAL ($1.44 < \abs{\eta} < 1.57$) has reduced efficiency, and
any electron candidates found in these regions are therefore excluded from consideration.
The reconstructed electron tracks are required to have hits
observed along their trajectories in all layers of the inner tracker.
Electron candidates must have $\pt > 35$ and ${>}20\GeV$
for the $\PW\gamma$ and $\cPZ\gamma$ analyses, respectively.

Particles misidentified as electrons are suppressed through the use of an
energy-weighted width quantity in pseudorapidity (\SEE)
that reflects the dispersion of energy in $\eta$ (``shower shape'')
in a $5 \times 5$
matrix of the 25 crystals centered about the crystal containing the largest
energy in the SC~\cite{photon_prod}. The \sihih parameter is defined through
a mean $\bar\eta = \sum \eta_i w_i/\sum w_i$ as follows:

\begin{equation}
    \label{eq:pho_showcov}
    \sigma_{\eta\eta}^2 = \frac
    { \sum \left( \eta_i - \bar \eta \right )^2  w_i,
    } {\sum w_i}, i = 1, \dots, 25,
\end{equation}

where the sum runs over all the elements of the $5 \times 5$ matrix, and
$\eta_i = 0.0174 \hat{\eta}_i$, with $\hat{\eta}_i$ denoting the $\eta$ index of the $i$th crystal;
the individual weights $w_i$ are given by $4.7 + \ln(E_i/\ET)$, unless
any of the $w_i$ are found to be negative, in which case they are set to zero.
In the ensuing analysis, the value of \sihih is required to be consistent
with expectations for electromagnetic showers, and the discriminant is used
to suppress background as well as to assess contribution from signal and
background in fits to the data discussed in Section~\ref{ssec:template_method}.

In addition, the $\eta$ and $\phi$ coordinates of the particle trajectories
extrapolated to the ECAL are required to match
those of the SC, and limits are imposed on the amount of
HCAL energy deposited within the spatial cone
$\Delta R = \sqrt{(\Delta\phi)^2 + (\Delta\eta)^2} < 0.15$
relative to the axis of the ECAL cluster.
To reduce background from $\gamma\to \Pep\Pem$ conversions in the
tracker material, the electron candidates are required to have no
``partner'' tracks
within 2\unit{mm} of the extrapolated point in the transverse plane
where both tracks are parallel to each other (near the hypothesized point
of the photon conversion), and the difference in the cotangents of their
polar angles must satisfy $\abs{\Delta \cot\theta} > 0.02$.
To ensure that an electron trajectory is consistent with originating
from the primary interaction vertex, taken to be the one with the largest
scalar sum of the $\pt^{2}$ of its associated tracks in the case of
multiple vertices, the distances of closest approach are required to be
$\abs{d_z} < 0.1\unit{cm}$ and $\abs{d_\mathrm{T}} < 0.02\unit{cm}$ for the
longitudinal and transverse coordinates, respectively.

To reduce background from jets misidentified as electrons, the electron
candidates are required to be isolated from other energy depositions in the detector.
The electron selection criteria are obtained by optimizing signal and background
levels using simulated samples. This optimization is done separately for the
EB and EE sections.
Different criteria are used for the $\PW\gamma \to \Pe\nu\gamma$ and
$\cPZ\gamma \to \Pe\Pe\gamma$ channels because of the different trigger requirements
and relative background levels.
For the $\cPZ\gamma$ analysis, a relative isolation parameter ($I_r$) is calculated
for each electron candidate through a separate sum of scalar $\pt$
in the ECAL, HCAL, and tracker (TRK), all defined relative to the axis
of the electron,
but without including its $\pt$, within a cone of $\Delta R < 0.3$.
In computing TRK isolation for electrons, each of the contributing tracks is required
to have $\pt > 0.7$\GeV and  be consistent with originating from within
$\abs{d_z} < 0.2\unit{cm}$ of the primary interaction vertex.
This sum, reduced by $\rho \times \pi \times 0.3^{2}$ to account for the pileup
contributions to the isolation parameter, and divided by the $\pt$ of the electron candidate,
defines the $I_r$ for each subdetector.
Here $\rho$ is the mean energy (in \GeVns{}) per unit area of ($\eta, \phi$)
for background from pileup, computed event by event using the \textsc{FastJet} package~\cite{fastjet}.

The $\PW\gamma$ analysis uses individual $I_r$ contributions from the three subdetectors.
Also, to minimize the contributions from $\cPZ\gamma$ events,
a less restrictive selection is applied to the additional electron.
The efficiencies for these criteria are measured in
$\cPZ\to \Pe\Pe$ data and in MC simulation,
using the ``tag-and-probe'' technique of Ref.~\cite{tagandprobe}.
An efficiency correction of ${\approx}3\%$ is applied to the MC
simulation to match the performance observed in data.

\subsection{Photon identification and selection}
\label{ss:PhotonSelection}

Photon candidates in the fiducial volume of the ECAL detector are
reconstructed as SC with efficiencies very close to 100\%
for $\pt^{\gamma} > 15\GeV$,
as estimated from MC simulation. The photon energy scale is measured using
$\cPZ\to\mu\mu\gamma$ events, following the ``PHOSPHOR'' procedure
described in Ref.~\cite{Veverka,HiggsLongPaper}.

As in the previous CMS analysis of $\vb\gamma$ final states~\cite{cms_vgamma},
we reduce the rate of jets misreconstructed as photons by using stringent
photon identification criteria, including isolation and
requirements on shapes of electromagnetic (EM) showers.
In particular, (i)~the ratio of HCAL to ECAL energies
deposited within a cone of $\Delta R = 0.15$ relative to the axis of the
seed ECAL crystal must be $<$0.05, (ii)~the value of \sihih must
be $<$0.011 in the barrel and $<$0.030 in the endcap, and
(iii)~to reduce background from misidentified electrons, photon candidates are
rejected if there are hits present in the first two inner layers of the silicon
pixel detector that can originate from an electron trajectory that extrapolates
to the location of the deposited energy in an ECAL SC of $\ET > 4$\GeV.
This requirement is referred to as the pixel veto.

However, unlike in the previous analysis~\cite{cms_vgamma}, the pileup conditions
during Run 2011 require modifications to photon isolation criteria to achieve
reliable modeling of pileup effects.
In particular, for photon candidates, the scalar sum of the $\pt$ for all
tracks originating from within $\abs{d_z} < 0.1\unit{cm}$ of the primary vertex,
that have $\abs{d_\mathrm{T}} < 0.02\unit{cm}$, and are located within a
$0.05 < \Delta R < 0.4$ annulus of the direction of each photon candidate,
are required to have
$\pt^\text{TRK}<2\GeV+0.001 \times \pt^{\gamma} + A_\text{eff} \times \rho$,
where $A_\text{eff}$ is the effective area used to correct each photon
shower for event pileup.
This procedure ensures that the isolation requirement
does not exhibit a remaining dependence on pileup.
For each photon candidate, the scalar sum of the $\pt$ deposited in the ECAL in an annulus
$0.06< \Delta R<0.40$, excluding a rectangular strip of
$\Delta\eta\times\Delta\phi=0.04\times 0.40$ to reduce the impact of energy leakage from
any converted $\gamma\to \EE$ showers, is computed.
The isolation in the ECAL is required to have
$\pt^\text{ECAL}<4.2\GeV+0.006\times \pt^\gamma+A_\text{eff}\times\rho$,
and, finally, the isolation criterion in the HCAL is
$\pt^\text{HCAL} < 2.2\GeV+0.0025\times \pt^\gamma+A_\text{eff}\times\rho$.
The expected values of $A_\text{eff}$ are defined by the ratio of slopes obtained in fits of
the isolation and $\rho$ parameters to a linear dependence on the number of vertices observed
in data. These are summarized in Table~\ref{tab:aeff}, separately for the three isolation
parameters, calculated for EM showers observed in the barrel and endcap regions of the ECAL.

\begin{table}[htb]
  \centering
    \topcaption{The values of $A_\text{eff}$, in units of $\Delta\eta\times\Delta\phi$, used
      to correct contributions from pileup to the summed $\pt$ accompanying photon
      candidates in the tracker and the two calorimeters, separately for photons observed in the
      barrel and endcap regions of the ECAL.}
    \label{tab:aeff}
    \begin{scotch}{lll}
      Isolation  & Barrel  & Endcap  \\ \hline
      Tracker    & 0.0167  & 0.032 \\
      ECAL       & 0.183   & 0.090 \\
      HCAL       & 0.062   & 0.180 \\
    \end{scotch}
\end{table}

\begin{figure}[!hbt]
\begin{center}
\includegraphics[width=0.49\textwidth]{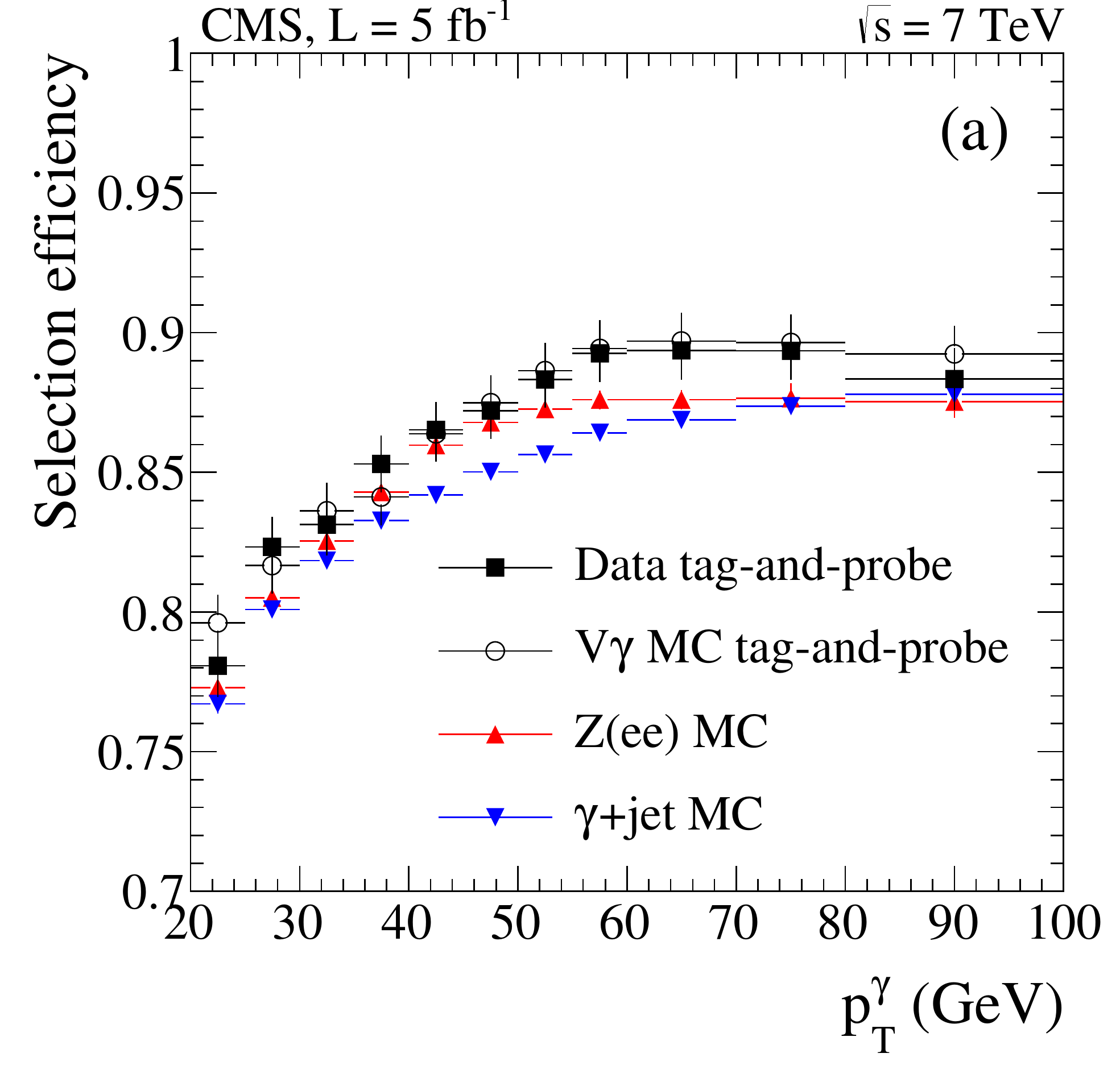}
\includegraphics[width=0.49\textwidth]{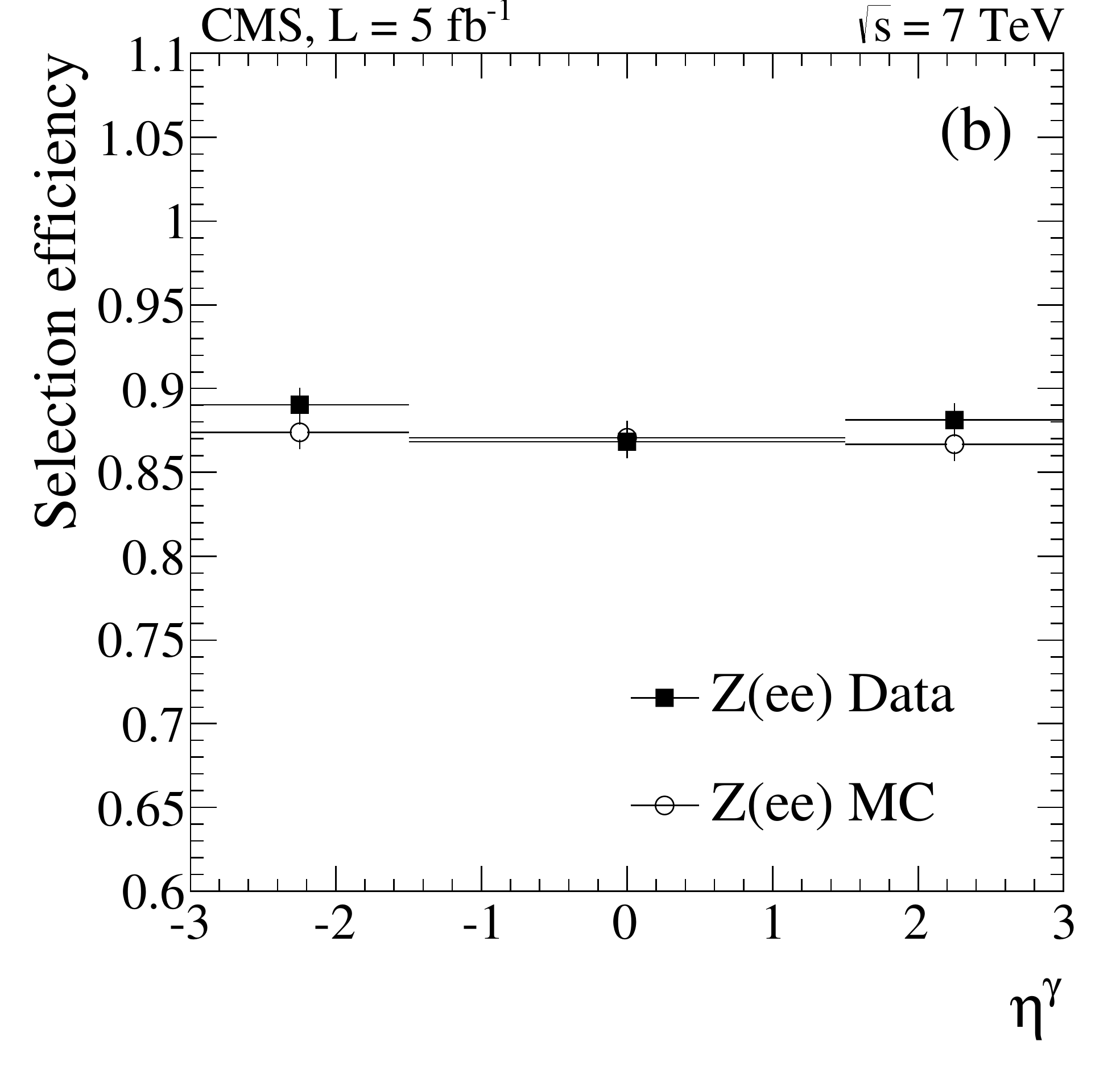}
\caption{Efficiency of photon selection, as a function of (a)~photon transverse momentum and (b)~photon pseudorapidity.}
\label{fig:photonID_ZeeFits}
\end{center}
\end{figure}

\begin{figure}[!hbt]
\begin{center}
\includegraphics[width=0.49\textwidth]{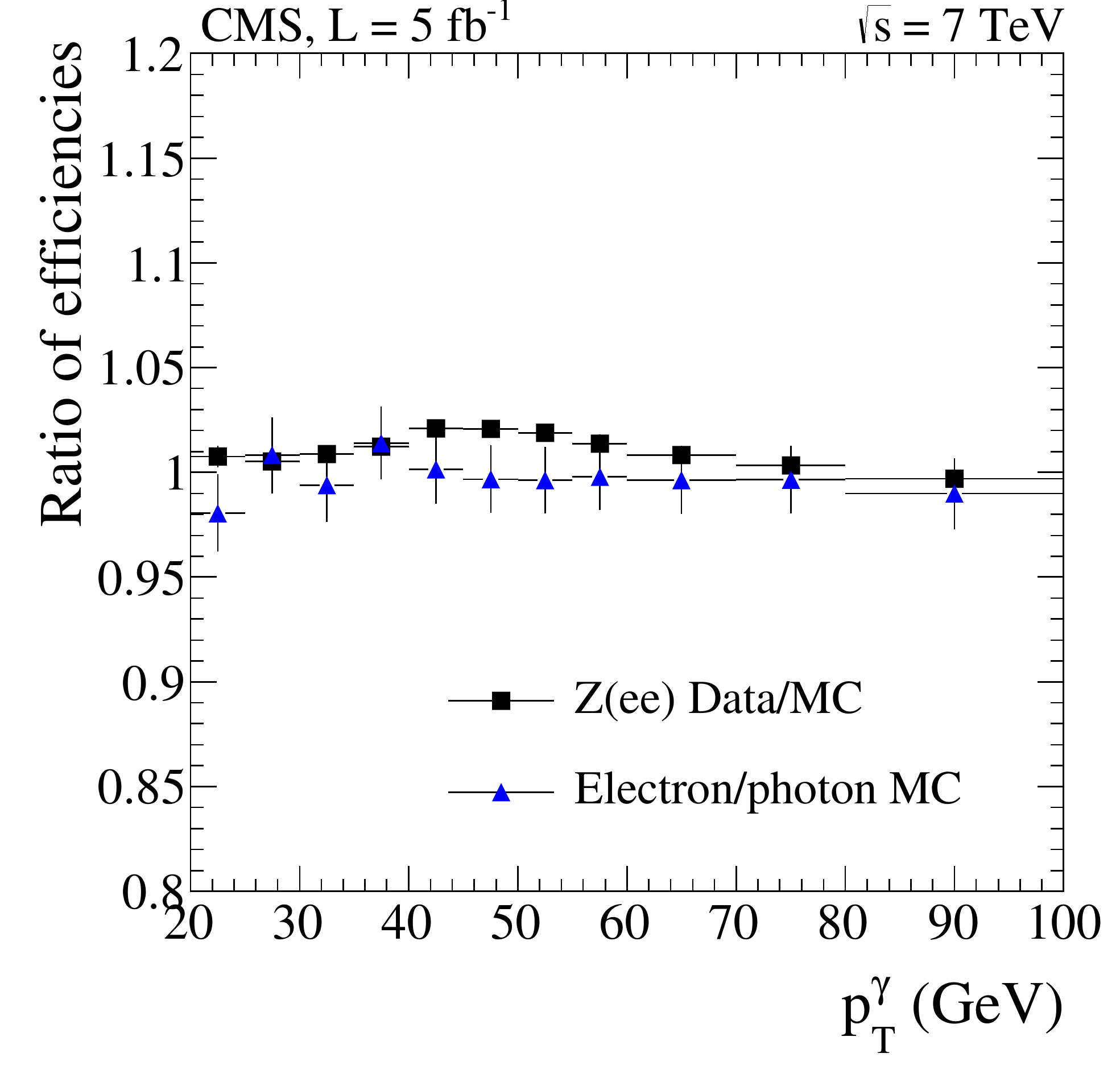}
\caption{Ratio of efficiencies for selecting photons in data relative to MC simulation,
obtained through the tag-and-probe method, and the ratio of electron to photon efficiencies,
obtained at the MC generator level, with both sets of ratios given as a function of the transverse momentum of the photon.}
\label{fig:photonID-SF}
\end{center}
\end{figure}

To estimate the efficiency of requirements on the shape and isolation of EM showers,
we use the similarity of photon and electron showers to select common shower parameters
based on electron showers, but use the isolation criteria that consider
differences between the photon and electron characteristics.
The results for data and MC simulation, as a function of $\pt^{\gamma}$
and $\eta^{\gamma}$, are shown in Fig.~\ref{fig:photonID_ZeeFits}.
The efficiencies obtained using generator-level information in $\cPZ\to \Pe\Pe$
and in $\gamma$+jets simulations are also shown in Fig.~\ref{fig:photonID_ZeeFits}.
The difference between these efficiencies is taken as an estimate of
systematic uncertainty in the photon-identification efficiency,
based on results from $\cPZ\to \Pe\Pe$ data.
The ratios of efficiency in data to that in simulation, both measured
by the tag-and-probe method (squares), and efficiency in
$\cPZ\to \Pe\Pe$ simulation to that in the $\gamma$+jets
simulation obtained from generator-level information (triangles), as a
function of $\pt$, integrated over the
full range of $\eta$, are shown in Fig.~\ref{fig:photonID-SF}.
We find that the efficiencies in data and MC simulation agree to within 3\%
accuracy. As for the case of electrons and muons, we reweight the simulated events to
reduce the residual discrepancy in modeling efficiency as a function of
$\pt^{\gamma}$ and $\eta^{\gamma}$.

The efficiency of the pixel veto is obtained from $\cPZ\to\mu\mu\gamma$
data, where the photon arises from final-state radiation. The purity of
such photon candidates is estimated to exceed 99.6\%, and they are therefore
chosen for checking photon-identification efficiency, energy scale,
and energy resolution. We find that the efficiency of the pixel veto
corresponds to 97\% and 89\% for photons in the barrel
and endcap regions of the ECAL, respectively.

\subsection{Muon identification and selection}
\label{sec:muonid}
Muons are reconstructed offline by matching particle trajectories in the
tracker and the muon system.
The candidates must have $\pt > 35$ and $> 20\GeV$
for the $\PW\gamma$ and $\cPZ\gamma$ analyses, respectively.
We require muon candidates to pass
the standard CMS isolated-muon selection criteria~\cite{muperf}, with minor
changes in requirements on the distance of closest approach of the muon
track to the primary vertex. We require $\abs{d_z} < 0.1\unit{cm}$, in the longitudinal
direction, and $\abs{d_\mathrm{T}} < 0.02\unit{cm}$, in the transverse plane.
The efficiencies for these criteria are measured in data and in MC simulation
using a tag-and-probe technique applied to $\cPZ\to\mu\mu$ events.
An efficiency correction of ${\approx}3\%$ is also applied to the MC
simulation to match the performance found in muon data.

\subsection{Reconstruction of \texorpdfstring{$\ETslash$}{missing ET}}
\label{ss:met}
Neutrinos from $\PW\to \ell\nu$ decay are not detected directly,
but give rise to an imbalance in reconstructed transverse momentum in an
event. This quantity is computed using objects reconstructed with the
particle-flow algorithm~\cite{PFMET}, which generates a list of
four-vectors of particles
based on information from all subsystems of the CMS detector.
The $\ETslash$ for each event is defined by the magnitude of the
vector sum of the transverse momenta of all the reconstructed particles.

\subsection{Trigger requirements}
\label{sec:triggers}

The $\PW\gamma \to \ell\nu\gamma$ and $\cPZ\gamma \to \ell\ell\gamma$
events are selected using unprescaled, isolated-lepton triggers.
The $\pt$ thresholds and isolation criteria imposed on
lepton candidates at the trigger level changed with time to accommodate
the instantaneous luminosity, and are less stringent than the offline requirements.

For the $\PW\gamma \to \Pe\nu\gamma$ channel, we use an isolated,
single-electron trigger, requiring electrons with $\abs{\eta} < 3$, and a $\pt$
threshold of 32\GeV, except for the
first part of Run 2011A (\rlumi = 0.2\fbinv), where the threshold is 27\GeV.
In addition, for the last part (\rlumi = 1.9\fbinv) of Run 2011A and the entire
Run 2011B, a selection is implemented on the transverse mass ($\MT^\PW$)
of the system consisting of the electron candidate and the $\ETslash $, requiring
$\MT^\PW = \sqrt{\smash[b]{2 \pt^{\ell} \ETslash
 (1 - \cos\Delta\phi(\ell,\ETslash))}} > 50\GeV$,
where $\Delta\phi$ is the angle between the $\pt^{\ell}$ and the
$\ETslash$ vectors.
The trigger used for the $\cPZ\gamma \to \Pe\Pe\gamma$ events requires two isolated electron candidates
with $\pt$ thresholds of 17\GeV on the leading (highest-$\pt$) candidate and 8\GeV
on the trailing candidate.

The trigger for $\PW\gamma\to\mu\nu\gamma$ events requires an isolated muon
with $\pt > 30\GeV$ and $\abs{\eta} < 2.1$. The dimuon trigger used to collect
$\cPZ\gamma\to\mu\mu\gamma$ events does not require the two muons to be isolated,
and has coverage for $\abs{\eta} < 2.4$.
For most of the data, the muon $\pt$ thresholds are
13\GeV for the leading and 8\GeV for the trailing candidates.
For the first part of Run 2011A (\rlumi= 0.2\fbinv) and for most of the remaining data,
these thresholds are 7\GeV for each muon candidate, except for the last part of
Run 2011B (\rlumi = 0.8\fbinv), where these increase to 17 and 8\GeV, respectively.

\subsection{\texorpdfstring{$\PW\gamma$}{W gamma} event selections}
\label{ss:WlnugammaSelection}

The $\PW\gamma\to \ell\nu\gamma$ process is characterized by a prompt, energetic,
and isolated lepton, a prompt isolated photon, and significant $\ETslash$ that
reflects the escaping neutrino. Both electrons and muons are required to have
$\pt > 35\GeV$, and photons to have $\pt > 15\GeV$.
The maximum allowed $\abs{\eta}$ values for electrons, photons, and muons
are 2.5, 2.5, and 2.1, respectively. We require the photon to be
separated from the lepton by $\Delta R(\ell,\gamma) > 0.7$.
To minimize contributions from $\cPZ\gamma\to \ell\ell\gamma$ production,
we reject events that have a second reconstructed lepton of the same flavor.
This veto is implemented only for electrons that have $\pt > 20\GeV$,
$\abs{\eta}<2.5$, and pass looser electron selections,
and for muons that have $\pt > 10\GeV$ and $\abs{\eta}<2.4$.

To suppress background processes without genuine $\ETslash$,
we require events to have $\MT^\PW > 70\GeV$.
We find that the $\ETslash $ distribution is well modeled,
but we apply a small efficiency correction to reduce the residual disagreement.
The efficiencies of the $\MT^\PW$ selection in data and simulation agree
at the 1\% level. The full set of $\ell\nu\gamma$ selections yield 7470 electron
and 10\,809 muon candidates in the data.
The selection criteria used to define the $\PW\gamma$ sample are summarized
in Table~\ref{tab:event_selections}.

\subsection{\texorpdfstring{$\cPZ\gamma$}{Z gamma} event selections}
\label{sec:Zgamma}
Accepted $\cPZ\gamma$ events are characterized by two prompt, energetic, and
isolated leptons, and an isolated prompt photon.
Both electrons and muons are required to have $\pt > 20\GeV$,
and the photons to have $\pt > 15\GeV$.
The maximum $\abs{\eta}$ values for accepted electrons, photons, and muons
are 2.5, 2.5, and 2.4, respectively.
We require photons to be separated from leptons by imposing
a $\Delta R(\ell,\gamma) > 0.7$ requirement.
Finally, the invariant mass of the two leptons is required to satisfy $m_{\ell\ell} > 50\GeV$.
Applying all these selections yields 4108 $\cPZ\gamma\to \Pe\Pe\gamma$
and 6463 $\cPZ\gamma \to \mu\mu\gamma$ candidates.
The selection criteria used to define the $\cPZ\gamma$ sample are summarized
in Table~\ref{tab:event_selections}.

\begin{table*}[hbt]
  \centering
  \topcaption{Summary of selection criteria used to define the $\PW\gamma$ and $\cPZ\gamma$ samples.}
  \label{tab:event_selections}
  \begin{scotch}{lcccc}
    Selection                 & $\PW\gamma\to \Pe\nu\gamma$ & $\PW\gamma\to \mu\nu\gamma$ & $\cPZ\gamma\to \Pe\Pe\gamma$ & $\cPZ\gamma \to \mu\mu\gamma$ \\
    \hline
    Trigger                   & single electron                     & single muon                     & dielectron                  & dimuon \\
    $\pt^{\ell}$~(\GeVns{}) & ${>}35$                              & ${>}35$                          & ${>}20$                      & ${>}20$ \\
    $|\eta^{\ell}|$           & EB or EE                            &  ${<}2.1$                        & EB or EE                    & ${<}2.4$ \\
    $\pt^{\gamma}$~(\GeVns{})    & ${>}15$                              & ${>}15$                          & ${>}15$                      & ${>}15$ \\
    $|\eta^{\gamma}|$         & EB or EE                            & EB or EE                        & EB or EE                    & EB or EE \\
    $\Delta R(\ell,\gamma)$   & ${>}0.7$                            & ${>}0.7$                        & ${>}0.7$                    & ${>}0.7$ \\
    $\MT^{\PW}$~(\GeVns{})       & ${>}70$                                & $> 70$   &   &  \\
    $m_{\ell\ell}$~(\GeVns{})    &  &      & ${>}50$ & ${>}50$ \\
    Other criterion         & only one lepton & only one lepton &  &  \\
  \end{scotch}
\end{table*}

\section{Background estimates}
\label{sec:dataDrivenBackgrounds}

The dominant background for both $\PW\gamma$ and $\cPZ\gamma$ production
arises from events in which jets, originating mostly from
\PW+jets and \cPZ+jets events, respectively, are misidentified as photons.
We estimate the background from these sources as a function of $\pt^{\gamma}$
using the two methods described in Section~\ref{ssec:jetmisid}.

For the $\PW\gamma$ channel, a second major background arises from Drell--Yan
($\cPq\cPaq\to\ell^{+}\ell^{-}$) and EW diboson production,
when one electron is misidentified as a photon.
This background is estimated from data as described in Section~\ref{sss:eltophoton}.

Other backgrounds to $\vb\gamma$ processes include
(i)~jets misidentified as leptons in $\gamma$+jet production,
(ii)~$\vb\gamma$ events, with \vb decaying into $\tau\nu$ or $\tau\tau$, and
subsequently $\tau\to\ell\nu\nu$,
(iii)~$\ttbar\gamma$ events, and (iv)~$\cPZ\gamma$ events, where one of the leptons from \cPZ\ decay is not reconstructed properly.
All these backgrounds are small relative to
the contribution from $\vb$+jets, and are estimated using MC simulation.

\subsection{Jets misidentified as photons}
\label{ssec:jetmisid}

\subsubsection{Template method}
\label{ssec:template_method}
The template method relies on a maximum-likelihood fit to the
distribution of \sihih in data to estimate the background
from misidentified jets in the selected $\vb\gamma$ samples.
The fit makes use of the expected distributions
(``templates'') for genuine photons and misidentified jets.
For isolated prompt photons, the \sihih distribution
is very narrow and symmetric, while for photons produced in hadron decays, the \sihih distribution
is asymmetric, with a slow falloff at large values.
The distribution in \sihih for signal photons is obtained from simulated $\PW\gamma$ events.
The \sihih distribution of electrons from \cPZ\ boson decays in data is observed to be shifted
to smaller values relative to simulated events. The shift is 0.9$\times10^{-4}$ and 2.0$\times 10^{-4}$
for the EB and EE regions, respectively, and corresponds to 1\% and 0.8\% shifts in the
average of the simulated photon \sihih values, which are corrected for the shift
relative to data.

The \sihih templates for background are defined by events in a background-enriched isolation sideband of data.
These photon candidates are selected using the photon identification criteria
described in Section~\ref{ss:PhotonSelection}, but without the \sihih selection,
and with inverted TRK isolation requirements:
(i)~$2\GeV < \pt^\text{TRK} - 0.001 \times
\pt^{\gamma}-0.0167 \times \rho < 5\GeV$, for $|\eta_\gamma| < 1.4442$,
and (ii)~$2\GeV < \pt^\text{TRK} - 0.001 \times \pt^{\gamma} - 0.0320 \times \rho < 3\GeV$,
for $1.566 < |\eta_\gamma| < 2.5$.
These requirements ensure that the contributions from genuine photons are negligible,
while the isolation requirements remain close to those used for selection of photons
and thereby provide jets with large EM energy fractions that have properties
similar to those of genuine photons. We observe that \sihih is
largely uncorrelated with the isolation parameter in simulated multijet
events, so that the
distribution observed for background from jets that are misidentified
as photons (\ie, with inverted tracker isolation criteria) is expected
to be the same as that for jets misidentified as isolated photons.

\begin{figure}[!hbt]
\begin{center}
\includegraphics[width=0.49\textwidth]{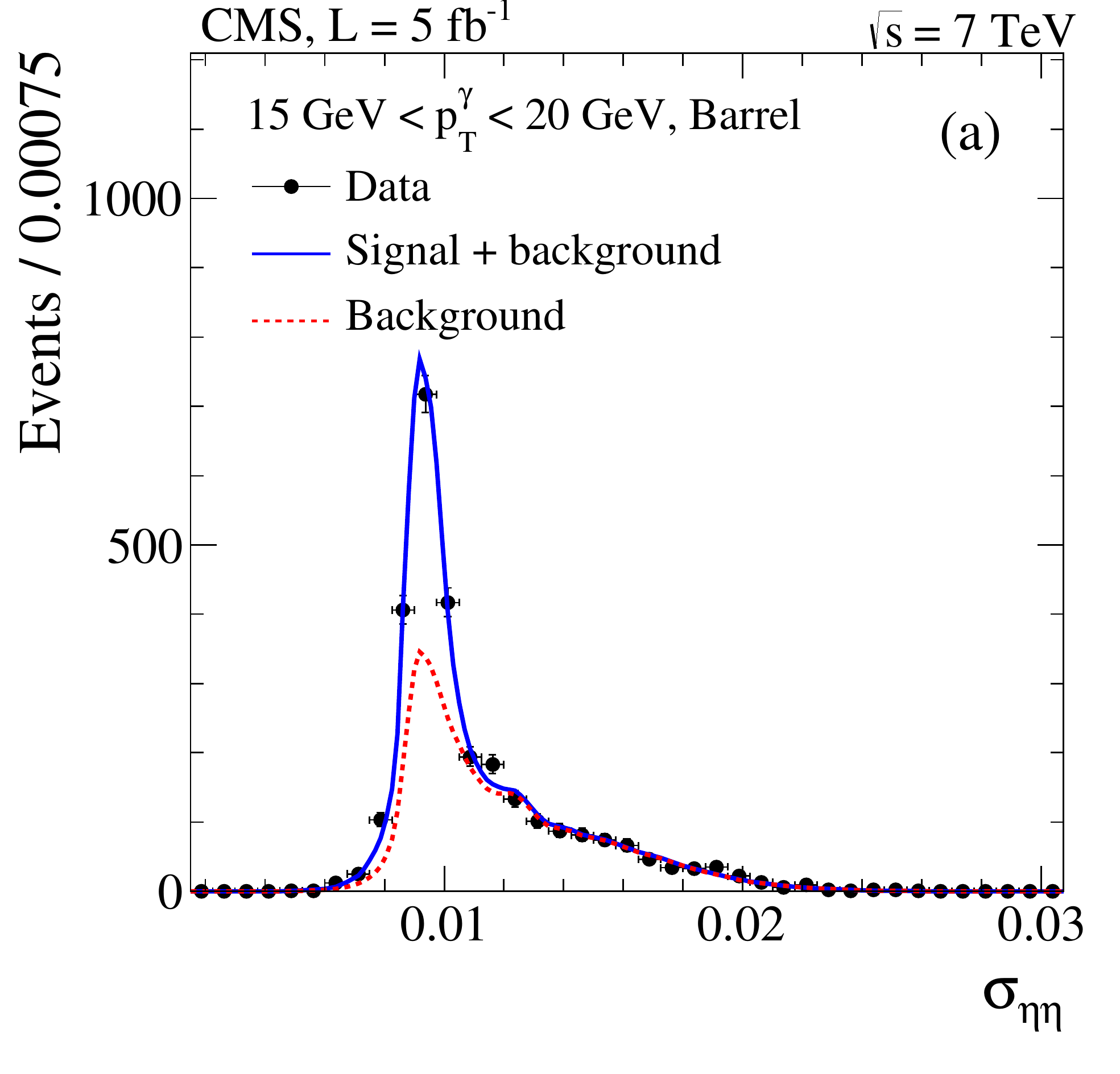}
\includegraphics[width=0.49\textwidth]{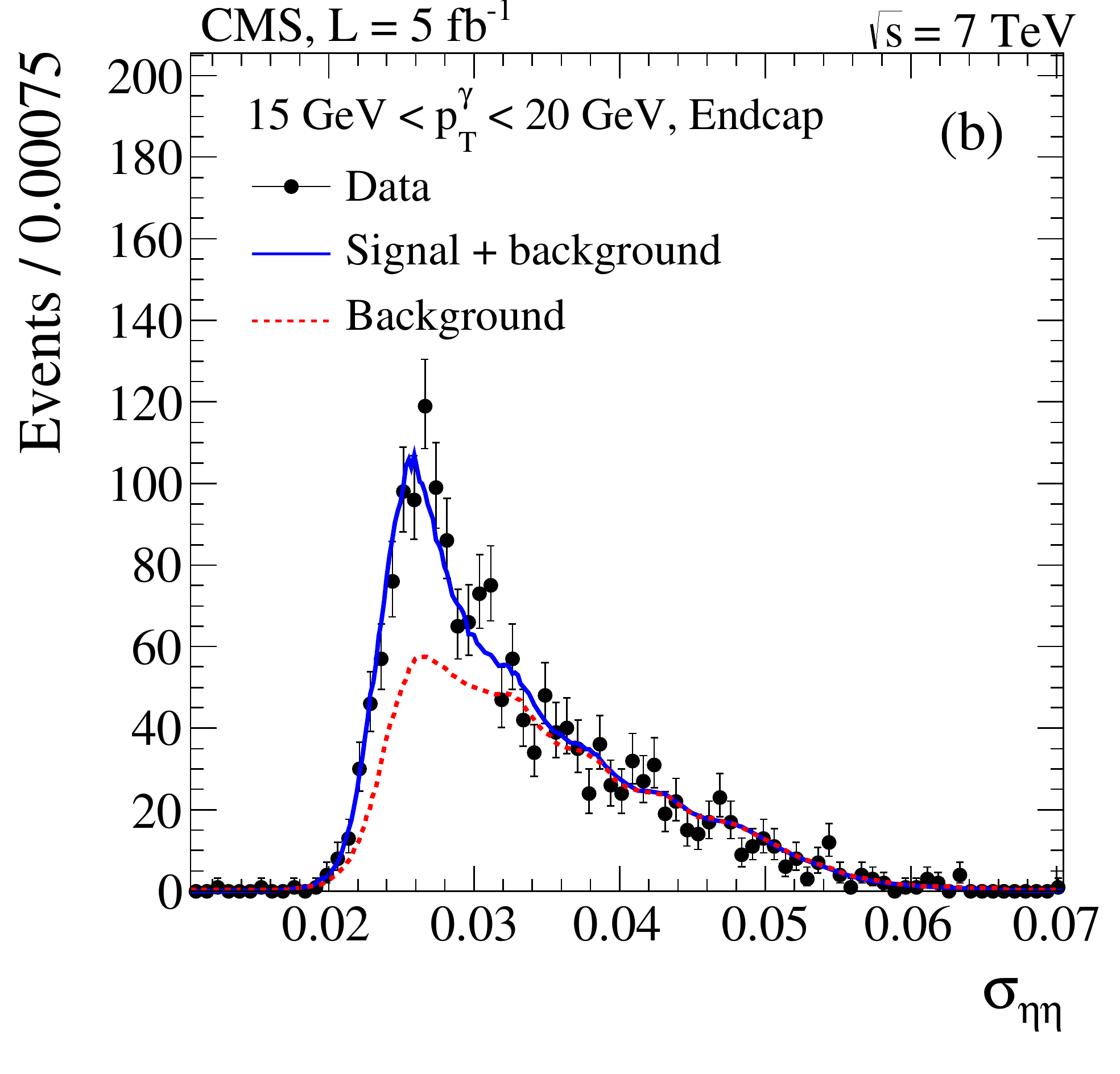}
\caption{Fit to the $\sihih$ distribution for photon candidates with
$15 < \pt^\gamma < 20\GeV$ in data with signal and background
templates in the (a)~barrel and (b)~endcaps.}
\label{fig:templateFits}
\end{center}
\end{figure}

Because of the $\MT^\PW$ requirement in selected $\PW\gamma$ events, the
presence of significant $\ETslash$ can bias the estimation of the background.
We therefore investigate possible correlations between the distribution
in \sihih for background events and the projection of $\ETslash$ along
the $\pt$ of jets misidentified as photons.
In particular, we define \sihih templates for background using events in data
with $\ETslash > 10\GeV$ and with the direction of the $\ETslash$ vector
along the photon-like jet. The estimated systematic uncertainty is
obtained from the smallest bin in $\pt^{\gamma}$ ($15 < \pt^{\gamma} < 20\GeV$),
as this is the bin that contains most of the background (Fig.~\ref{fig:templateFits})
and corresponds to the largest control sample for input to the \sihih template representing the background.
Based on the modified templates, we assign a systematic uncertainty that reflects
the largest discrepancy relative to the nominal yield, which is found
to be 13\% and 7\% for the barrel and endcap, respectively. A more detailed
discussion of systematic uncertainties in the background estimate is given
in Section~\ref{sec:systematics}.

The systematic uncertainty in electron misidentification
is estimated through changes made in the modeling of signal and background,
the electron and photon energy resolutions, and the distributions for pileup in MC
simulations.

The function fitted to the observed distribution of \sihih is
the sum of contributions from signal ($S$) and background ($B$):
\ifthenelse{\boolean{cms@external}}{
\begin{multline}
N_{S}S(\sihih) + N_{B}B(\sihih) =
N \Bigg[\frac{N_S}{N}S(\sihih)+\\ \left(1-\frac{N_S}{N}\right)B(\sihih)\Bigg],
\end{multline}
}{
\begin{equation}
N_{S}S(\sihih) + N_{B}B(\sihih) =
N \left[\frac{N_S}{N}S(\sihih)+\left(1-\frac{N_S}{N}\right)B(\sihih)\right],
\end{equation}
}
where $N$, $N_{S}$, and $N_{B}$ are the total number of events and
the numbers of signal and background candidates in data for any
given bin of $\pt^\gamma$, respectively. The $S(\sihih)$ and $B(\sihih)$
represent the expected signal and background distributions in \SEE.
These distributions are smoothed using a kernel-density
estimator~\cite{Cranmer:2000du}, or
through direct interpolation when the statistical uncertainties are small, which
makes it possible to use unbinned fits to the data in regions where statistics are poor,
while preserving the good performance of the fit.
The fit uses an unbinned extended likelihood~\cite{pdg2012} function ($L$)
to minimize $-\ln L$ as a function of the signal fraction $f_S = N_S/N$:
\begin{equation}
-\ln{L} = (N_S + N_B) -
 \ln[f_S S(\sihih)  + (1 - f_S) B(\sihih) ].
\label{eq:logl}
\end{equation}

\subsubsection{Ratio method}
\label{ssec:ratio_method}

We use a second method, referred to as the ``ratio method,'' to infer the V+jets
background as a cross-check of the results obtained with the template method at
large $\pt^\gamma$, where the template method is subject to larger
statistical uncertainties. The ratio method uses $\gamma$+jets and multijet data
to extract the misidentification rate, taking into account the quark/gluon composition
of the jets in $\vb$+jets events.

The ratio method exploits a category of jets that have properties similar
to electromagnetic objects in the ECAL, and are called photon-like jets.
Photon-like jets are jets selected through the presence of photons that pass
all photon selection criteria, but fail either the photon isolation
or $\sihih$ requirements. However, these kinds of jets are still more isolated and
have higher EM fractions than most generic jets.

The ratio method provides a ratio $R_p$ of the probability for a jet to pass
photon selection criteria to that of passing photon-like requirements.
Once $R_p$ is known, the number of jets that satisfy the final photon selection
criteria ($N_{\vb+\text{jets}}$) can be estimated as the product of $R_p$ and the number
of photon-like jets in data.

\begin{figure}[!hbt]
\begin{center}
  \includegraphics[width=0.5\textwidth]{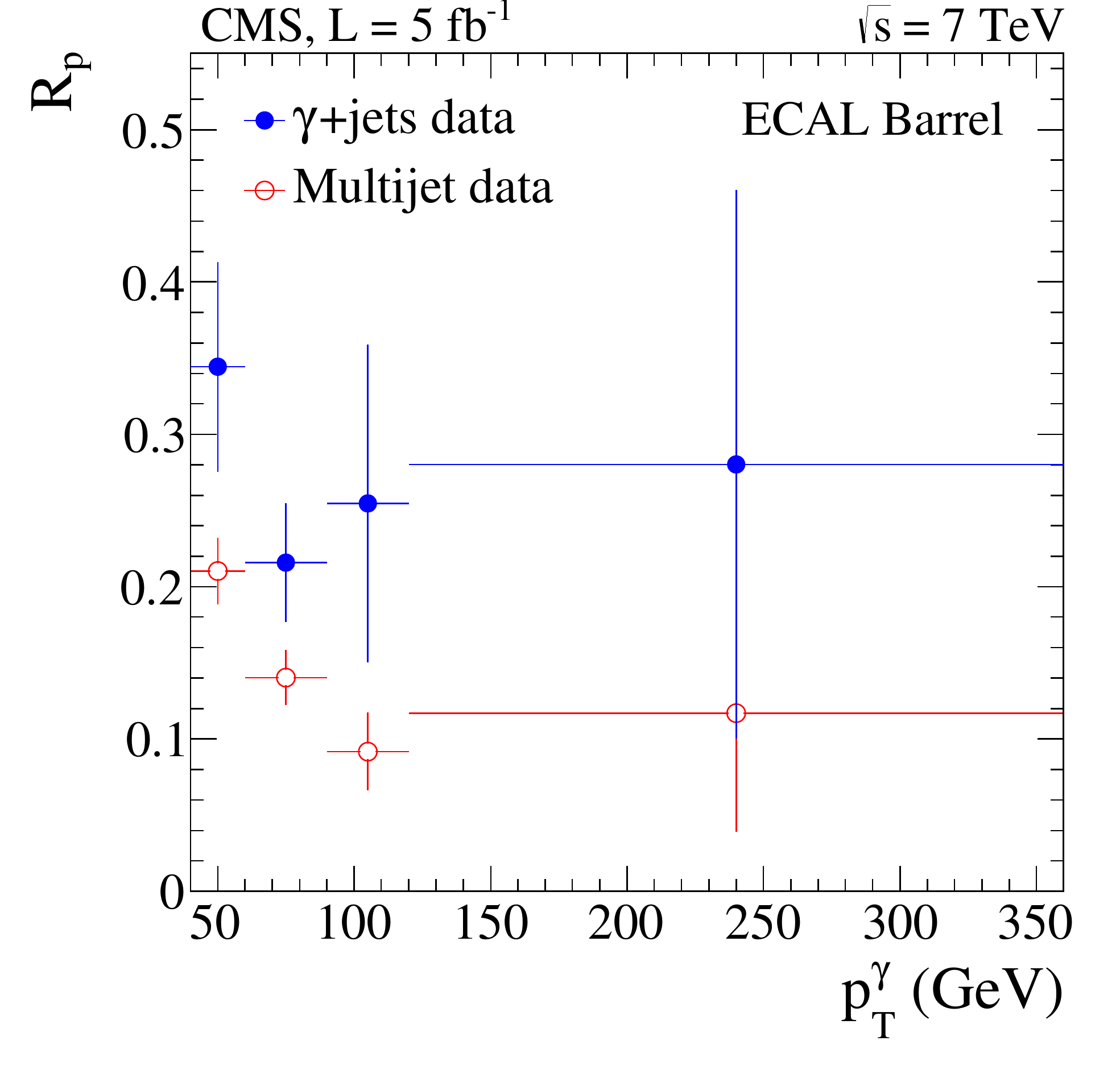}
  \caption{The $R_p$ ratio (described in text) as a function of the $\pt$ of photon candidates for the barrel region
of the ECAL in $\gamma$+jets and multijet data.
The difference in $R_p$ values for the two processes is attributed to the fact that jets in
$\gamma$+jets events
are dominated by quark fragmentation, while jets in multijet events are dominated by gluon fragmentation.}
  \label{fig:FoverE2011ABbarrel}
\end{center}
\end{figure}

We measure $R_p$ separately for each $\pt^\gamma$ bin of the analysis
both for the barrel and endcap regions of the ECAL, using ``diphoton'' events,
defined by the presence of either two photon candidates that pass the final photon
selections, or that have one photon candidate that passes
the final selections and one that passes only photon-like jet selections.
To reduce correlations induced by the diphoton production kinematics, we require
that the photons corresponding to each diphoton candidate be in  the same $\eta$ region
and $\pt^\gamma$ bin.
A two-dimensional fit is performed based on templates of distributions in $\sihih$ of each
photon candidate to estimate $R_p$, and thereby subtract the contribution from genuine photons
to the photon-like jet yield. As only 5--10\% of
genuine photons in multijet events pass photon-like jet requirements,
we correct the distribution in $R_p$ using MC simulation of multijet events, and check
the correction through $\cPZ\to \Pe\Pe$ data and simulation.

The observed $R_p$ values for the barrel region of the ECAL are given in
Fig.~\ref{fig:FoverE2011ABbarrel} as a function of $\pt^\gamma$.
The difference between the two sets of $R_p$ values extracted in different ways indicates the
sensitivity of the method to whether the photon-like jet originates from hadronization
of a quark or a gluon. We use the simulation
of the gluon-to-quark jet ratio in $\PW$+jets and $\cPZ$+jets events to correct $R_p$
as a function of the $\pt$ of the photon-like jet. We find the predictions from the ratio method
to be consistent with those from the template method, and consider their difference as
an additional source of systematic uncertainty in the analysis.

\subsection{Background from electrons misidentified as photons in \texorpdfstring{$\ell\nu\gamma$}{l nu gamma} events}
\label{sss:eltophoton}

The criterion that differentiates electrons from photons is the presence in the
pixel detector of a track that is associated with a shower in the ECAL.
We use $\cPZ\to \Pe\Pe$ data to measure the probability ($P_{\Pe\to\gamma}$)
for an electron not to have a matching track by requiring one of the electrons to pass
stringent electron identification criteria, and then by checking how often the other
electron passes the full photon selection criteria, including the requirement
of having no associated track in the pixel detector. Fitting to the $m_{\ell\ell}$
distribution using a convolution of Breit--Wigner and
``Crystal Ball''~\cite{CB} functions to describe the signal and a falling exponential
function for background, we obtain the probability for an electron to have no associated
track as $P_{\Pe\to\gamma} = 0.014 \pm 0.003 \syst$ and $0.028 \pm 0.004 \syst$
for the barrel and the endcap regions, respectively.

To estimate the background from sources where an electron is misidentified
as a photon in the $\mu\nu\gamma$ channel, we select events that pass all event selection
criteria except that the presence of a track in the pixel detector associated with
the photon candidate is ignored.
The contribution from genuine electrons misidentified as photons can therefore be calculated as
\begin{equation}
N_{\Pe\to\gamma} = N_{\mu \nu \Pe} \times
                 \frac{P_{\Pe\to\gamma}}{1 - P_{\Pe\to\gamma}},
\end{equation}

where $N_{\Pe\to\gamma}$ is the background from misidentified electrons
and $N_{\mu \nu \Pe}$ is the number of events selected without any requirement
on the pixel track.
The systematic
uncertainties associated with this measurement are discussed in detail
in Section~\ref{sec:systematics}.

The background in the $\Pe\nu\gamma$ channel is dominated by \cPZ+jets events,
where one of the electrons from $\cPZ\to \Pe\Pe$ decays is misidentified
as a photon. To estimate the $\cPZ\to \Pe\Pe$ contribution to the
$\PW\gamma\to \Pe\nu\gamma$ signal, we apply the full selection
criteria and fit the invariant mass of the photon and electron candidates
with a Breit--Wigner function convolved with a Crystal Ball function for
the \cPZ\ boson, and an exponential form for the background.
Contributions to $\Pe\nu\gamma$ events from other sources with genuine electrons
misidentified as photons (\eg, $\ttbar$+jets and diboson
processes) are estimated using MC simulation, in which a photon candidate
is matched spatially to the generator-level electron.

\subsection{Total background}

The background from jets that are misidentified as photons is summarized
as a function of $\pt$ of the photon in
Table~\ref{tab:wmng_templ_syst_2011AB} for $\ell\nu\gamma$ events
and in Table~\ref{tab:zg_templ_syst_2011AB} for $\ell\ell\gamma$ events,
and the sums are listed as $N_B^{\PW+\text{jets}}$ in
Table~\ref{tab:wengParameters} and as $N_B^{\cPZ+\text{jets}}$ in
Table~\ref{tab:zg_results}. The background from electrons in selected
$\ell\nu\gamma$ events that are misidentified as
photons, $N_B^{\Pe\Pe X}$, is summarized in Table~\ref{tab:wmng_templ_syst_2011AB}
for both $\Pe\nu\gamma$ and $\mu\nu\gamma$ channels. The $N_B^{\text{other}}$
in Tables~\ref{tab:wengParameters} and~\ref{tab:zg_results} indicates the rest
of the background contributions estimated from simulation.
For the $\Pe\nu\gamma$ channel, the largest contribution to $N_B^{\text{other}}$ (53\%)
is from $\cPZ\gamma$ events, and the next largest is from $\gamma$+jets with a contribution of 33\%.
For the $\mu\nu\gamma$ channel, the dominant background to $N_B^{\text{other}}$ is
from $\cPZ\gamma$, with a contribution of 84\%. All the specific parameters
will be discussed in more detail in Sections~\ref{sec:systematics}--\ref{ssec:zgammaXS}.

\begin{table*}[hbt]
  \scriptsize
  \centering
  \topcaption{Yield of misidentified photons from jets in W+jets events and their symmetrized associated
    systematic uncertainties as a function of $\pt^{\gamma}$
    in the $\PW\gamma\to \ell\nu\gamma$ analyses. The results
    are specified in the second column by the numbers of events expected
    in the $\Pe\nu\gamma$ and $\mu\nu\gamma$ channels, and by the uncertainties
    from each of the sources in the rest of the columns.}
  \label{tab:wmng_templ_syst_2011AB}
  \begin{scotch}{lcccccc}
    & & \multicolumn{5}{c}{Systematic uncertainties on yields ($\Pe\nu\gamma$/$\mu\nu\gamma$)} \\ \hline
    $\pt^{\gamma}$ &  Yield from       & Shape of        & Shape of    & Sampling of   & Correlation of    & Diff. between \\
    (\GeVns{})          &  W+jets events & $\gamma$ shower & jet shower  & distributions & $\gamma$ and \ETslash & $\text{jet}\to\gamma$ predictions \\ \hline
    15--20          & 1450 / 2760    &  9.3 / 21     & 83 / 159  & 19 / 36   & 130 / 250   &  \\
    20--25          &  650 / 1100    &  5.2 / 20     & 37 / 63   & 11 / 19   & 54 / 94     &  \\
    25--30          &  365 / 520     &  3.7 / 9.4    & 21 / 30   & 9.4 / 14  & 33 / 43     &  \\
    30--35          &  220 / 330     & 10.5 / 3.3    & 12 / 19   & 7.5 / 11  & 19 / 29     &  \\
    35--40          &  160 / 200     &  3.4 / 2.8    & 10 / 12   & 6.2 / 7.9 & 14 / 16     &  \\
    40--60          &  220 / 270     &  3.5 / 0.7    & 19 / 23   & 5.1 / 6.3 & 19 / 24     & 22  / 4.4 \\
    60--90          &   77 / 100     &  1.4 / 0.9    & 10 / 13   & 3.0 / 3.8 & 6.6 / 8.5   & 7.7 / 1.6 \\
    90--120         &   26 / 21      &  2.0 / 2.3    & 5.3 / 4.1 & 0.9 / 0.9 & 2.4 / 1.8   & 2.6 / 0.4 \\
    120--500        &   15 / 38      &  4.3 / 2.1    & 7.6 / 26  & 1.1 / 0.7 & 1.0 / 3.9   & 1.5 / 0.6 \\
    \hline
    \multirow{2}{*}{Totals}   & \multirow{2}{*}{3180 / 5350} & 17 / 30 & 98 / 179  & 27 / 45 & 280 / 470 & 34 / 7.0 \\ \cline{3-7}
    &                         & \multicolumn{4}{c}{300 / 510}  \\
  \end{scotch}
\end{table*}

\begin{table*}[hbt]
  \centering
  \topcaption{Yield of misidentified photons from jets in \cPZ+jets events and their symmetrized
    associated systematic uncertainties as a function of $\pt^{\gamma}$
    in the $\cPZ\gamma\to \ell\ell\gamma$ analyses. The results are specified
    by the numbers of events expected in the $\Pe\Pe\gamma$ and $\mu\mu\gamma$ channels,
    and by the uncertainties from each of the sources.}
  \label{tab:zg_templ_syst_2011AB}
    \begin{scotch}{lccccc}
      & & \multicolumn{4}{c}{Systematic uncertainties on yields ($\Pe\Pe\gamma$/$\mu\mu\gamma$)} \\
      $\pt^{\gamma}$ &  Yield from       & Shape of        & Shape of    & Sampling of   & Diff. between \\
      (\GeVns{})          &  \cPZ+jets events  & $\gamma$ shower & jet shower  & distributions & $\text{jet}\to\gamma$ prediction \\ \hline
      15--20          & 460 / 710     & 11 / 50       & 27 / 41     &  6.4 / 16     &   \\
      20--25          & 200 / 310     &  6.8 / 23     & 11 / 18     &  3.7 / 6.7    &   \\
      25--30          &  82 / 130     &  3.7 / 7.6    &  4.7 / 7.6  &  2.3 / 3.0    &   \\
      30--35          &  51 / 82      &  2.8 / 10     &  2.9 / 4.7  &  1.9 / 1.8    &   \\
      35--40          &  46 / 54      &  3.0 / 4.0    &  2.6 / 3.6  &  1.8 / 1.2    &   \\
      40--60          &  40 / 72      &  3.8 / 11     &  2.3 / 5.8  &  0.9 / 1.5    & 11  / 9.5 \\
      60--90          &  18 / 25      &  3.0 / 6.5    &  1.1 / 3.6  &  0.7 / 0.6    & 4.8 / 3.2 \\
      90--120         &   0.0 / 14    &  0.0 / 3.8    &  0.0 / 1.9  &  0.0 / 0.3    & 0.0 / 4.4 \\
      120--500        &   5.3 / 6.6   &  4.6 / 13     &  0.4 / 1.4  &  0.1 / 0.2    & 1.4 / 3.6 \\ \hline
      \multirow{2}{*}{Totals}  & \multirow{2}{*}{ 910 / 1400 } & 16  / 59 & 30 / 46 & 8.3 / 18 & 17  / 12 \\ \cline{3-6}
                               &                         & \multicolumn{4}{c}{38 / 77 }  \\
    \end{scotch}
\end{table*}

\begin{table*}[!hbt]
  \centering
    \topcaption{Summary of parameters used in the measurement of the $\PW\gamma$ cross section.}
    \label{tab:wengParameters}
    \begin{scotch}{lcc}
      Parameter               & $\Pe\nu\gamma$ channel                            & $\mu\nu\gamma$ channel \\ \hline
      $N^{\ell\nu\gamma}$     & $7470$                                                & $10\,809$ \\
      $N_B^{\PW+\text{jets}}$   & $3180 \pm  50\stat\pm 300\syst$ & $5350 \pm  60\stat\pm 510\syst$ \\
      $N_B^{\Pe\Pe X}$       & $ 690 \pm  20\stat\pm  50\syst$ & $  91 \pm   1\stat\pm   5\syst$ \\
      $N_B^{\text{other}}$    & $ 410 \pm  20\stat\pm  30\syst$ & $ 400 \pm  20\stat\pm  30\syst$ \\
      $N_S^{\ell\nu\gamma}$   & $3200 \pm 100\stat\pm 320\syst$ & $4970 \pm 120\stat\pm 530\syst$ \\
      $A_S$                   & $0.108 \pm 0.001\stat$                     & $0.087 \pm 0.001\stat$ \\
      $A_S \cdot \epsilon_\text{MC}$ ($\PW\gamma\to \ell\nu\gamma$) & $0.0187 \pm 0.0010\syst$ & $0.0270 \pm 0.0014\syst$ \\
      $\rho_\text{eff}$        & $0.940 \pm 0.027\syst$                     & $0.990 \pm 0.025\syst$ \\
      $\rlumi$ (\fbinv)              & $5.0 \pm 0.1\syst$                         & $5.0  \pm 0.1\syst$ \\
  \end{scotch}
\end{table*}

\begin{table*}[hbt]
  \centering
    \topcaption{Summary of parameters used in the measurement of the $\cPZ\gamma$ cross section.}
    \label{tab:zg_results}
    \begin{scotch}{lcc}
      Parameter               & $\Pe\Pe\gamma$ channel                            & $\mu\mu\gamma$ channel \\ \hline
      $N^{\ell\ell\gamma}$    & $4108$                                              & $6463$ \\
      $N_B^{\cPZ+\text{jets}}$   & $ 910 \pm 50\stat \pm 40\syst$  & $1400 \pm  60\stat\pm  80\syst$ \\
      $N_B^\text{other}$      & $  40 \pm  3\stat$                       & $  24 \pm   2\stat$  \\
      $N_S^{\ell\ell\gamma}$  & $3160 \pm 80\stat\pm 90\syst$ & $5030 \pm 100\stat\pm 210\syst$ \\
      $A_S$                   & $0.249 \pm 0.001\stat$                   & $0.286 \pm 0.001\stat$ \\
      $A_S \cdot \epsilon_\text{MC}$ ($\cPZ\gamma\to \ell\ell\gamma$) & $0.1319 \pm 0.0018\syst$ & $0.1963 \pm 0.0013\stat$   \\
      $\rho_\text{eff}$        & $0.929 \pm 0.047\syst$                   & $0.945 \pm 0.016\syst$ \\
      $\rlumi$ (\fbinv)              & $5.0 \pm 0.1\syst$                       & $5.0 \pm 0.1\syst$   \\
  \end{scotch}
\end{table*}

\section{Results}
\label{sec:Results}
\subsection{The \texorpdfstring{$\PW\gamma$}{W gamma} process and radiation-amplitude zero}
\begin{figure}[hbt]
 \begin{center}
   \includegraphics[width=0.49\textwidth]{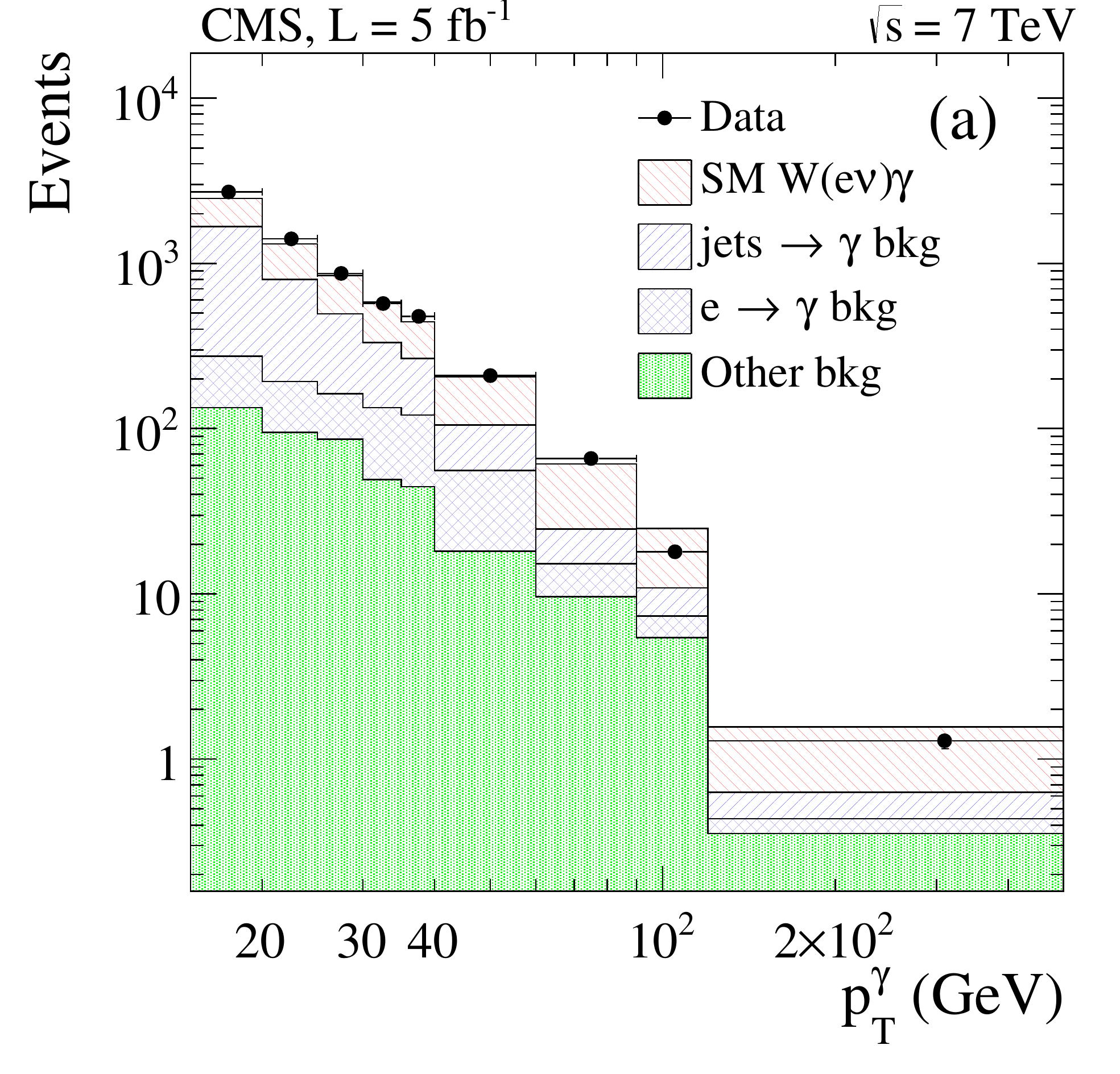}
   \includegraphics[width=0.49\textwidth]{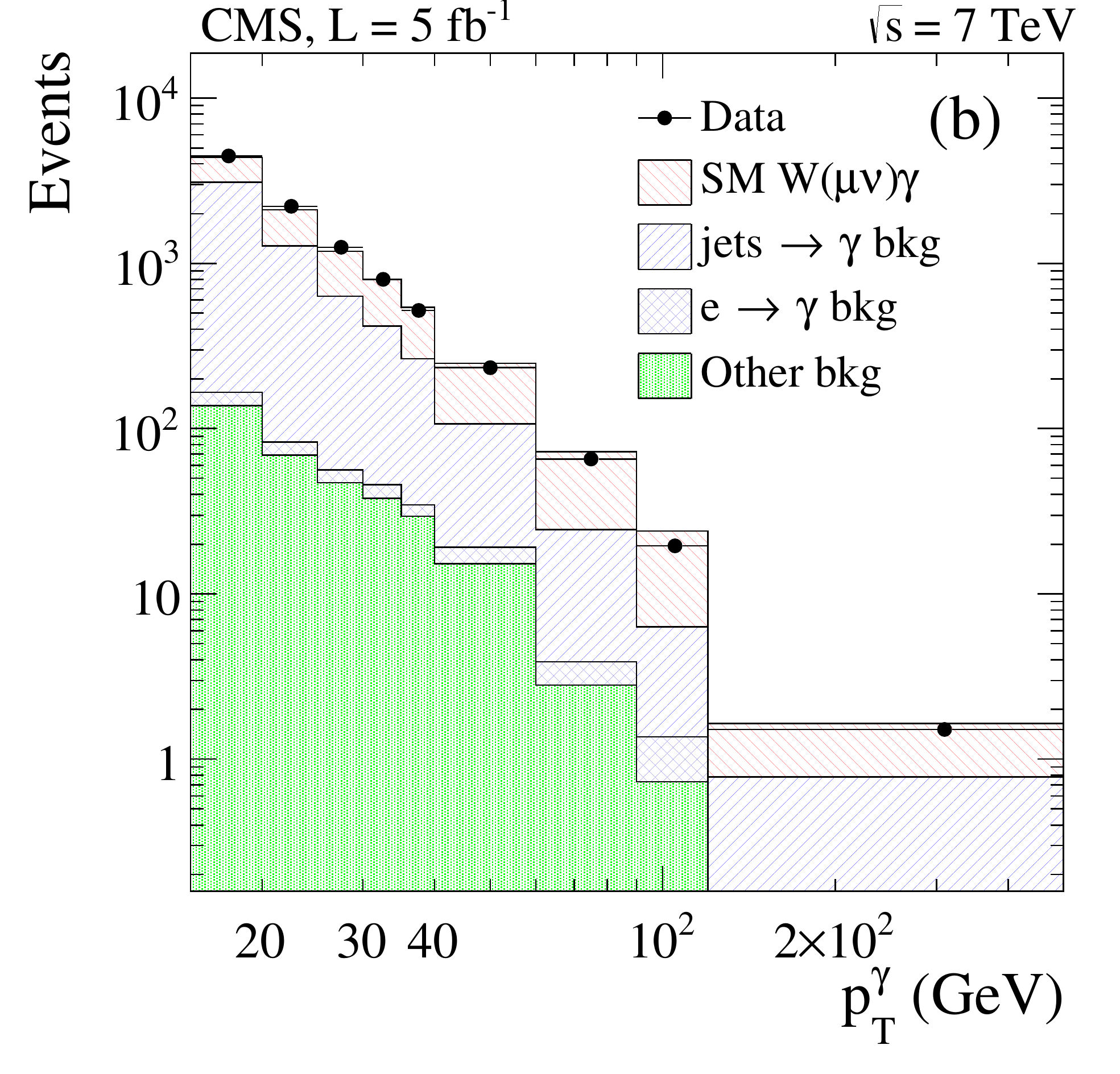}
 \end{center}
 \caption{
   Distributions in $\pt^\gamma$ for $\PW\gamma$ candidate events in data,
   with signal and background MC simulation contributions to
   (a)~$\PW\gamma\to \Pe\nu\gamma$ and (b)~$\PW\gamma\to \mu\nu\gamma$
   channels shown for comparison.
 }
\label{fig:WmunuGammaDataMC_2011AB}
\end{figure}

For photon transverse momenta ${>}15\GeV$ and angular separations between the charged leptons and photons
of $\Delta R > 0.7$, the $\PW\gamma$ production cross section at NLO for each leptonic decay channel
is expected to be $31.8 \pm 1.8\unit{pb}$~\cite{mcfm1,mcfm2}.
This cross section point is used to normalize the $\pt^\gamma$ distributions
for the signal in Fig.~\ref{fig:WmunuGammaDataMC_2011AB}, which shows good agreement of the data with
the expectations from the SM.

\begin{figure}[hbt]
 \begin{center}
   \includegraphics[width=0.49\textwidth]{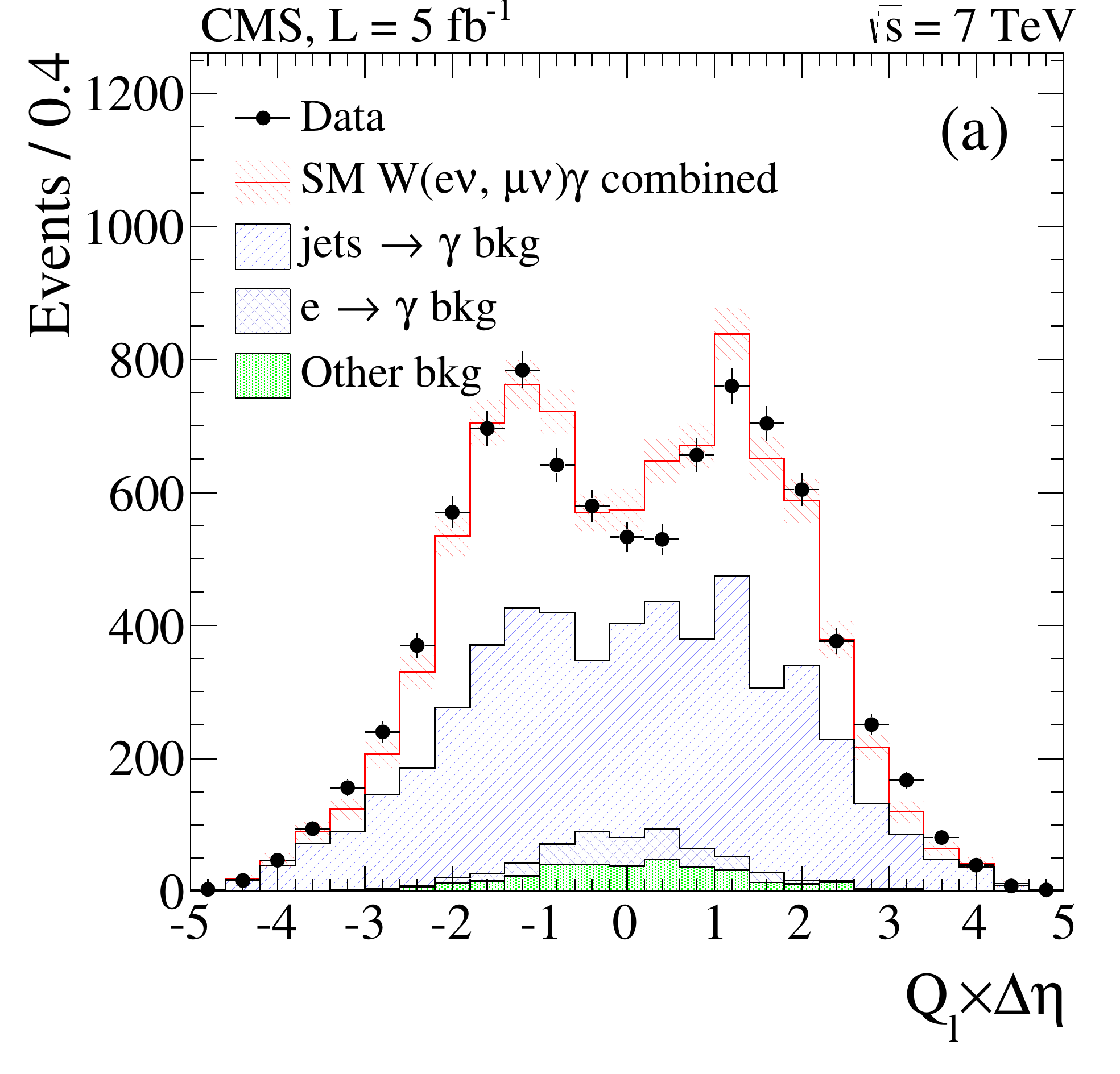}
   \includegraphics[width=0.49\textwidth]{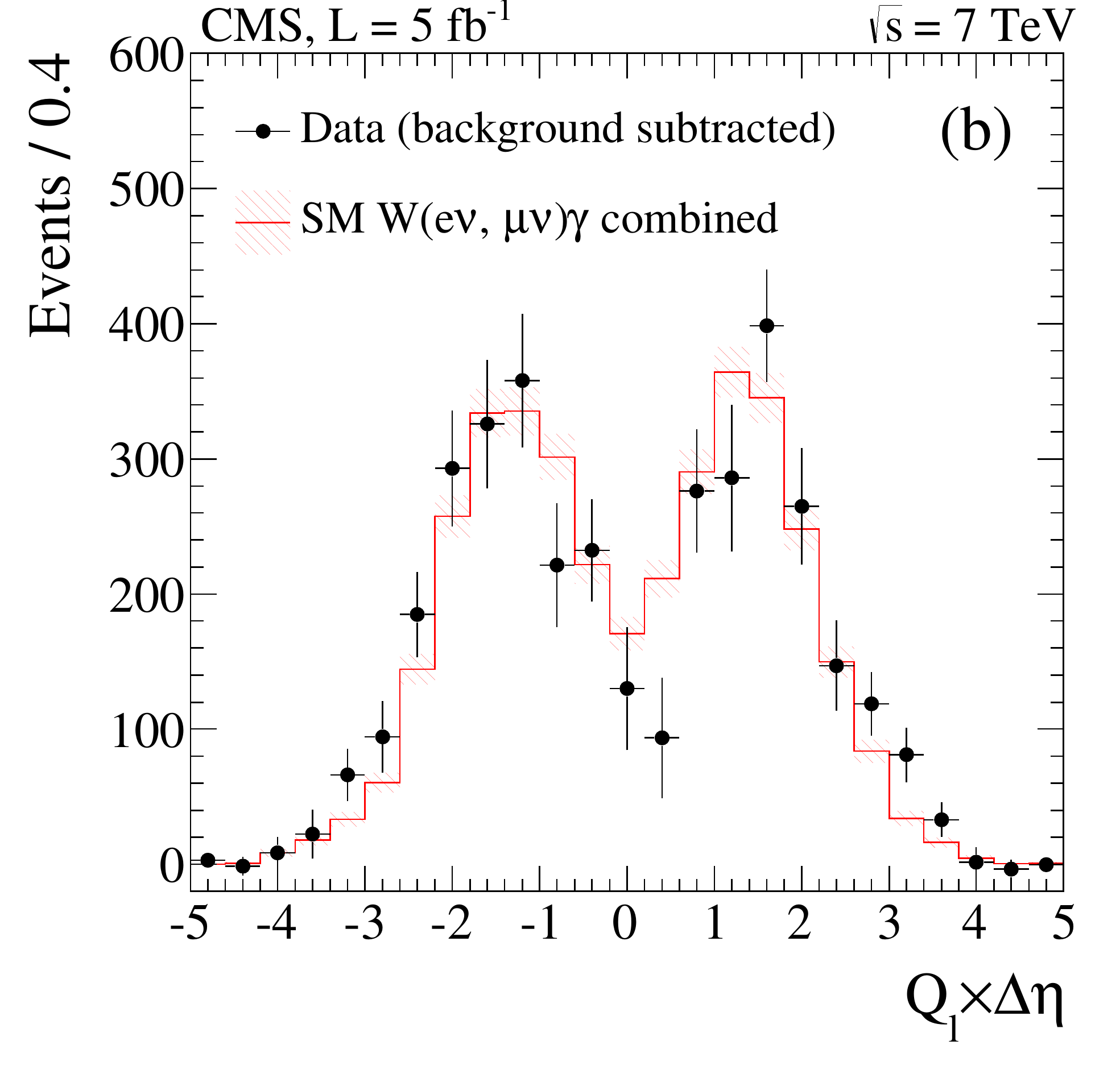}
 \caption{Charge-signed rapidity difference $Q_\ell \times \Delta\eta$
   between the photon candidate and a lepton for $\PW\gamma$ candidates in
   data (filled circles) and expected SM signal and backgrounds
   (shaded regions) normalized to (a)~data, and (b)~background-subtracted
   data. The hatched bands illustrate the full uncertainty in the MC
   prediction.
   \label{fig:RAZ_2011AB_emu}
}
\end{center}
\end{figure}

The three leading-order $\PW\gamma$ production diagrams in Fig.~\ref{fig:feynman_vg}
interfere with each other, resulting in a vanishing of the yield at specific regions of phase space.
Such phenomena are referred to as radiation-amplitude zeros
(RAZ)~\cite{razTheory1,razTheory2,razTheory3,razTheory4,razTheory5},
and the effect was first observed by the D0 Collaboration~\cite{tevatron_wg2} using the
charge-signed rapidity
difference $Q_\ell \times \Delta\eta$ between the photon candidate and the
charged lepton candidate from $\PW\to \ell\nu$ decays~\cite{raz1}. In the
SM, the minimum is at $Q_\ell \times \Delta\eta = 0$
for $\Pp\Pp$ collisions. Anomalous $\PW\gamma$ contributions can affect the
distribution in $Q_\ell \times \Delta\eta$ and make the minimum less pronounced.
The differential yield as a function of charge-signed rapidity difference, shown in
Fig.~\ref{fig:RAZ_2011AB_emu}(a) for $\PW\gamma$ events normalized to the yield of signal in data,
is obtained with the additional requirements of having no accompanying jets with $\pt > 30\GeV$ and
a transverse three-body mass, or cluster mass~\cite{raz1}, of the photon, lepton, and $\ETslash$ system
$> 110\GeV$. The three body mass $\MT(\ell\gamma\ETslash )$ is calculated as
\ifthenelse{\boolean{cms@external}}{
\begin{multline*}
\MT(\ell\gamma\ETslash{})^2 = \left[(M_{\ell\gamma}^2 + \abs{\mathbf{p}_\mathrm{T}(\gamma) + \mathbf{p}_\mathrm{T}(\ell)|^2)^{1/2} + \ETslash{} \right]^2\\
  - |\mathbf{p}_\mathrm{T}(\gamma) + \mathbf{p}_\mathrm{T}(\ell) + \mathbf{\ETslash}{}}^2,
\end{multline*}
}{
\begin{equation}
\MT(\ell\gamma\ETslash{})^2 = \left[(M_{\ell\gamma}^2 + \abs{\mathbf{p}_\mathrm{T}(\gamma) + \mathbf{p}_\mathrm{T}(\ell)|^2)^{1/2} + \ETslash{} \right]^2
  - |\mathbf{p}_\mathrm{T}(\gamma) + \mathbf{p}_\mathrm{T}(\ell) + \mathbf{\ETslash}{}}^2,
\end{equation}
}
where $M_{\ell\gamma}$ denotes the invariant mass of the $\ell\gamma$ system, and
$\mathbf{p}_\mathrm{T}(i)$, $i = \gamma$, $\ell$, and $\ETslash$
are the projections of the photon, lepton, and $\ETslash$ vectors on the transverse plane, respectively.
Figure~\ref{fig:RAZ_2011AB_emu}(b) shows the background-subtracted data.
The shaded bars indicate statistical and systematic
uncertainties on the MC prediction.
The distributions demonstrate the characteristic RAZ expected for $\PW\gamma$ production.
Both figures indicate no significant difference between data and
expectations from SM MC simulations.

\subsection{The \texorpdfstring{$\cPZ\gamma$}{Z gamma} process}
\begin{figure}[hbt]
  \begin{center}
  \includegraphics[width=0.49\textwidth]{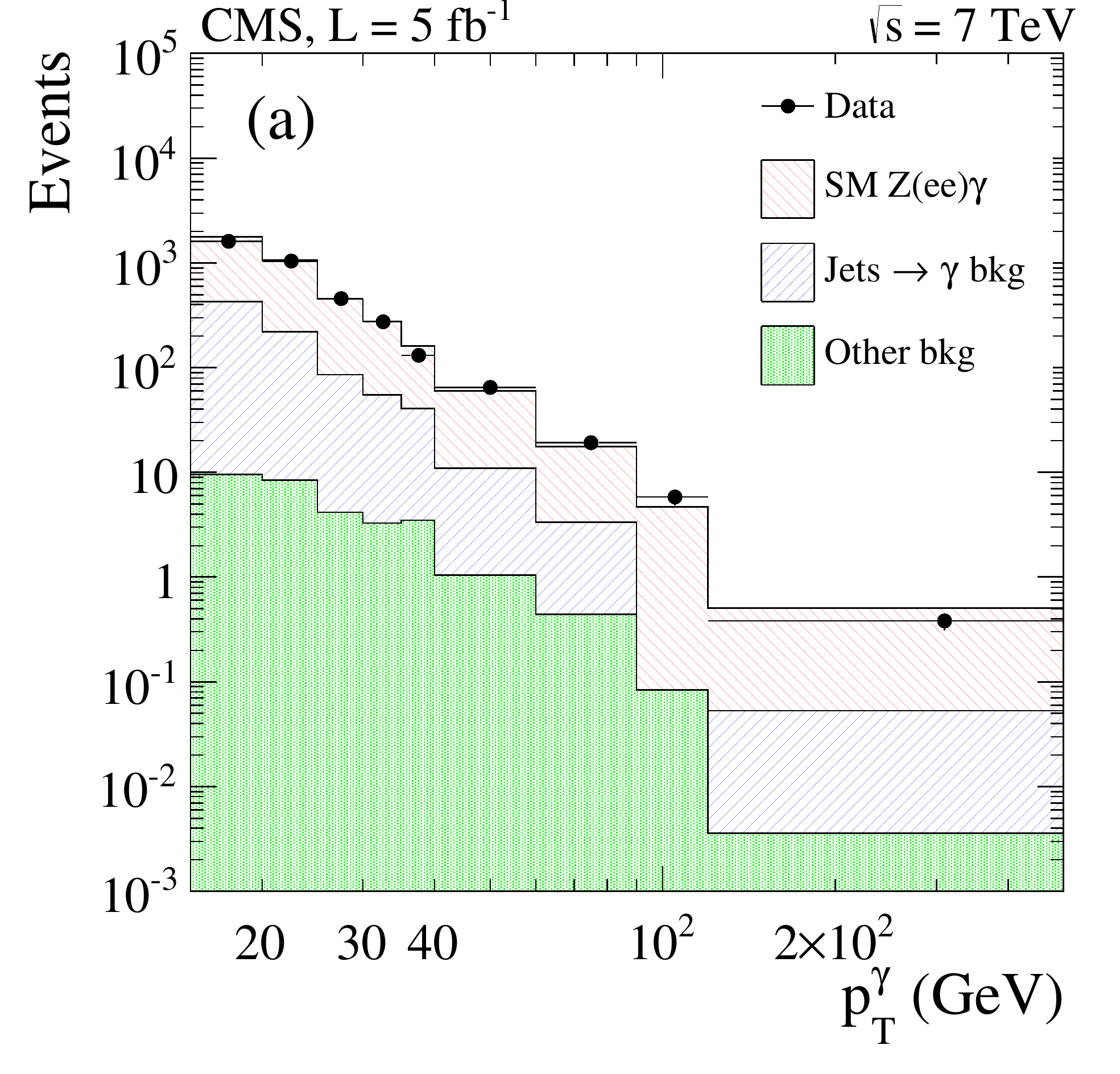}
  \includegraphics[width=0.49\textwidth]{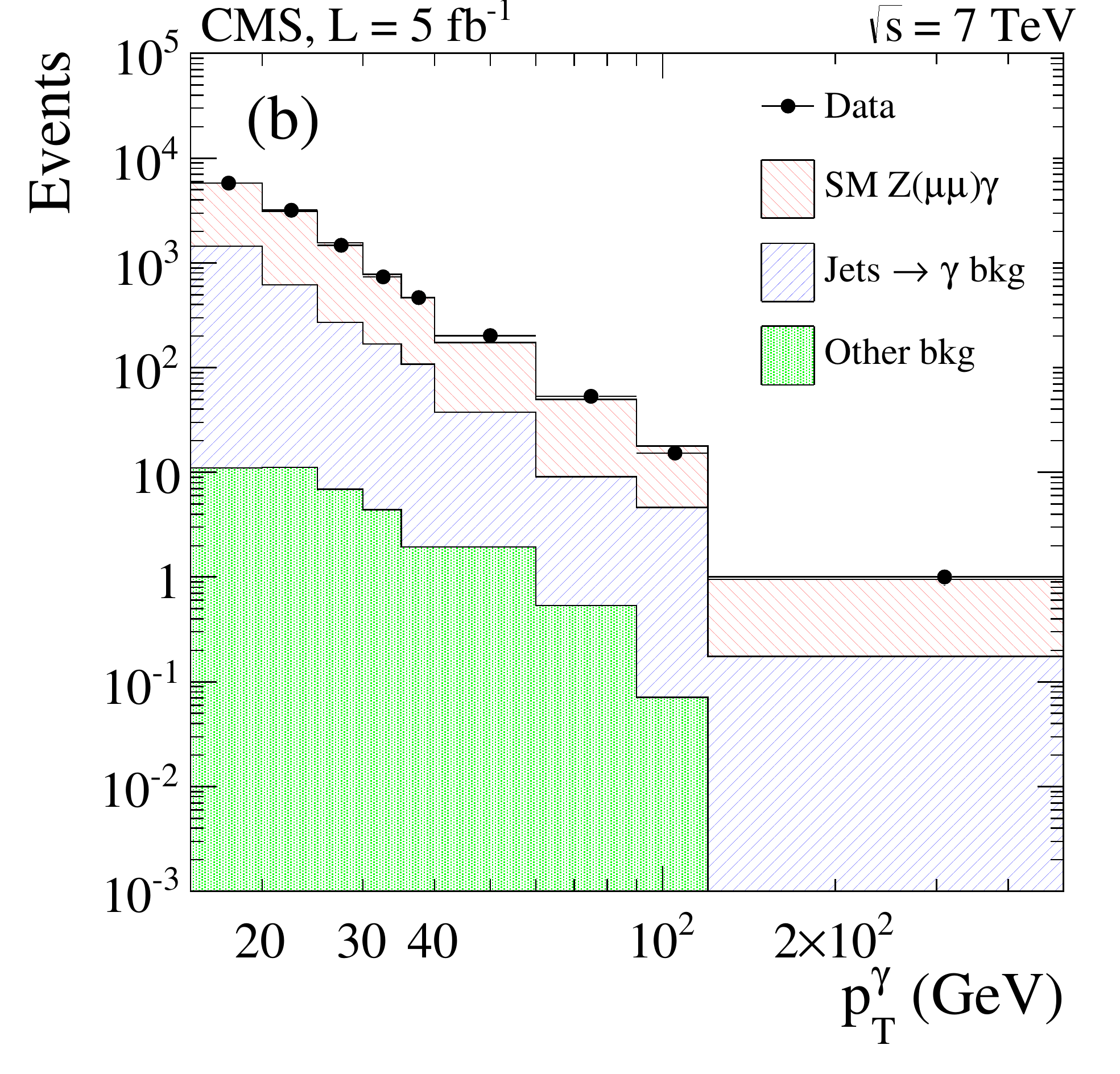}
    \caption{
   Distributions in $\pt^\gamma$ for $\cPZ\gamma$ candidate events in data,
   with signal and background MC simulation contributions to
   (a)~$\cPZ\gamma\to \Pe\Pe\gamma$ and (b)~$\cPZ\gamma\to \mu\mu\gamma$
   channels shown for comparison.
    }
    \label{fig:ZGtoEEG_Result}
  \end{center}
\end{figure}

The cross section for $\cPZ\gamma$ production at NLO in the SM, for
$\pt^\gamma > 15\GeV$,
$\Delta R(\ell,\gamma) > 0.7$ between the photon and either of the charged leptons
from the $\cPZ\to \ell^{+}\ell^{-}$ decay, and $m_{\ell\ell} > 50\GeV$,
is predicted to be $5.45 \pm 0.27\unit{pb}$~\cite{mcfm1,mcfm2}.
After applying all selection criteria, the $\pt^\gamma$ distributions for data and contributions
expected from MC simulation are shown for $\Pe\Pe\gamma$ and $\mu\mu\gamma$ final states in
Figs.~\ref{fig:ZGtoEEG_Result}(a) and (b), respectively.
Again, good agreement is found between data and the SM predictions.

\subsection{Production cross sections}
\label{sec:csHowto}
The cross section for any signal process of interest can be written as
\begin{equation}
\sigma_S = \frac{N_S}{A_S \cdot \epsilon_S \cdot \text{L}}\ \ ,
\label{eq:x-sectionDef}
\end{equation}
where $N_S$ is the number of observed signal events, $A_S$ is the geometric and kinematic
acceptance of the detector, $\epsilon_S$ is the selection efficiency for signal events
in the region of acceptance, and L is the integrated luminosity.
The value of $A_S$ in our analyses is calculated through MC simulation, and is affected by the choice of PDF and
other uncertainties of the model, while the value of $\epsilon_S$ is sensitive to uncertainties
in the simulation, triggering, and reconstruction. To reduce uncertainties in efficiency, we apply
corrections to the efficiencies obtained from MC simulation, which
reflect ratios of efficiencies $\rho_\text{eff} = \epsilon_\text{data} / \epsilon_\text{MC}$
obtained by measuring the efficiency in the same way for data and simulation.
The product
 $A_S \times \epsilon_S$ can then be replaced by the product
$\mathcal{F}_S \times \rho_\text{eff}$,
where  $\mathcal{F}_S \equiv  A_S \times \epsilon_\text{MC}$
corresponds to the fraction of generated signal events selected in the simulation.

Equation (\ref{eq:x-sectionDef}) can therefore be rewritten as
\begin{equation}
\sigma_S = \frac{N-N_B}{\mathcal{F}_S \cdot \rho_\text{eff} \cdot \text{L}},
\label{eq:x-sectionDef2}
\end{equation}
in which we replace the number of signal events $N_S$ by subtracting
the estimated number of background events $N_B$ from the observed
number of selected events $N$.

We calculate $\mathcal{F}_S$ using MC simulation, with $\mathcal{F}_S$ defined by
$N_\text{accept} / N_\text{gen}$, where $N_\text{accept}$ is the number of
signal events that pass all selection requirements in the MC simulation of signal,
and $N_\text{gen}$ is the number of MC generated events restricted to
$\pt^{\gamma} > 15\GeV$ and $\Delta R(\ell,\gamma) > 0.7$, for $\PW\gamma$,
and with an additional requirement, $m_{\ell\ell} > 50\GeV$, for $\cPZ\gamma$ .

\subsection{Systematic uncertainties}
\label{sec:systematics}

Systematic uncertainties are grouped into five categories. The first group includes
uncertainties that affect the signal, such as uncertainties on lepton and photon energy scales.
We assess the systematic uncertainties in the electron and photon energy scales separately to
account for the differences in the clustering procedure, the response of the ECAL,
and calibrations between the electrons and photons.
The estimated uncertainty is 0.5\% in the EB and 3\% in the EE for electrons, and 1\%
in the EB and 3\% in the EE for photons.
The uncertainties in the two scales are conservatively treated as fully correlated.
For the muon channel, the muon momentum is changed by 0.2\%.
The systematic effect on the measured cross
section is obtained by reevaluating $N_S$ for such changes in each source of systematic uncertainty.
To extract the systematic effect of the energy scale on the signal yield, the data-driven background
estimation is performed using signal and background templates modified to use the varied energy scale.
This ensures that migrations of photons and misidentified photon-like jets
across the low-$\pt^\gamma$ boundaries are properly taken into account for
this systematic uncertainty.

In the second group, we combine uncertainties that affect the product of the acceptance,
reconstruction, and identification efficiencies of final state objects, as determined from
simulation. These include uncertainties in the lepton and photon energy resolution,
effects from pileup, and uncertainties in the PDF. The uncertainty in the product of acceptance
and efficiency ($A_S \times \epsilon_S$) is determined from
MC simulation of the $\vb\gamma$ signal and is affected by the lepton and photon energy
resolution through the migration of events in and out of the acceptance.
The electron energy resolution is determined from data using the observed
width of the \cPZ\ boson peak in the $\cPZ\to \Pe\Pe$ events, following the same procedure
as employed in Ref.~\cite{Hgg}.
To estimate the effect of electron resolution on $A_S \times \epsilon_S$, each electron
candidate's energy is smeared randomly by the energy resolution determined from data,
before applying the standard selections.
The photon energy resolution is determined simultaneously with the photon
energy scale from data, following the description in Refs.~\cite{Veverka,HiggsLongPaper}.
The systematic effect of photon resolution on $A_S \times \epsilon_S$ is calculated by
smearing the reconstructed photon energy in simulation to match that in data.

The number of pileup interactions per event is estimated from data using a convolution
procedure that extracts the estimated pileup from the instantaneous bunch luminosity.
The total inelastic $\Pp\Pp$ scattering cross section
is used to estimate the number of pileup interactions expected in a given bunch crossing,
with a systematic uncertainty from modeling of the pileup interactions obtained by changing
the total inelastic cross section within its uncertainties~\cite{ppinelastic}
to determine the impact on $A_S \times \epsilon_S$.
The uncertainties from the choice of PDF are estimated using the CTEQ6.6 PDF set~\cite{cteq66}.
The uncertainty in the modeling of the signal is taken from the difference in acceptance
between \MCFM and \MADGRAPH predictions.

The third group of uncertainties includes the systematic sources affecting the relative $\rho_\text{eff}$
correction factors for efficiencies of the trigger, reconstruction, and identification requirements
in simulations and data. Among these sources are the uncertainties in lepton triggers, lepton and photon
reconstruction and identification, and $\ETslash$ for the $\PW\gamma$ process.
The uncertainties in lepton and photon efficiencies are estimated by changing the modeling of background
and the range of the fits used in the tag-and-probe method.

\begin{figure}[!hbt]
\centering
      \includegraphics[width=0.49\textwidth]{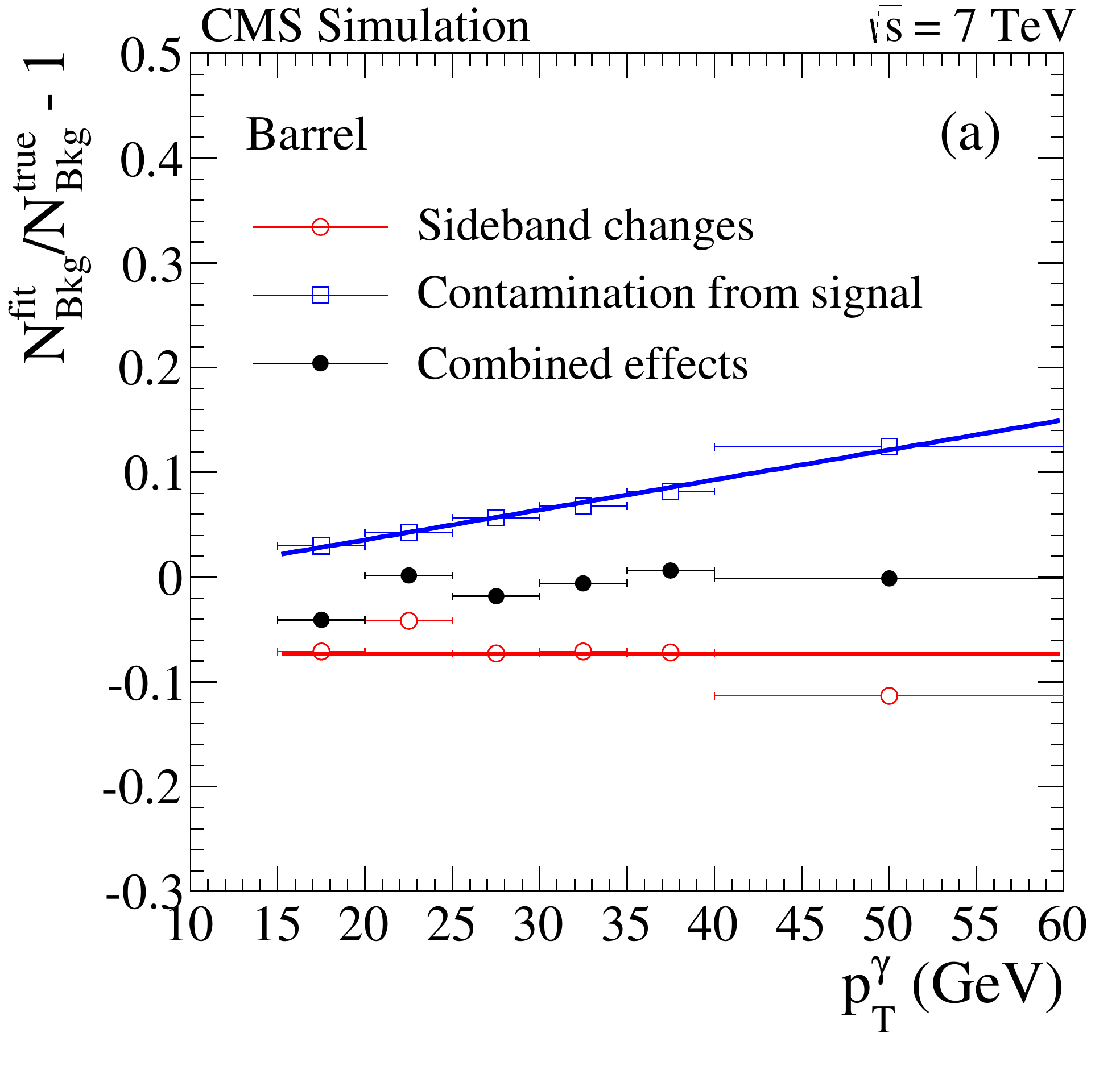}
      \includegraphics[width=0.49\textwidth]{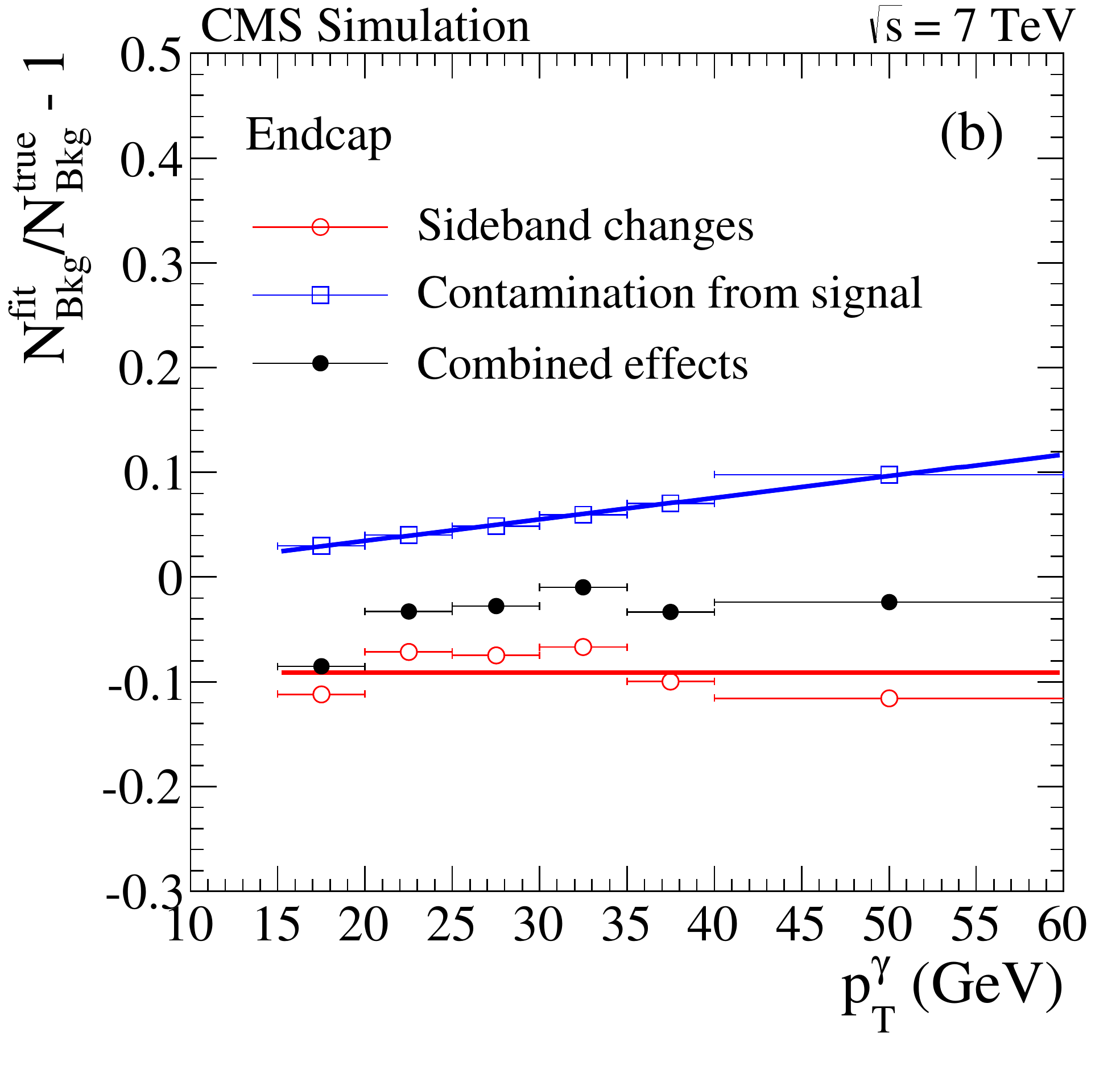}
    \caption{Bias in the background contamination related to the background templates for
    $\sihih$, as a function of $\pt^{\gamma}$, for the (a)~barrel and (b)~endcap regions of ECAL.}
    \label{fig:MCBkg_Uncertainty}
\end{figure}

The fourth category of uncertainties comprises the contributions from background.
These are dominated by uncertainties in estimating the \PW+jets and \cPZ+jets backgrounds
from data. The difference in \sihih distributions between data and simulated events
(Section~\ref{ssec:template_method}) is attributed to systematic uncertainties in signal templates,
which are used to calculate the background estimate and measure its effect on the final result.
To infer the background from photon-like jets that pass full photon-isolation criteria,
we use the \sihih distributions obtained by reversing the original isolation requirement for the tracker.
The possible correlation of \sihih with tracker isolation, and a contribution
from genuine photons that pass the reversed isolation requirement, can cause bias
in the estimation of background. The first issue is investigated by comparing the
sideband and true \sihih distributions in simulated multijets events, where genuine
photons can be distinguished from jets. The resulting bias on the background estimation is
shown by the open circles in Fig.~\ref{fig:MCBkg_Uncertainty}. The second issue, concerning
the contamination of the background template by signal, is investigated by
comparing the sideband \sihih distributions of simulated samples, both with and without
admixtures of genuine photons. The results of the bias studies are shown by the open
squares in Fig.~\ref{fig:MCBkg_Uncertainty}, and the overall effect, given by the
filled black circles, is found to be small.

Since smoothing is used to define a continuous function for describing the \sihih
distribution for background, the effect of statistical sampling of the
background probability density requires an appreciation of the features of the underlying
distribution. This is studied as follows. The simulation is used to generate
a distribution for background, which can be used to generate a template.
These new distributions are also smoothed, and used to fit the background
fraction in data. The results of fits using each such distribution are saved,
and the standard deviation associated with the statistical fluctuation in the template is
taken as a systematic uncertainty.
The systematic uncertainties from different inputs in the estimation of background
from \PW+jets and \cPZ+jets events were shown in Table~\ref{tab:wmng_templ_syst_2011AB}
and~\ref{tab:zg_templ_syst_2011AB}, respectively.

The uncertainties in background from electrons misidentified as photons in $\PW\gamma$
candidate events are estimated by taking the difference in $P_{\Pe\to\gamma}$ between the measurement
described in Section~\ref{sss:eltophoton} and that obtained using a simple counting method.
The uncertainties for lesser contributions to background are defined by the
statistical uncertainties in the samples used for their simulation.
Finally, the systematic uncertainty in the measured integrated luminosity is 2.2\%~\cite{lumiPAS}.

\subsection{\texorpdfstring{$\PW\gamma$}{W gamma} cross section}

In the summary of parameters used in the measurement of the
$\Pp\Pp \to \PW\gamma$ cross sections listed in
Table~\ref{tab:wengParameters}, $N^{\ell\nu\gamma}$ is the number of
observed events, $N_S^{\ell\nu\gamma}$ is the number of observed
signal events after background subtraction, and
$A_S \times \epsilon_S$, $\rho_\text{eff}$, and \rlumi
are described in Section~\ref{sec:csHowto}.
A summary of all systematic uncertainties in the measured $\PW\gamma$ cross sections
is given in Table~\ref{tab:wgSystematicsAB}, separately for electron and muon channels.

\begin{table*}[hbt]
  \topcaption{Summary of systematic uncertainties in the measurement of
    $\PW\gamma$ cross section, separated into the main groups of sources for the
    $\Pe\nu\gamma$ and $\mu\nu\gamma$ channels. ``n/a'' stands for ``not applicable''.}
  \label{tab:wgSystematicsAB}
  \begin{scotch}{lccc}
    &                             & $\Pe\nu\gamma$    &  $\mu\nu\gamma$       \\ \hline
    Source (Group 1)              & Uncertainties         & \multicolumn{2}{c}{Effect from \nsig}           \\ \hline
    \Pe/$\gamma$ energy scale       & (\Pe: 0.5\%; $\gamma$: 1\% (EB), 3\% (EE))       & 2.9\%     & n/a   \\
    $\gamma$ energy scale         & (1\% (EB), 3\% (EE))                           & n/a     	 & 2.9\% \\
    $\mu$ $\pt$ scale       & (0.2\%)                                        & n/a     	 & 0.6\% \\
    \multicolumn{2}{l}{Total uncertainty in \nsig}  & 2.9\%    	 & 3.0\% \\ \hline
    Source (Group 2)              & Uncertainties         & \multicolumn{2}{c}{Effect from $\mathcal{F}_S = A_S \cdot \epsilon_S$} \\ \hline
    \Pe/$\gamma$ energy resolution  & (1\% (EB), 3\% (EE))  & 0.3\%    	 & n/a    \\
    $\gamma$ energy resolution	  & (1\% (EB), 3\% (EE))  & n/a    	 & 0.1\%    \\
    $\mu$ $\pt$ resolution  & (0.6\%)               & n/a     	 & 0.1\%  \\
    Pileup                        & (Shift pileup distribution by $\pm$ 5\%) & 2.4\%   	 & 0.8\%  \\
    PDF                           &                       & 0.9\%   	 & 0.9\%  \\
    Modeling of signal            &              	  & 5.0\%   	 & 5.0\%  \\
    \multicolumn{2}{l}{Total uncertainty in $\mathcal{F}_S = A_S \cdot \epsilon_S$}                               & 5.6\%    	 & 5.1\%  \\ \hline
    Source (Group 3)              & Uncertainties         & \multicolumn{2}{c}{Effect from $\rho_\text{eff}$} \\ \hline
    Lepton reconstruction 	  &                       & 0.4\%    	 & 1.5\%   \\
    Lepton trigger		  &			  & 0.1\%    	 & 0.9\%   \\
    Lepton ID and isolation       &                       & 2.5\%    	 & 0.9\%   \\
    \ETslash selection                &                       & 1.4\%  	 & 1.5\%   \\
    $\gamma$ identification and isolation & (0.5\% (EB), 1.0\% (EE)) & 0.5\%  	 & 0.5\% \\
    \multicolumn{2}{l}{Total uncertainty in $\rho_\text{eff}$}        & 2.9\%  	 & 2.5\%   \\ \hline
    Source (Group 4)              &                       & \multicolumn{2}{c}{Effect from background yield} \\ \hline
    Template method               &  		          & 9.3\% 	 & 10.2\%   \\
    Electron misidentification	  &			  & 1.5\%  	 &  0.1\%  \\
    MC prediction		  &			  & 0.8\%   	 &  0.5\%  \\
    \multicolumn{2}{l}{Total uncertainty due to background}   & 9.5\%  	 & 10.2\%   \\ \hline
    Source (Group 5)              &                       & \multicolumn{2}{c}{}   \\ \hline
    Luminosity                    &                       & 2.2\%        & 2.2\%   \\
  \end{scotch}
\end{table*}

The measured cross sections are
\ifthenelse{\boolean{cms@external}}{
\begin{multline*}
\sigma(\Pp\Pp \to \PW\gamma) \times \mathcal{B}(\PW\to \Pe\nu) =\\ 36.6 \pm 1.2\stat \pm 4.3\syst \pm 0.8\lum\unit{pb},
\end{multline*}
\begin{multline*}
\sigma(\Pp\Pp \to \PW\gamma) \times \mathcal{B}(\PW\to \mu\nu) =\\ 37.5 \pm 0.9\stat \pm 4.5\syst \pm 0.8\lum\unit{pb}.
\end{multline*}
}{
\begin{equation*}
\sigma(\Pp\Pp \to \PW\gamma) \times \mathcal{B}(\PW\to \Pe\nu) =\\ 36.6 \pm 1.2\stat \pm 4.3\syst \pm 0.8\lum\unit{pb},
\end{equation*}
\begin{equation*}
\sigma(\Pp\Pp \to \PW\gamma) \times \mathcal{B}(\PW\to \mu\nu) =\\ 37.5 \pm 0.9\stat \pm 4.5\syst \pm 0.8\lum\unit{pb}.
\end{equation*}
}
The mean of these cross sections, obtained using a best linear unbiased estimator (BLUE)~\cite{blue},
is
\ifthenelse{\boolean{cms@external}}{
\begin{multline*}
\sigma(\Pp\Pp\to \PW\gamma) \times \mathcal{B}(\PW \to \ell\nu) =\\ 37.0 \pm 0.8\stat \pm 4.0\syst \pm 0.8\lum\unit{pb}.
\end{multline*}
}{
\begin{equation*}
\sigma(\Pp\Pp\to \PW\gamma) \times \mathcal{B}(\PW \to \ell\nu) = 37.0 \pm 0.8\stat \pm 4.0\syst \pm 0.8\lum\unit{pb}.
\end{equation*}
}
All three results are consistent within uncertainties with the NLO prediction of $31.8 \pm 1.8\unit{pb}$,
computed with \MCFM.
The uncertainty on the prediction is obtained using the CTEQ6.6 PDF set~\cite{cteq66}.

\subsection{\texorpdfstring{$\cPZ\gamma$}{Z gamma} cross section}
\label{ssec:zgammaXS}

In the summary of parameters used in the measurement of the
$\Pp\Pp \to \cPZ\gamma$ cross section listed in Table~\ref{tab:zg_results},
$N^{\ell\ell\gamma}$ is the number of observed
events, and $N_S^{\ell\ell\gamma}$ is the number of observed
signal events after background subtraction.
The systematic uncertainties for the measurement of the $\cPZ\gamma$ cross sections are
listed in Table~\ref{tab:zgSystematics}. The cross sections for the two channels are
\ifthenelse{\boolean{cms@external}}{
\begin{multline*}
\sigma(\Pp\Pp \to \cPZ\gamma) \times \mathcal{B}(\cPZ \to \Pe\Pe) = \\5.20 \pm 0.13\stat\pm 0.32\syst \pm 0.11\lum\unit{pb},
\end{multline*}
\begin{multline*}
\sigma(\Pp\Pp \to \cPZ\gamma) \times \mathcal{B}(\cPZ \to \mu\mu) = \\5.43 \pm 0.10 \stat \pm 0.29 \syst \pm 0.12 \lum\unit{pb},
\end{multline*}
}{
\begin{equation*}
\sigma(\Pp\Pp \to \cPZ\gamma) \times \mathcal{B}(\cPZ \to \Pe\Pe) = 5.20 \pm 0.13\stat\pm 0.32\syst \pm 0.11\lum\unit{pb},
\end{equation*}
\begin{equation*}
\sigma(\Pp\Pp \to \cPZ\gamma) \times \mathcal{B}(\cPZ \to \mu\mu) = 5.43 \pm 0.10 \stat \pm 0.29 \syst \pm 0.12 \lum\unit{pb},
\end{equation*}
}
and their mean, extracted using the BLUE method is
\ifthenelse{\boolean{cms@external}}{
\begin{multline*}
\sigma(\Pp\Pp \to \cPZ\gamma) \times \mathcal{B}(\cPZ \to \ell\ell) =\\ 5.33 \pm 0.08 \stat \pm 0.25 \syst \pm 0.12 \lum\unit{pb}.
\end{multline*}
}{
\begin{equation*}
\sigma(\Pp\Pp \to \cPZ\gamma) \times \mathcal{B}(\cPZ \to \ell\ell) = 5.33 \pm 0.08 \stat \pm 0.25 \syst \pm 0.12 \lum\unit{pb}.
\end{equation*}
}
All three results are also consistent within the uncertainties with the theoretical NLO cross section of
$5.45 \pm 0.27\unit{pb}$, computed with \MCFM.
The uncertainty on the prediction is obtained using the CTEQ6.6 PDF set~\cite{cteq66}.

\begin{table*}[hbt]
  \begin{center}
    \caption{Summary of systematic uncertainties for the measurement of the $\cPZ\gamma$ cross section. ``n/a'' stands for ``not applicable''.}
    \label{tab:zgSystematics}
    \begin{scotch}{lccc}
                                                     &                                              & $\Pe\Pe\gamma$    &  $\mu\mu\gamma$   \\ \hline
      Source (Group 1)                               & Uncertainties                                & \multicolumn{2}{c}{Effect from \nsig} \\ \hline
      \Pe/$\gamma$ energy scale         & (\Pe: 0.5\%; $\gamma$: 1\% (EB), 3\% (EE))       & 3.0\%     & n/a   \\
      $\mu$ $\pt$ scale                        & (0.2\%)                                      & n/a             & 0.6\% \\
      $\gamma$ energy scale                          & (1\% (EB), 3\% (EE))                         & n/a             & 4.2\%  \\
      \multicolumn{2}{l}{Total uncertainty in \nsig}                                                & 3.0\%           & 4.2\% \\ \hline
      Source (Group 2)                               & Uncertainties & \multicolumn{2}{c}{Effect from $\mathcal{F}_S = A_S \cdot \epsilon_S$} \\ \hline
      \Pe/$\gamma$ energy resolution                   & (1\% (EB), 3\% (EE))                         & 0.2\%           & n/a    \\
      $\gamma$ energy resolution                     & (1\% (EB), 3\% (EE))                         & n/a             & 0.1\%    \\
      $\mu$ $\pt$ resolution                   & (0.6\%)                                      & n/a             & 0.2\%  \\
      Pileup                                         & Shift pileup distribution by $\pm$ 5\%       & 0.6\%           & 0.4\%  \\
      PDF                                            &                                              & 1.1\%           & 1.1\%  \\
      Modeling of signal                             &                                              & 0.6\%           & 0.5\%  \\
      \multicolumn{2}{l}{Total uncertainty in $\mathcal{F}_S = A_S \cdot \epsilon_S$}               & 1.4\%           & 1.3\%  \\ \hline
      Source (Group 3)                               & Uncertainties & \multicolumn{2}{c}{Effect from $\rho_\text{eff}$}        \\ \hline
      Lepton reconstruction                          &                                              & 0.8\%           & 1.0\%  \\
      Lepton trigger                                 &                                              & 0.1\%           & 1.0\%  \\
      Lepton ID and isolation                        &                                              & 5.0\%           & 1.8\%  \\
      Photon ID and isolation                        & (0.5\% (EB), 1.0\% (EE))                     & 0.5\%           & 1.0\%  \\
      \multicolumn{2}{l}{Total uncertainty in $\rho_\text{eff}$}                                     & 5.1\%           & 2.5\%  \\ \hline
      Source (Group 4)                               &               & \multicolumn{2}{c}{Effect from background yield}         \\ \hline
      Template method                                &                                              & 1.2\%           & 1.5\%  \\
      \multicolumn{2}{l}{Total uncertainty due to background}                                       & 1.2\%           & 1.5\%  \\ \hline
      Source (Group 5)                               &               & \multicolumn{2}{c}{} \\ \hline
      Luminosity                                     &                                              & 2.2\%           & 2.2\%  \\
    \end{scotch}
  \end{center}
\end{table*}

\subsection{Ratio of \texorpdfstring{$\PW\gamma$ and $\cPZ\gamma$}{W gamma and Z gamma} production cross sections}

We calculate the ratio of the $\PW\gamma$ and $\cPZ\gamma$ cross
sections using the BLUE method to account for correlated systematic uncertainties
between individual channels for both measurements and predictions.
The \MCFM prediction of $5.8 \pm 0.1$ is consistent with
the measured ratio, $6.9 \pm 0.2\stat\pm 0.5\syst$.

\subsection{Comparisons to \MCFM predictions}

\begin{figure*}[!hbt]
 \centering
   \includegraphics[width=0.45\textwidth]{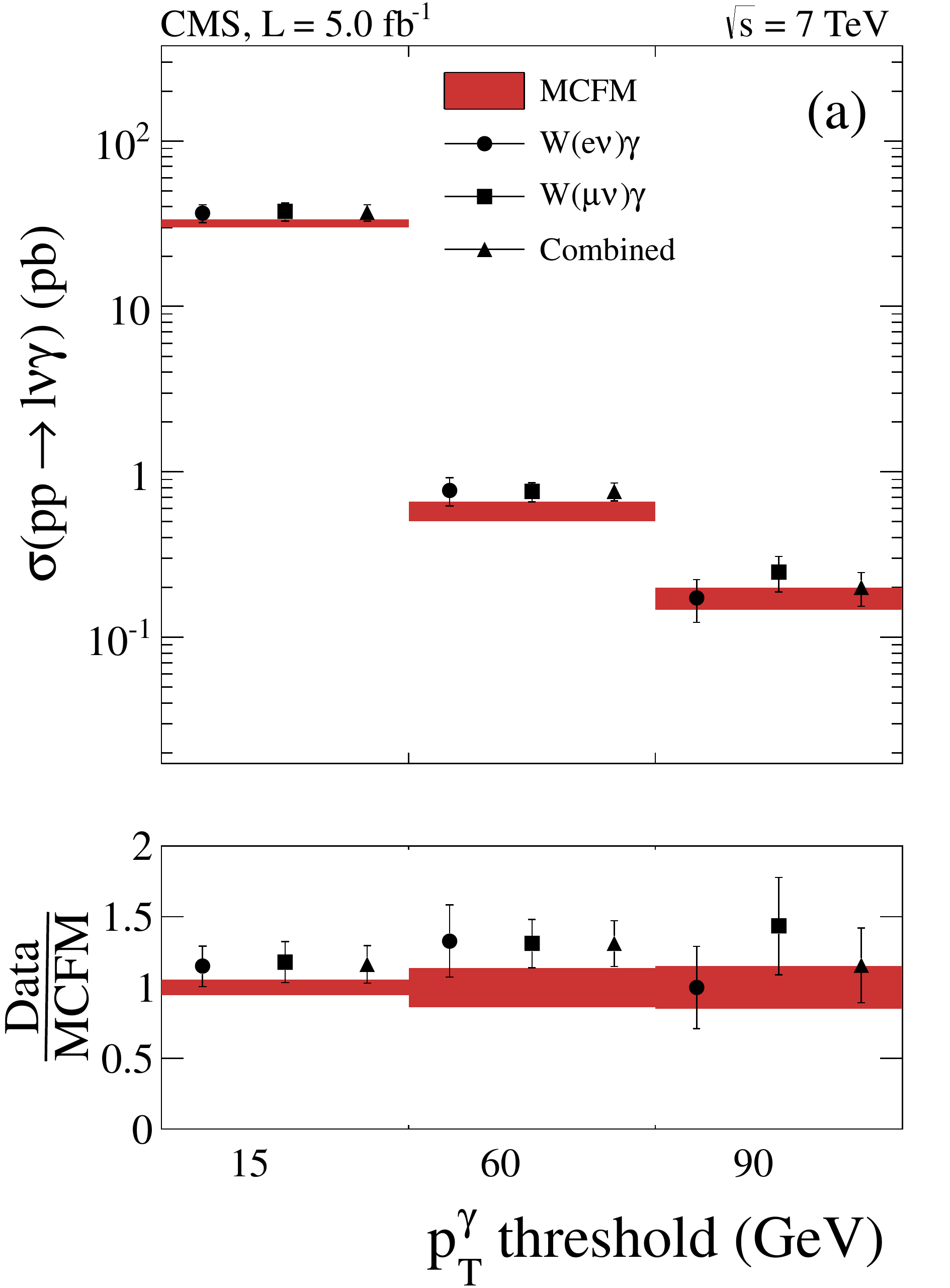}
   \includegraphics[width=0.45\textwidth]{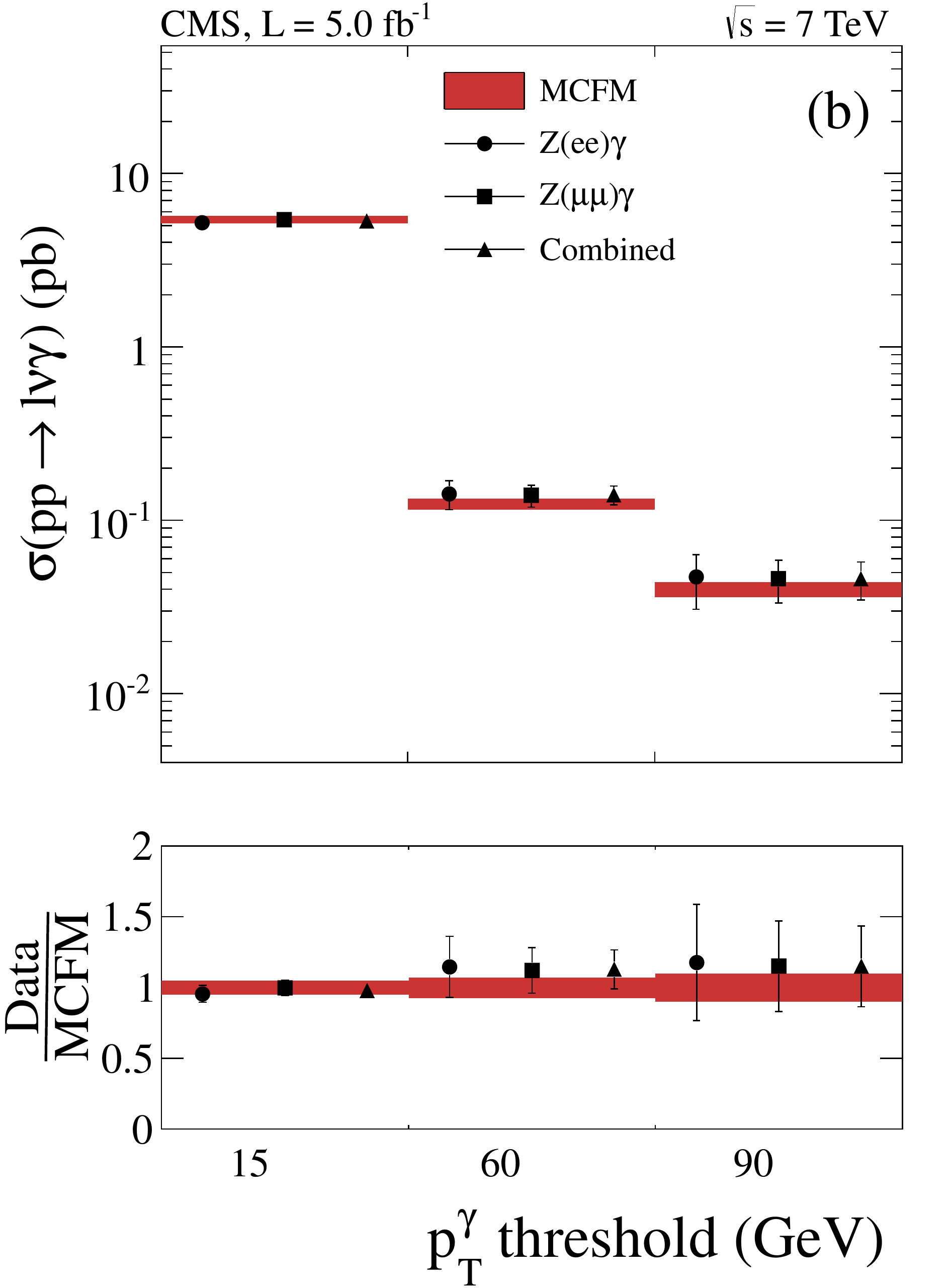}
 \caption{A summary of measured cross sections for three $\pt^\gamma$ thresholds, compared to SM
predictions for (a)~$\PW\gamma$ and (b)~$\cPZ\gamma$ production.}
 \label{fig:XS_summary}
\end{figure*}

Finally, we present a summary of the $\PW\gamma$ and $\cPZ\gamma$ cross sections
measured with larger requirements on the minimum photon $\pt^\gamma$.
After accounting for all systematic uncertainties for $\pt^\gamma > 60$ and ${>}90\GeV$,
we find no significant disagreement with the \MCFM predictions for $\vb\gamma$
processes. These cross sections, predictions, and their uncertainties are
summarized in Table~\ref{tab:XS_summary} and in Fig.~\ref{fig:XS_summary}.

\begin{table*}[!hbt]
\centering
\caption{Summary of the measured cross sections and predictions
    for $\pt^\gamma > 60$ and $> 90\GeV$ for $\PW\gamma$ and $\cPZ\gamma$ production.}
  \label{tab:XS_summary}
  \begin{scotch}{lccc}
Process                          & $\pt^\gamma$ (\GeVns{}) & $\sigma\times \mathcal{B}$ (pb) & Theory (pb) \\ \hline
$\PW\gamma\to \Pe\nu\gamma$    & $>60$ & $0.77 \pm 0.07\stat \pm 0.13\syst \pm 0.02\lum$ & $0.58\pm0.08$ \\
$\PW\gamma\to \mu\nu\gamma$  & $>60$ & $0.76 \pm 0.06\stat \pm 0.08\syst \pm 0.02\lum$ & $0.58\pm0.08$ \\
$\PW\gamma\to \ell\nu\gamma$ & $>60$ & $0.76 \pm 0.05\stat \pm 0.08\syst \pm 0.02\lum$ & $0.58\pm0.08$ \\ \hline
$\PW\gamma\to \Pe\nu\gamma$    & $>90$ & $0.17 \pm 0.03\stat \pm 0.04\syst \pm 0.01\lum$ & $0.17\pm0.03$ \\
$\PW\gamma\to \mu\nu\gamma$  & $>90$ & $0.25 \pm 0.04\stat \pm 0.05\syst \pm 0.01\lum$ & $0.17\pm0.03$ \\
$\PW\gamma\to \ell\nu\gamma$ & $>90$ & $0.20 \pm 0.03\stat \pm 0.04\syst \pm 0.01\lum$ & $0.17\pm0.03$ \\ \hline

$\cPZ\gamma\to \Pe\Pe\gamma$   & $>60$ & $0.14 \pm 0.02\stat \pm 0.02\syst \pm 0.01\lum$ & $0.12\pm0.01$ \\
$\cPZ\gamma\to \mu\mu\gamma$   & $>60$ & $0.14 \pm 0.01\stat \pm 0.02\syst \pm 0.01\lum$ & $0.12\pm0.01$ \\
$\cPZ\gamma\to \ell\ell\gamma$ & $>60$ & $0.14 \pm 0.01\stat \pm 0.01\syst \pm 0.01\lum$ & $0.12\pm0.01$ \\ \hline
$\cPZ\gamma\to \Pe\Pe\gamma$   & $>90$ & $0.047 \pm 0.013\stat \pm 0.010\syst \pm 0.001\lum$ & $0.040\pm0.004$ \\
$\cPZ\gamma\to \mu\mu\gamma$   & $>90$ & $0.046 \pm 0.008\stat \pm 0.010\syst \pm 0.001\lum$ & $0.040\pm0.004$ \\
$\cPZ\gamma\to \ell\ell\gamma$ & $>90$ & $0.046 \pm 0.007\stat \pm 0.009\syst \pm 0.001\lum$ & $0.040\pm0.004$ \\ \hline
\end{scotch}
\end{table*}

\section{Anomalous triple gauge couplings in \texorpdfstring{$\PW\gamma$ and $\cPZ\gamma$}{W gamma and Z gamma} production}
\label{sec:ATGC}

\subsection{\texorpdfstring{$\PW\PW\gamma$}{WW gamma} coupling}
The most general Lorentz invariant, effective Lagrangian that describes
$\PW\PW\gamma$ and $\PW\PW\cPZ$ couplings has 14 independent
parameters~\cite{hagiwara1,hagiwara2},
seven for each triple-boson vertex. Assuming charge conjugation ($C$)
and parity ($P$) invariance for the effective EW Lagrangian
($\mathcal{L}_{\PW\PW\vb}$), normalized by its EW coupling strength
($g_{\PW\PW\vb}$), leaves only six independent couplings for
describing the $\PW\PW\gamma$ and $\PW\PW\cPZ$ vertices:
\ifthenelse{\boolean{cms@external}}{
\begin{multline}
\label{eq:atgc_lagrangian}
\frac{\mathcal{L}_{\PW\PW\vb}}{g_{\PW\PW\vb}} =
  ig_1^V(W_{\mu\nu}^\dagger W^\mu V^\nu - W^\dagger_\mu V_\nu W^{\mu\nu})
 +\\ i\kappa_V W^\dagger_\mu W_\nu V^{\mu\nu} +
   \frac{i\lambda_V}{M_W^2}W^\dagger_{\delta\mu}W^\mu _\nu V^{\nu\delta},
\end{multline}
}{
\begin{equation}
\label{eq:atgc_lagrangian}
\frac{\mathcal{L}_{\PW\PW\vb}}{g_{\PW\PW\vb}} =
  ig_1^V(W_{\mu\nu}^\dagger W^\mu V^\nu - W^\dagger_\mu V_\nu W^{\mu\nu})
 + i\kappa_V W^\dagger_\mu W_\nu V^{\mu\nu} +
   \frac{i\lambda_V}{M_W^2}W^\dagger_{\delta\mu}W^\mu _\nu V^{\nu\delta},
\end{equation}
}
where $\vb = \gamma$ or \cPZ, $W^\mu$ are the \Wpm fields,
$W_{\mu\nu} = \partial_\mu W_\nu - \partial_\nu W_\mu$, with
the overall couplings given by
$g_{\PW\PW\gamma} = -e$, and $g_{\PW\PW\cPZ} = -e \cot\theta_W$,
where $\theta_W$ is the weak mixing angle. Assuming electromagnetic
gauge invariance, $g_1^\gamma = 1$; the remaining parameters
that describe the $\PW\PW\gamma$ and $\PW\PW\cPZ$ couplings are
$g_1^\cPZ$, $\kappa_\cPZ$, $\kappa_\gamma$, $\lambda_\cPZ$, and
$\lambda_\gamma$. In the SM, $\lambda_\cPZ = \lambda_\gamma = 0$ and
$g_1^\cPZ = \kappa_\cPZ = \kappa_\gamma = 1$.
In this analysis, we follow the convention that describes
the couplings in terms of their deviation from the SM values:
$\Delta g_1^\cPZ \equiv g_1^\cPZ - 1$,
$\Delta \kappa_\cPZ \equiv \kappa_\cPZ - 1$,
and $\Delta \kappa_\gamma \equiv \kappa_\gamma -1$.
Invariance under $SU(2)_{L}\times U(1)_{Y}$ transformations reduces these
to three independent couplings:

\begin{equation}
\label{eq:hisz}
\Delta\kappa_\cPZ = \Delta g_1^\cPZ -
\Delta \kappa_\gamma \cdot \tan^2\theta_\PW,\quad
\lambda = \lambda_\gamma = \lambda_\cPZ,
\end{equation}
where $\Delta\kappa_\gamma$ and $\lambda_\gamma$ are determined from
$\PW\gamma$ production.

\subsection{\texorpdfstring{$\cPZ\cPZ\gamma$ and $\cPZ\gamma\gamma$}{ZZ gamma and Z gamma gamma} couplings}
The most general vertex function for $\cPZ\cPZ\gamma$~\cite{baur} can be written as
\begin{equation}
  \begin{aligned}
    \Gamma^{\alpha\beta\mu}_{\cPZ\cPZ\gamma}(q_1,q_2,p) &=
    \ \frac{p^2 - q_1^2}{\ m^2_Z}\Bigg[
    \quad h_1^\cPZ(q^\mu_2 g^{\alpha\beta} - q^\alpha_2 g^{\mu\beta}) \\
    & \qquad+\ \frac{h^\cPZ_2}{m^2_\cPZ}p^\alpha\Big[(p\cdot q_2)g^{\mu\beta} -
    g^\mu_2 p^\beta\Big] \\
    & \qquad +\ h^\cPZ_3\epsilon^{\mu\alpha\beta\rho}q_{2\rho} \\
    & \qquad  +\ \frac{h^\cPZ_4}{m^2_\cPZ}p^\alpha\epsilon^{\mu\beta\rho\sigma}
    p_\rho q_{2\sigma}
    \Bigg],
  \end{aligned}
\end{equation}
with the $\cPZ\gamma\gamma$ vertex obtained by the replacements
\begin{equation}
\label{eq:zgg}
\frac{p^2 - q_1^2}{m_\cPZ^2}\to \frac{p^2}{m_\cPZ^2}
\text{ and } h_i^\cPZ \to h_i^\gamma,~i = 1, \ldots, 4.
\end{equation}
The couplings $h_i^\vb$ for $\vb = \cPZ$ or $\gamma$, and $i = 1, 2,$ violate
$CP$ symmetry, while those with $i = 3, 4$ are $CP$-even. Although,
at tree level, all these couplings vanish in the SM,
at the higher, one-loop level, the $CP$-conserving couplings are
${\approx}10^{-4}$. As the sensitivity to $CP$-odd and $CP$-even
couplings is the same when using $\pt^\gamma$ to check for the presence of
contributions from ATGCs, we interpret the results as limits on $h_3^\vb$
and $h_4^\vb$ only.

\subsection{Search for anomalous couplings in \texorpdfstring{$\PW\gamma$ and $\cPZ\gamma$}{W gamma and Z gamma} production}
To extract limits on the ATGCs, we simply count the
yield of events in bins of $\pt^\gamma$. The 95\% confidence level
(\CL) upper limits on values of ATGCs are set
using the modified frequentist CL$_s$ method~\cite{CLs_asym}.

As the simulation of the ATGC signal is not available in \MADGRAPH,
the signals are generated using the \SHERPA MC program~\cite{sherpa}
to simulate $\PW\gamma$+jets and
$\cPZ\gamma$+jets with up to two jets in the final state.

For the $\PW\gamma$ analysis, we set one and two-dimensional limits
on each ATGC parameter $\Delta\kappa_\gamma$ and $\lambda_\gamma$,
while $g_1^Z$ is set to the SM value, assuming the ``equal couplings''
scenario of the LEP parameterization~\cite{LEPparam}.

For the $\cPZ\gamma$ analysis, we set $h_1^\vb$ and $h_2^\vb$
to the SM values, and set two-dimensional limits on the $h_3^\vb$ and
$h_4^\vb$ anomalous couplings, with $\vb = \cPZ$ or $\gamma$.
For limits set on the \cPZ-type couplings, the $\gamma$ couplings
are set to their SM values, \ie, to zero, and vice versa.
In this study, we follow the CMS convention of not suppressing
the anomalous TGCs by an energy-dependent form factor.

\begin{figure}[hbt]
\begin{center}
\includegraphics[width=0.49\textwidth]{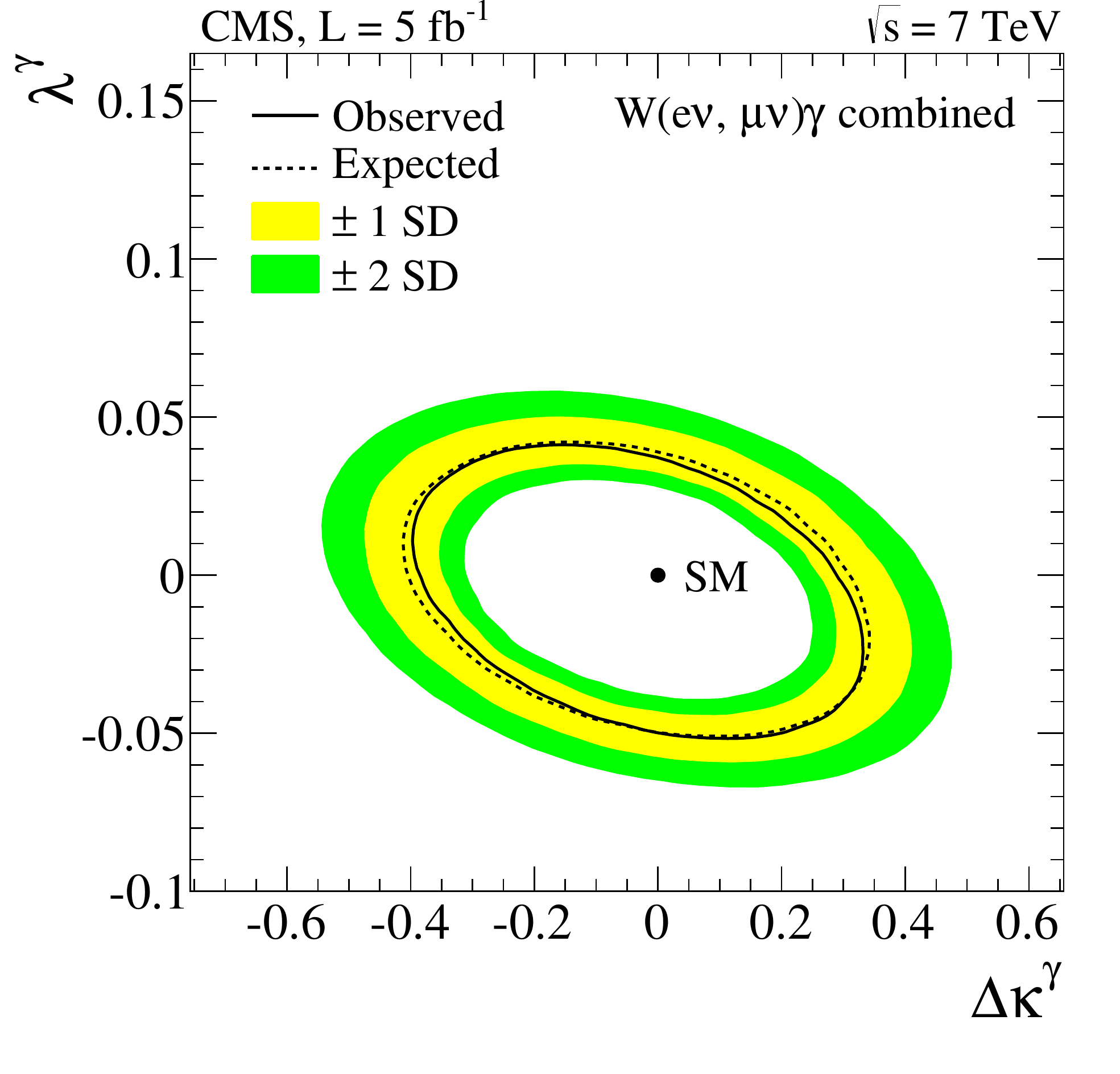}
\caption{Observed (solid curve) and expected (dashed curve) 95\% \CL exclusion contours for
anomalous $\PW\PW\gamma$ couplings, with ${\pm}1$ and ${\pm}2$ standard deviation contours from uncertainties
in the measurements indicated by light and dark shaded bands, respectively.}
\label{fig:Wg_2Dlimits}
\end{center}
\end{figure}

\begin{figure}[hbtp]
  \begin{center}
    \includegraphics[width=0.49\textwidth]{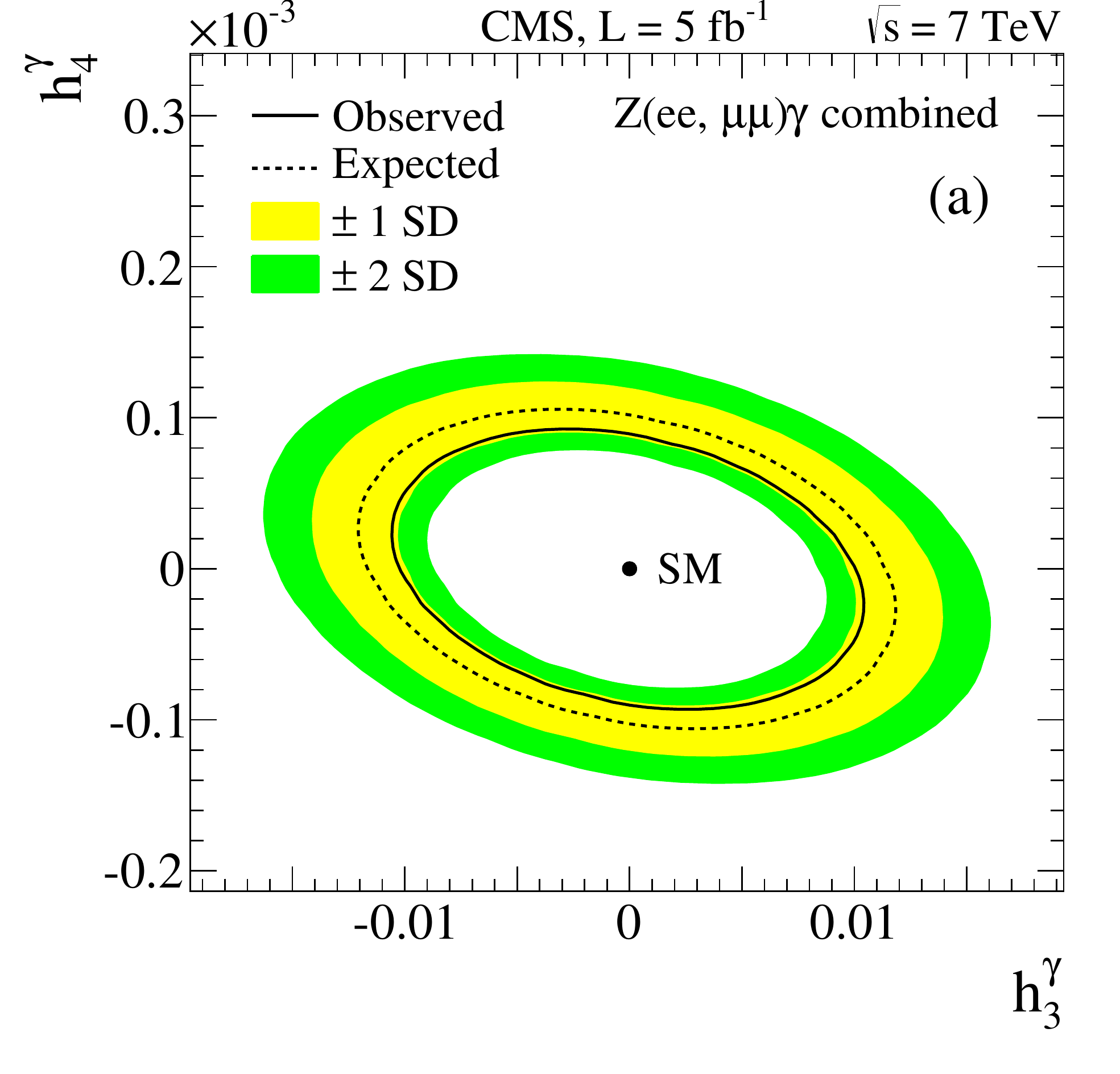}
    \includegraphics[width=0.49\textwidth]{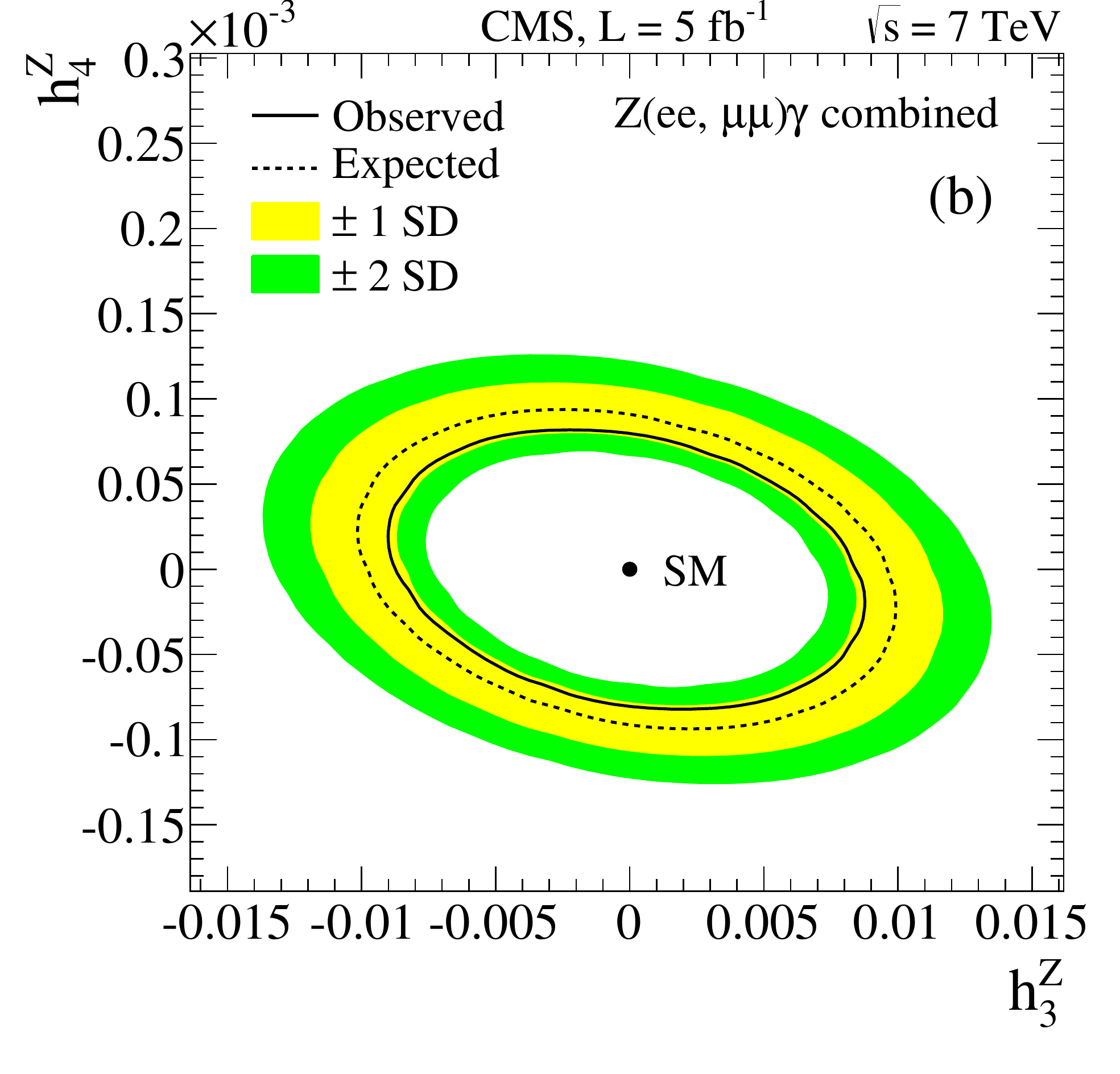}
    \caption{Observed (solid curves) and expected (dashed curves) 95\% \CL exclusion contours for
      anomalous (a)~$\cPZ\gamma\gamma$ and (b)~$\cPZ\cPZ\gamma$ couplings, with ${\pm}1$ and ${\pm}2$
      standard deviation contours indicated by light and dark shaded bands.}
    \label{fig:ZggZZgAtgc2Dlimits}
  \end{center}
\end{figure}

The two-dimensional contours for upper limits at the 95\% confidence level
are given in Fig.~\ref{fig:Wg_2Dlimits}
for the $\PW\gamma$, and Fig.~\ref{fig:ZggZZgAtgc2Dlimits} for the
$\cPZ\gamma$ channels, with the corresponding one-dimensional limits listed
in Table~\ref{tab:WgammaAtgcLimitsN0} for
$\PW\gamma$, and Table~\ref{tab:ZgammaAtgcLimitsN0} for $\cPZ\gamma$.

\begin{table}[hbtp]
\begin{center}
\topcaption{One-dimensional 95\% \CL limits on ATGCs for
$\PW\gamma \to \Pe\nu\gamma$, $\PW\gamma \to \mu\nu\gamma$,
and for the combined analyses. The intervals shown represent the allowed ranges of the coupling parameters.}
\label{tab:WgammaAtgcLimitsN0}
\begin{scotch}{lcc}
        &    $\Delta \kappa_\gamma$ & $\lambda_\gamma$ \\
  \hline
  $\PW\gamma\to \Pe\nu\gamma$ & $[-0.45, 0.36]$ & $[-0.059, 0.046]$ \\
  $\PW\gamma\to \mu\nu\gamma$     & $[-0.46, 0.34]$ & $[-0.057, 0.045]$ \\
  $\PW\gamma\to \ell\nu\gamma$    & $[-0.38, 0.29]$ & $[-0.050, 0.037]$ \\
\end{scotch}
\end{center}
\end{table}

\begin{table}[hbtp]
  \centering
    \topcaption{One-dimensional 95\% \CL limits on ATGCs for
$\cPZ\gamma\to \Pe\Pe\gamma$, $\cPZ\gamma\to\mu\mu\gamma$, and for the combined
analyses. The intervals shown represent the allowed ranges of the coupling parameters.
\label{tab:ZgammaAtgcLimitsN0}}
\setlength{\extrarowheight}{0.2ex}
\renewcommand{\arraystretch}{1.1}
\begin{scotch}{lcccc}
      & $h^\gamma_{3}$ $[10^{-2}]$ & $h^\gamma_{4}$ $[10^{-4}]$ & $h^\cPZ_{3}$ $[10^{-2}]$  & $h^\cPZ_{4}$ $[10^{-4}]$ \\ \hline
      $\cPZ\gamma\to \Pe\Pe\gamma$ & $[-1.3, 1.3]$ & $[-1.1, 1.1]$ & $[-1.1, 1.1]$ & $[-1.0, 1.0]$   \\
      $\cPZ\gamma\to \mu\mu\gamma$   & $[-1.3, 1.3]$ & $[-1.1, 1.2]$ & $[-1.1, 1.1]$ & $[-1.0, 1.1]$   \\
      $\cPZ\gamma\to \ell\ell\gamma$ & $[-1.0, 1.0]$ & $[-0.9, 0.9]$ & $[-0.9, 0.9]$ & $[-0.8, 0.8]$   \\
    \end{scotch}
\end{table}

\section{Summary}
\label{sec:Summary}
We have presented updated measurements of the $\vb\gamma$ inclusive production cross sections
in $\Pp\Pp$ collisions at $\sqrt{s} = 7\TeV$, based on leptonic decays of EW vector bosons
$\PW\to \Pe\nu$, $\PW\to \mu\nu$, $\cPZ\to \Pe\Pe$, and $\cPZ\to \mu\mu$.
The data were collected by the CMS experiment at the LHC in 2011 and correspond to an
integrated luminosity of 5.0\fbinv.
A separation is required between the photon and the charged leptons in ($\eta, \phi$) space of $\Delta R > 0.7$,
and an additional requirement of $m_{\ell\ell} > 50\GeV$ is placed on $\cPZ\gamma$ candidates.
The measured cross sections for $\pt^{\gamma} > 15\GeV$,
$\sigma(\Pp\Pp \to \PW\gamma) \times \mathcal{B}(\PW\to\ell\nu) = 37.0 \pm 0.8\stat \pm 4.0\syst \pm 0.8 \lum\unit{pb}$
and
$\sigma(\Pp\Pp \to \cPZ\gamma) \times \mathcal{B}(\cPZ\to\ell\ell) = 5.33 \pm 0.08\stat \pm 0.25\syst \pm 0.12\lum\unit{pb}$,
are consistent with predictions of the SM; the ratio of these measurements, $6.9 \pm 0.2\stat\pm 0.5\syst$,
is also consistent with the SM value of $5.8 \pm 0.1$ predicted by \MCFM.
Measured cross sections for $\pt^{\gamma} > 60$ and ${>}90$ GeV also agree with the SM.
With no evidence observed for physics beyond the SM, we set
the limits on anomalous $\PW\PW\gamma$, $\cPZ\cPZ\gamma$, and $\cPZ\gamma\gamma$ couplings given in
Tables~\ref{tab:WgammaAtgcLimitsN0} and~\ref{tab:ZgammaAtgcLimitsN0}.

\section*{Acknowledgements}
\label{sec:Acknowledgements}
\hyphenation{Bundes-ministerium Forschungs-gemeinschaft Forschungs-zentren} We congratulate our colleagues in the CERN accelerator departments for the excellent performance of the LHC and thank the technical and administrative staffs at CERN and at other CMS institutes for their contributions to the success of the CMS effort. In addition, we gratefully acknowledge the computing centers and personnel of the Worldwide LHC Computing Grid for delivering so effectively the computing infrastructure essential to our analyses. Finally, we acknowledge the enduring support for the construction and operation of the LHC and the CMS detector provided by the following funding agencies: the Austrian Federal Ministry of Science and Research and the Austrian Science Fund; the Belgian Fonds de la Recherche Scientifique, and Fonds voor Wetenschappelijk Onderzoek; the Brazilian Funding Agencies (CNPq, CAPES, FAPERJ, and FAPESP); the Bulgarian Ministry of Education, Youth and Science; CERN; the Chinese Academy of Sciences, Ministry of Science and Technology, and National Natural Science Foundation of China; the Colombian Funding Agency (COLCIENCIAS); the Croatian Ministry of Science, Education and Sport; the Research Promotion Foundation, Cyprus; the Ministry of Education and Research, Recurrent financing contract SF0690030s09 and European Regional Development Fund, Estonia; the Academy of Finland, Finnish Ministry of Education and Culture, and Helsinki Institute of Physics; the Institut National de Physique Nucl\'eaire et de Physique des Particules~/~CNRS, and Commissariat \`a l'\'Energie Atomique et aux \'Energies Alternatives~/~CEA, France; the Bundesministerium f\"ur Bildung und Forschung, Deutsche Forschungsgemeinschaft, and Helmholtz-Gemeinschaft Deutscher Forschungszentren, Germany; the General Secretariat for Research and Technology, Greece; the National Scientific Research Foundation, and National Office for Research and Technology, Hungary; the Department of Atomic Energy and the Department of Science and Technology, India; the Institute for Studies in Theoretical Physics and Mathematics, Iran; the Science Foundation, Ireland; the Istituto Nazionale di Fisica Nucleare, Italy; the Korean Ministry of Education, Science and Technology and the World Class University program of NRF, Republic of Korea; the Lithuanian Academy of Sciences; the Mexican Funding Agencies (CINVESTAV, CONACYT, SEP, and UASLP-FAI); the Ministry of Science and Innovation, New Zealand; the Pakistan Atomic Energy Commission; the Ministry of Science and Higher Education and the National Science Centre, Poland; the Funda\c{c}\~ao para a Ci\^encia e a Tecnologia, Portugal; JINR (Armenia, Belarus, Georgia, Ukraine, Uzbekistan); the Ministry of Education and Science of the Russian Federation, the Federal Agency of Atomic Energy of the Russian Federation, Russian Academy of Sciences, and the Russian Foundation for Basic Research; the Ministry of Science and Technological Development of Serbia; the Secretar\'{\i}a de Estado de Investigaci\'on, Desarrollo e Innovaci\'on and Programa Consolider-Ingenio 2010, Spain; the Swiss Funding Agencies (ETH Board, ETH Zurich, PSI, SNF, UniZH, Canton Zurich, and SER); the National Science Council, Taipei; the Thailand Center of Excellence in Physics, the Institute for the Promotion of Teaching Science and Technology of Thailand and the National Science and Technology Development Agency of Thailand; the Scientific and Technical Research Council of Turkey, and Turkish Atomic Energy Authority; the Science and Technology Facilities Council, UK; the US Department of Energy, and the US National Science Foundation.
Individuals have received support from the Marie-Curie programme and the European Research Council and EPLANET (European Union); the Leventis Foundation; the A. P. Sloan Foundation; the Alexander von Humboldt Foundation; the Belgian Federal Science Policy Office; the Fonds pour la Formation \`a la Recherche dans l'Industrie et dans l'Agriculture (FRIA-Belgium); the Agentschap voor Innovatie door Wetenschap en Technologie (IWT-Belgium); the Ministry of Education, Youth and Sports (MEYS) of Czech Republic; the Council of Science and Industrial Research, India; the Compagnia di San Paolo (Torino); and the HOMING PLUS programme of Foundation for Polish Science, cofinanced from European Union, Regional Development Fund.

\bibliography{auto_generated}   

\newpage
\appendix
\section{Various kinematic distributions of V gamma candidate events}

\begin{figure*}[hbtp]
  \begin{center}
    \includegraphics[width=0.42\textwidth]{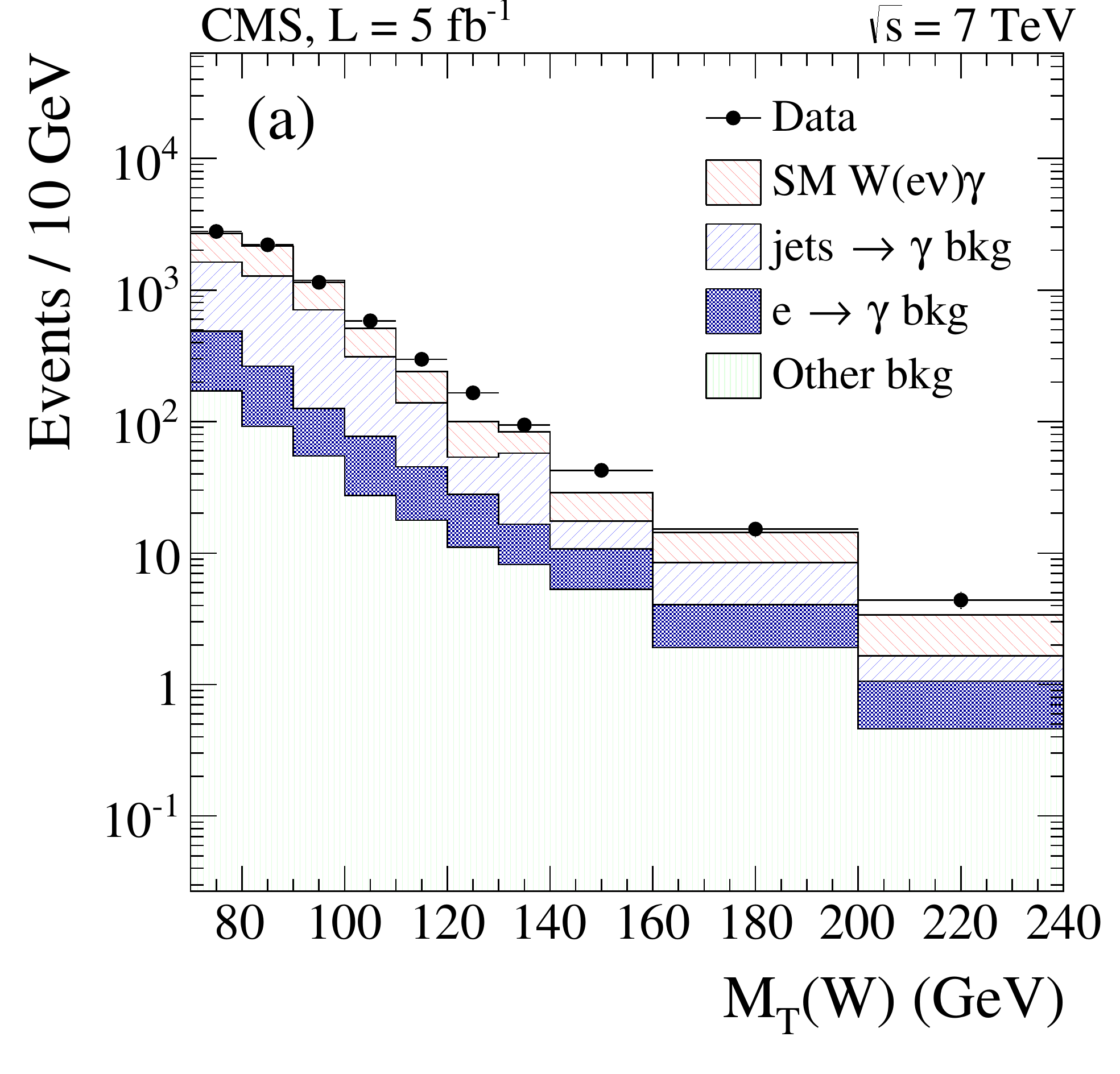}
    \includegraphics[width=0.42\textwidth]{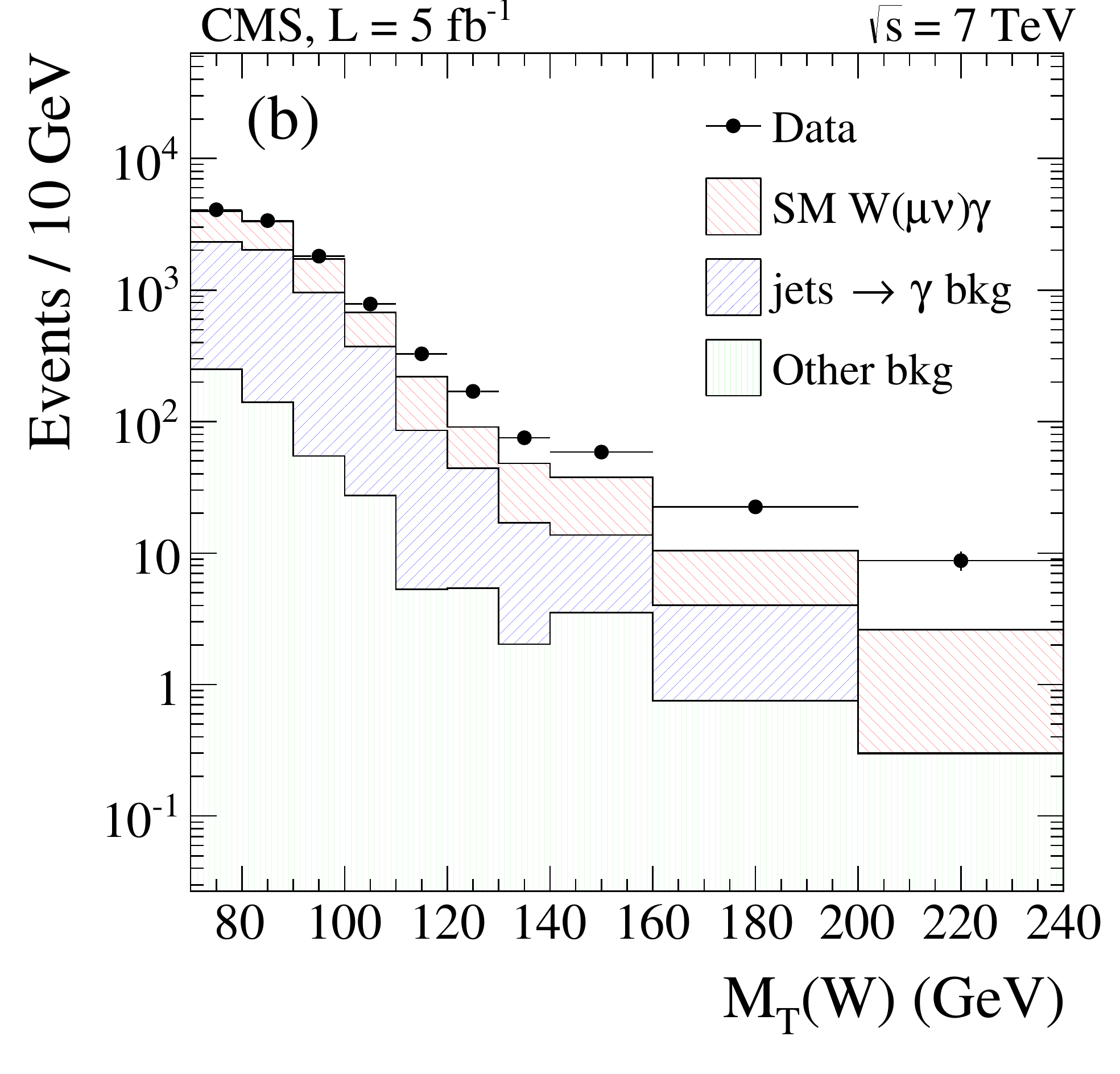}
    \includegraphics[width=0.42\textwidth]{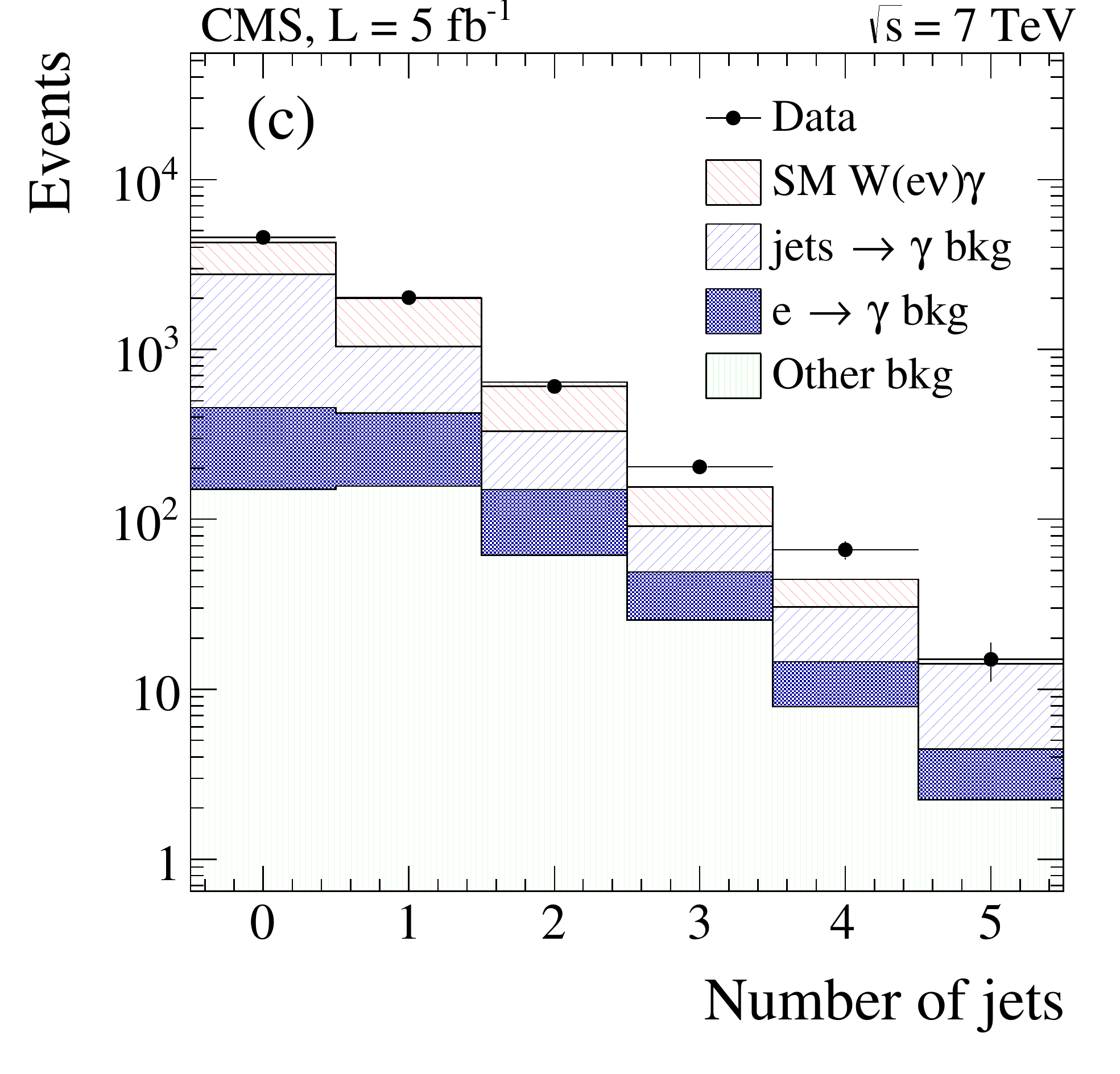}
    \includegraphics[width=0.42\textwidth]{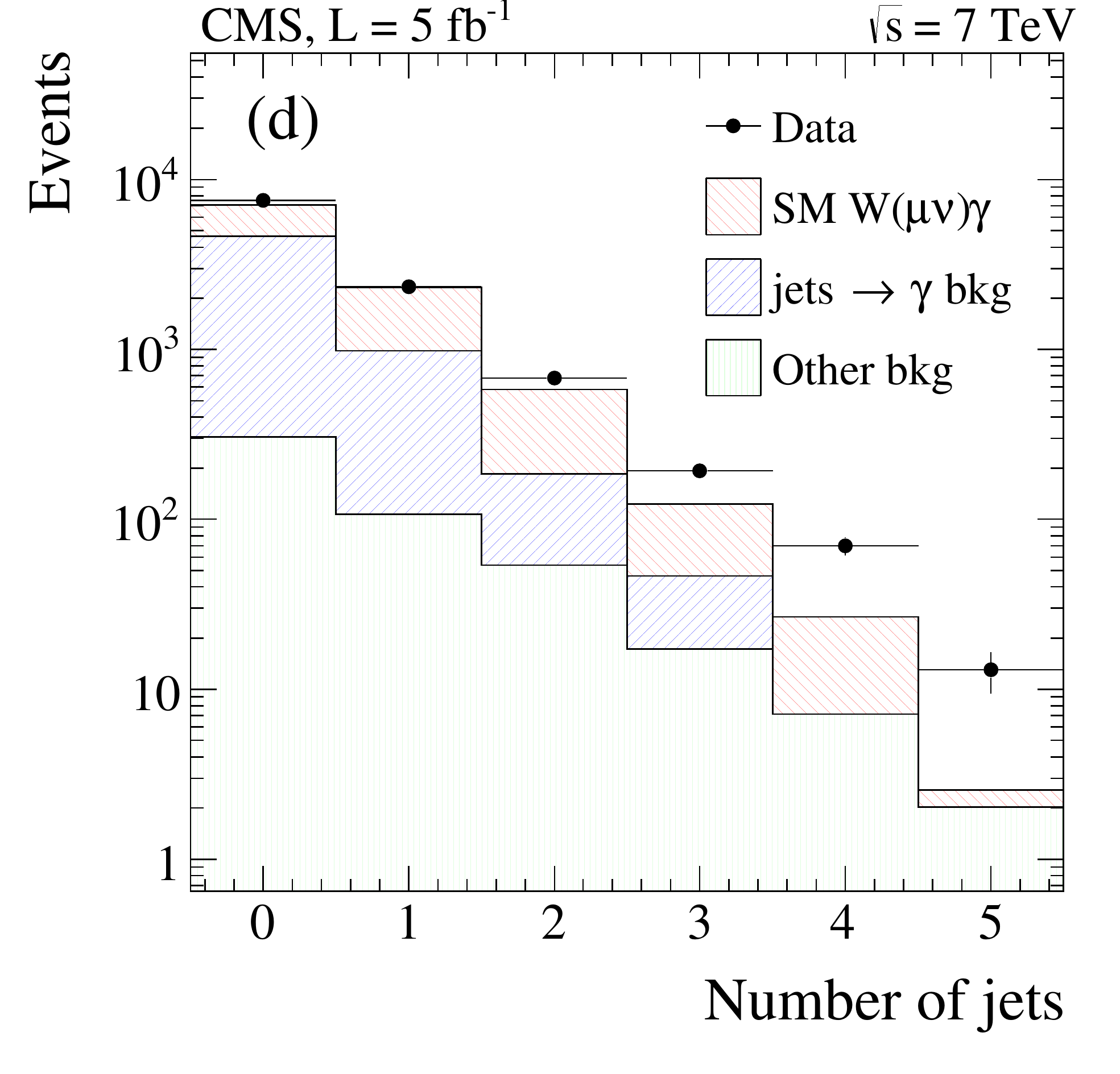}
    \includegraphics[width=0.42\textwidth]{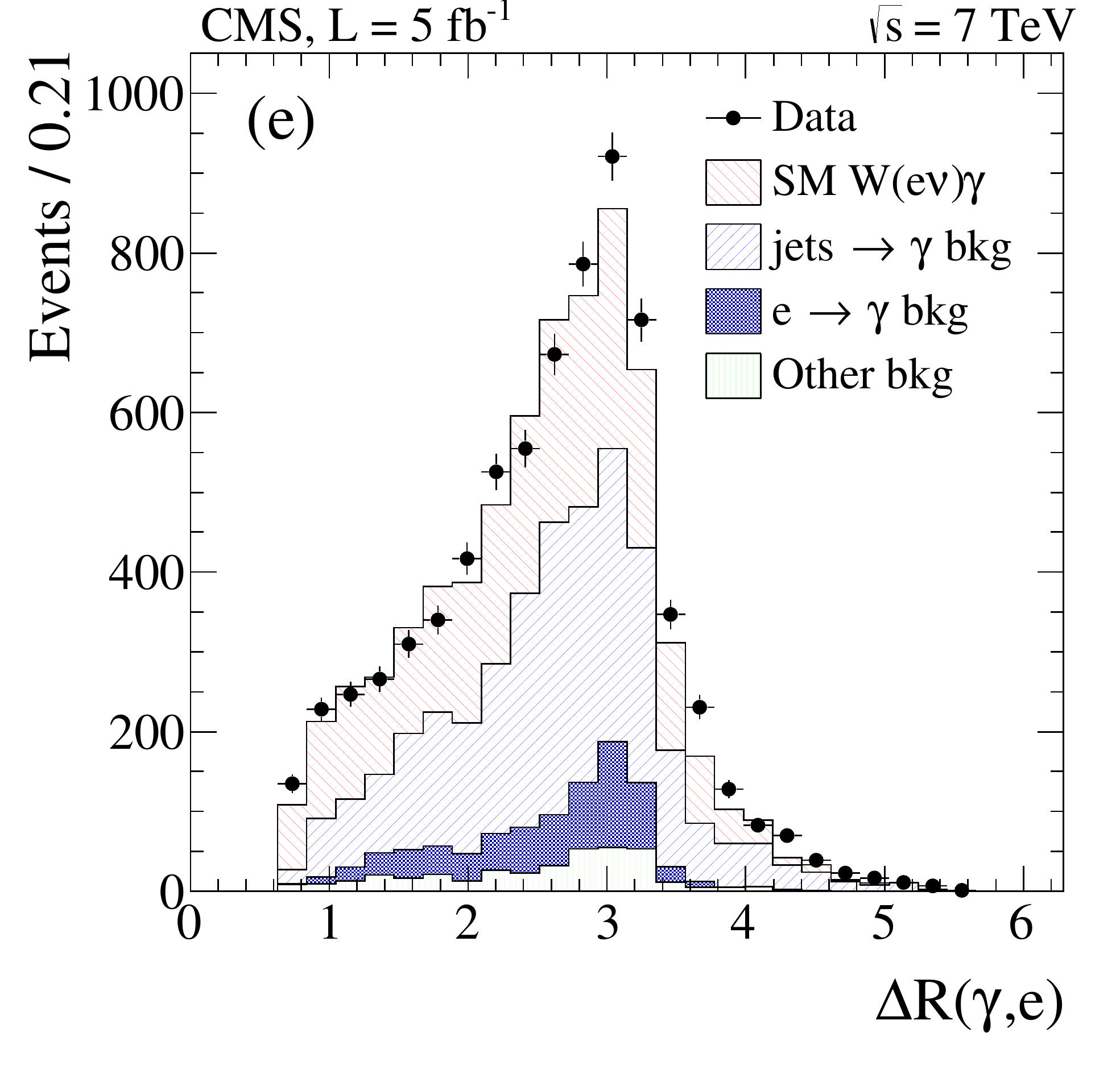}
    \includegraphics[width=0.42\textwidth]{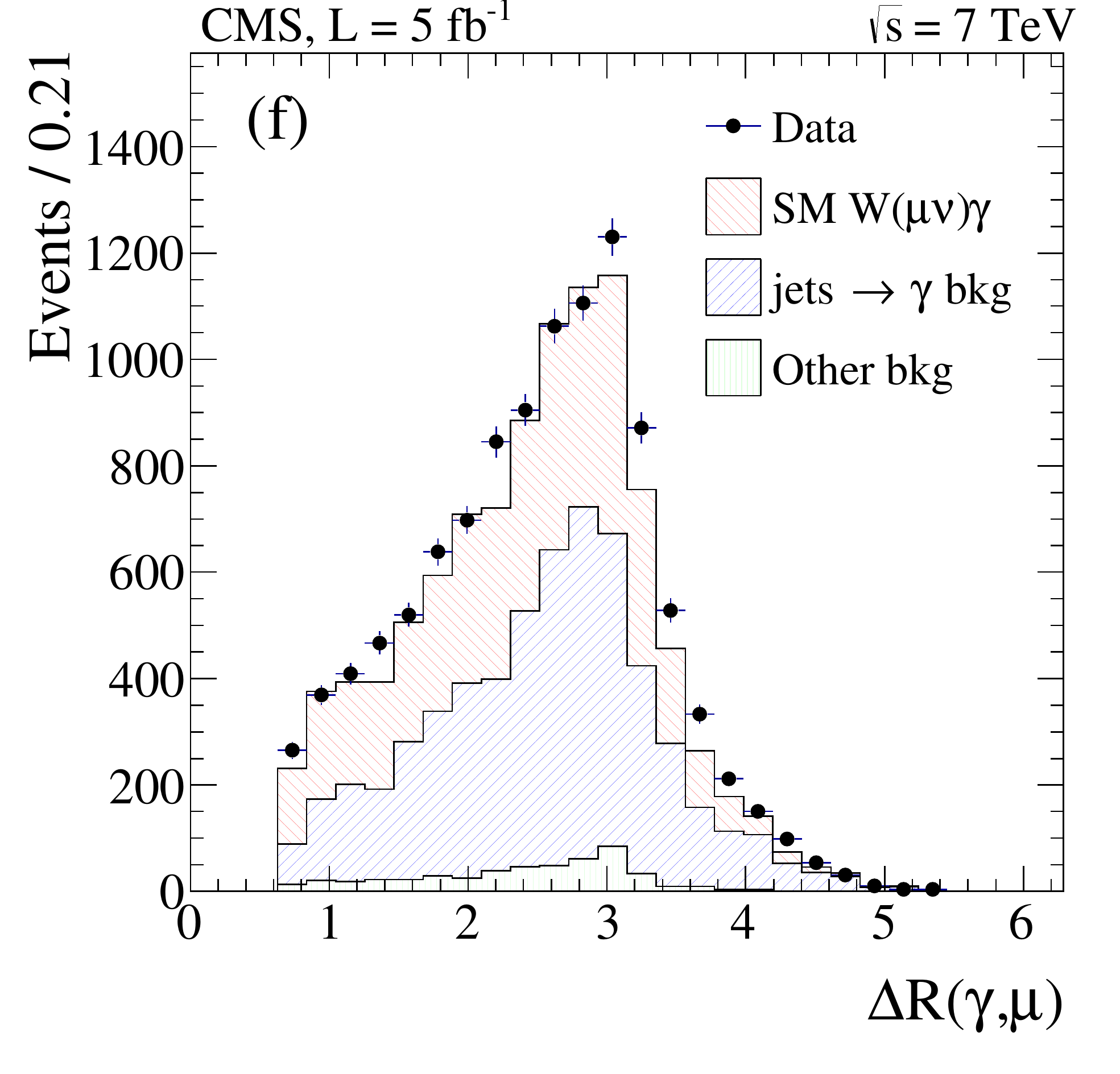}
    \end{center}
    \caption{
      Distributions in the \PW\ boson transverse invariant mass for $\PW\gamma$
      candidate events in data, with signal and background MC simulation
      contributions to (a)~$\PW\gamma\to \Pe\nu\gamma$ and
      (b)~$\PW\gamma\to\mu\nu\gamma$ channels shown for comparison. That for
      number of jets with $\pt > 30\GeV$ is given in (c) and (d),
      respectively. The separation in $R$ between the charged lepton and the
      photon is given in (e) for the electron channel, and that for muon
      channel is illustrated in (f).
    }
    \label{fig:app1}
\end{figure*}

\begin{figure*}[hbtp]
  \begin{center}
    \includegraphics[width=0.42\textwidth]{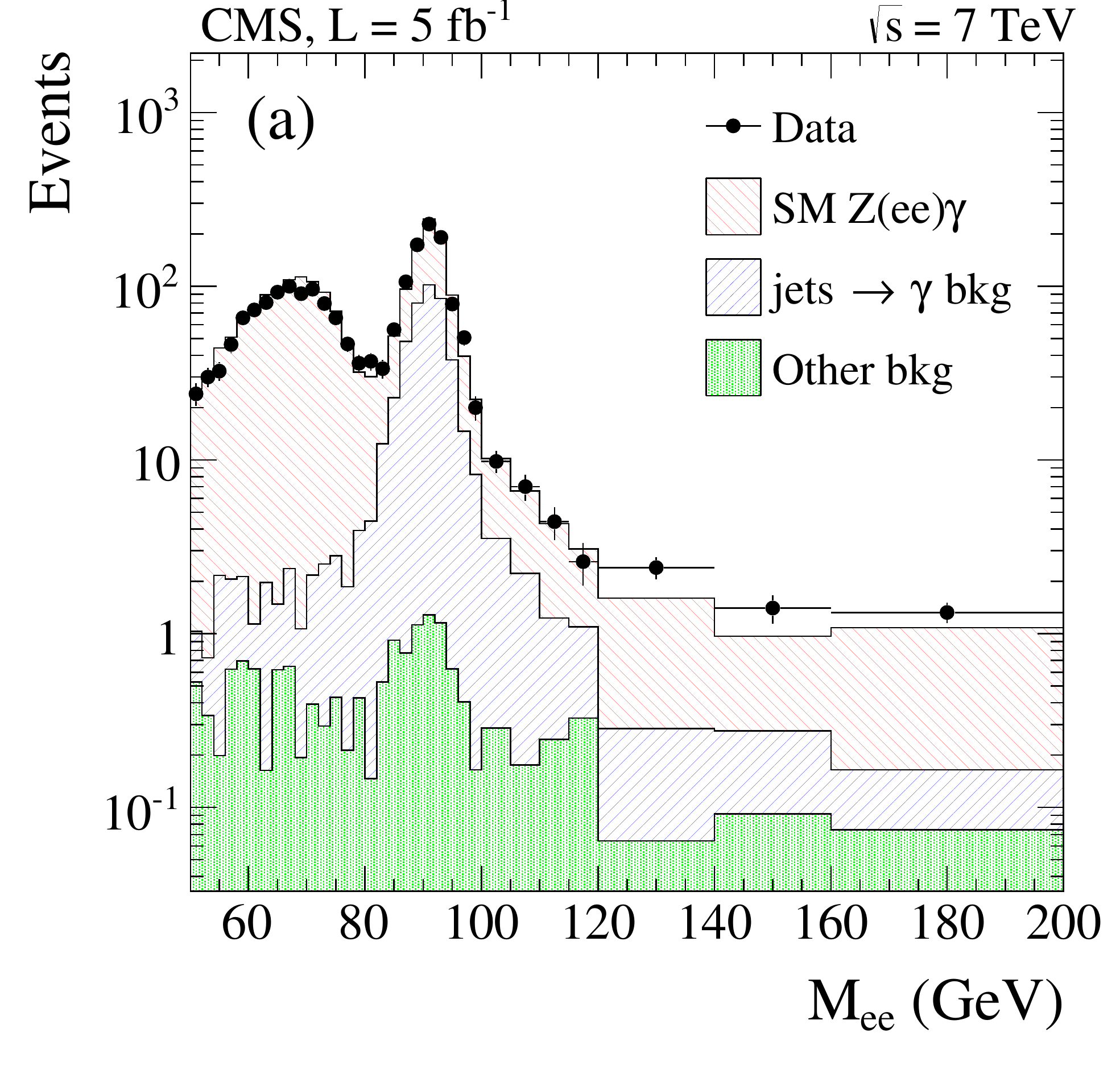}
    \includegraphics[width=0.42\textwidth]{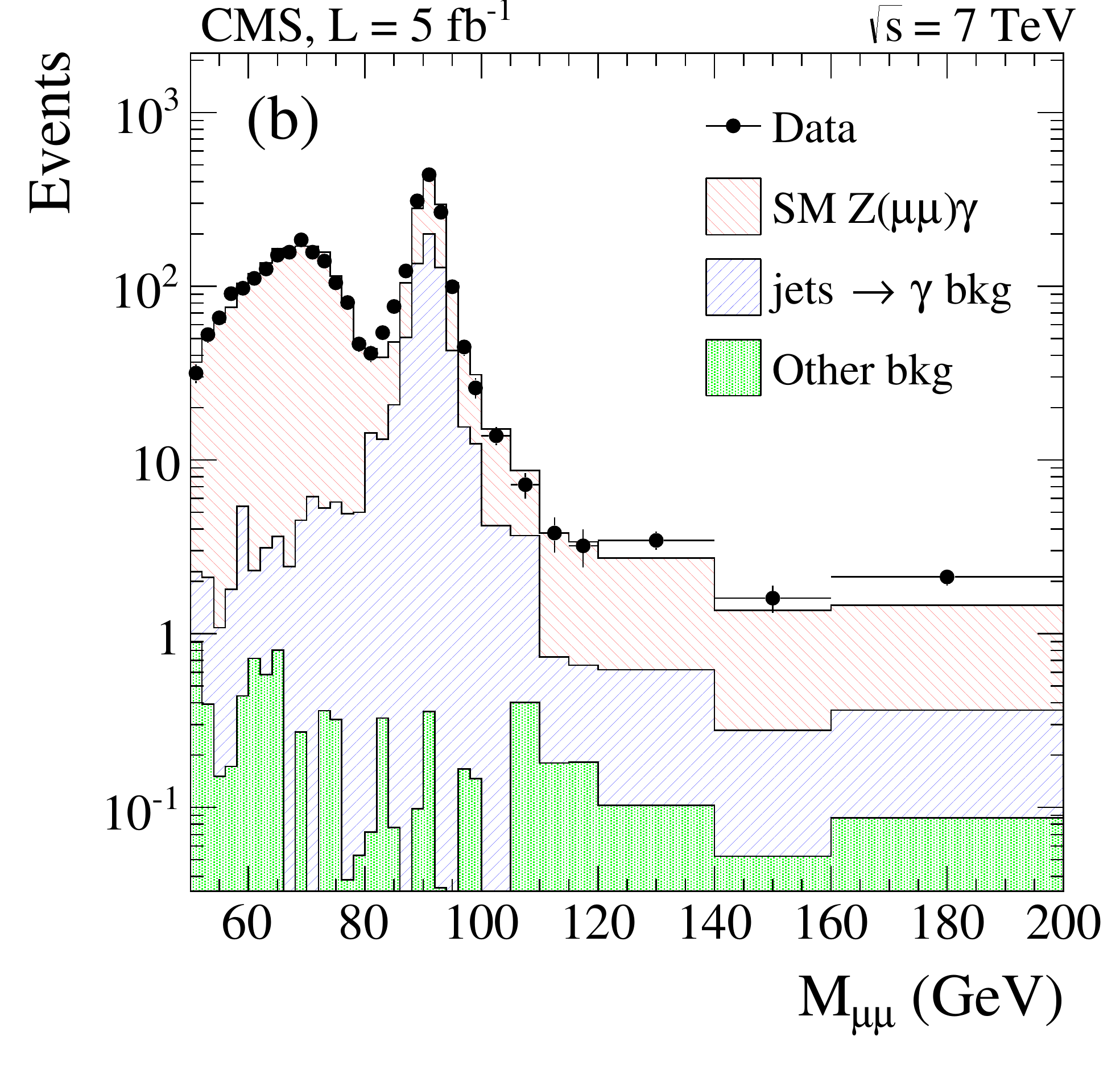}
    \includegraphics[width=0.42\textwidth]{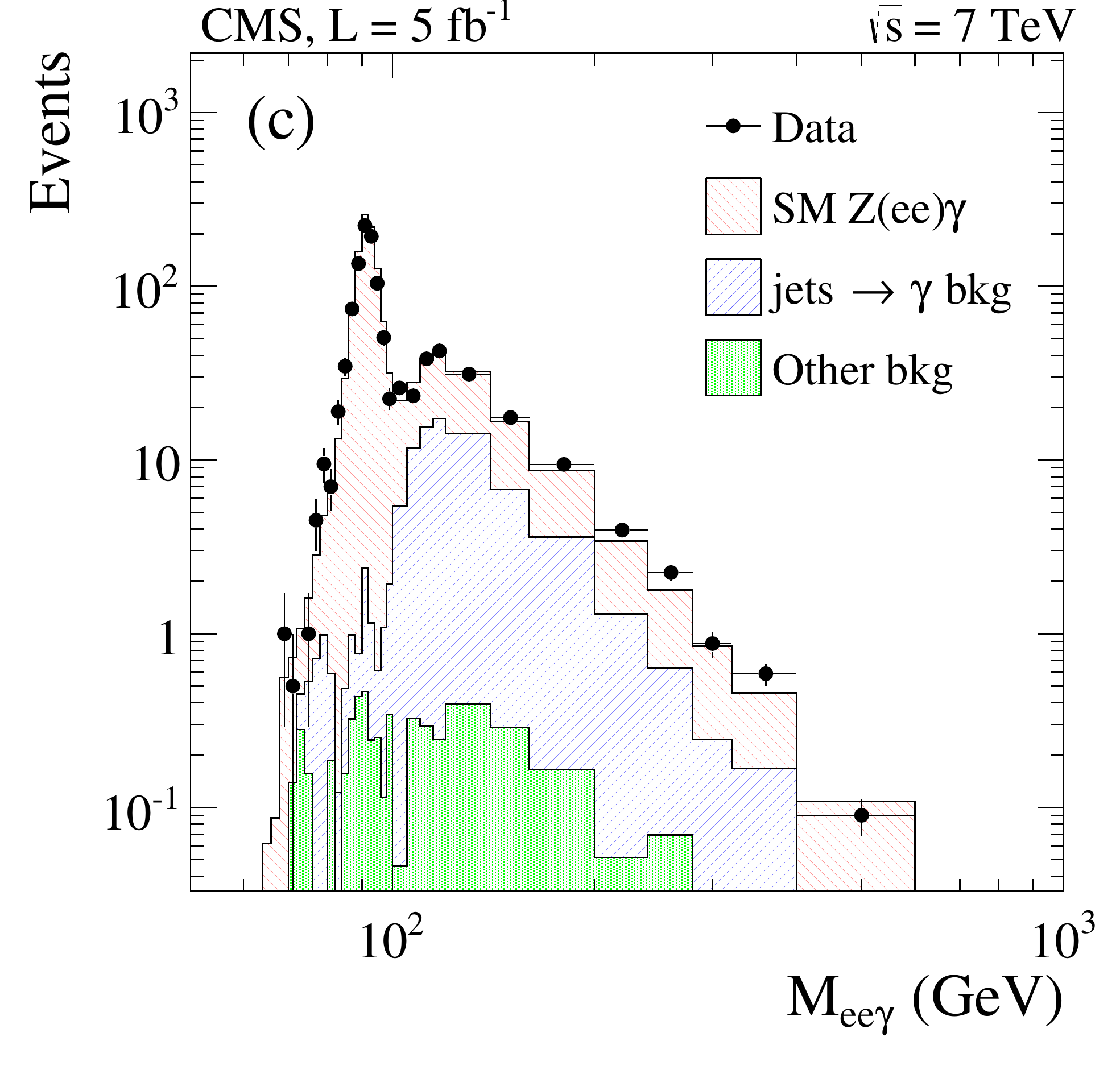}
    \includegraphics[width=0.42\textwidth]{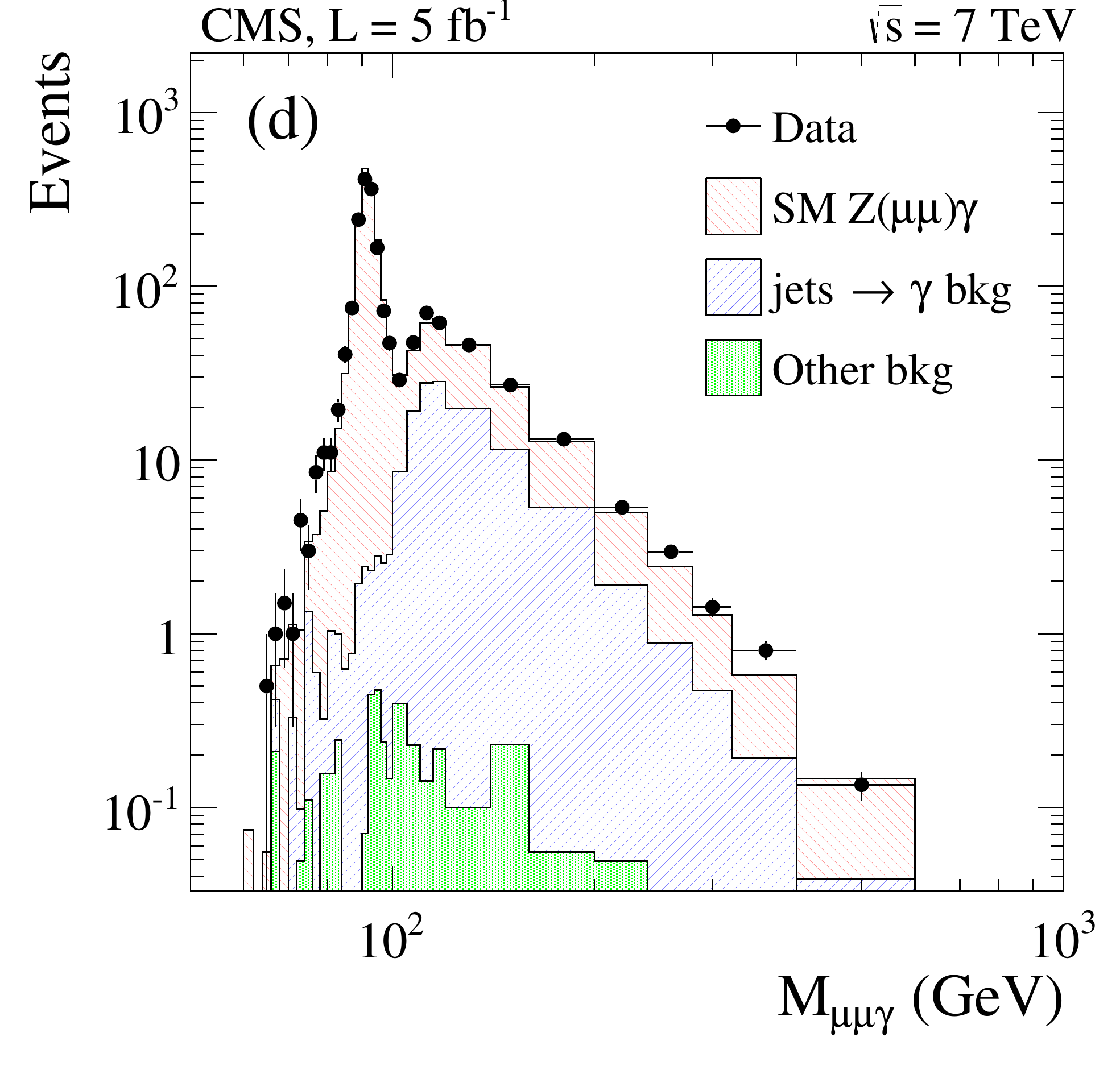}
    \includegraphics[width=0.42\textwidth]{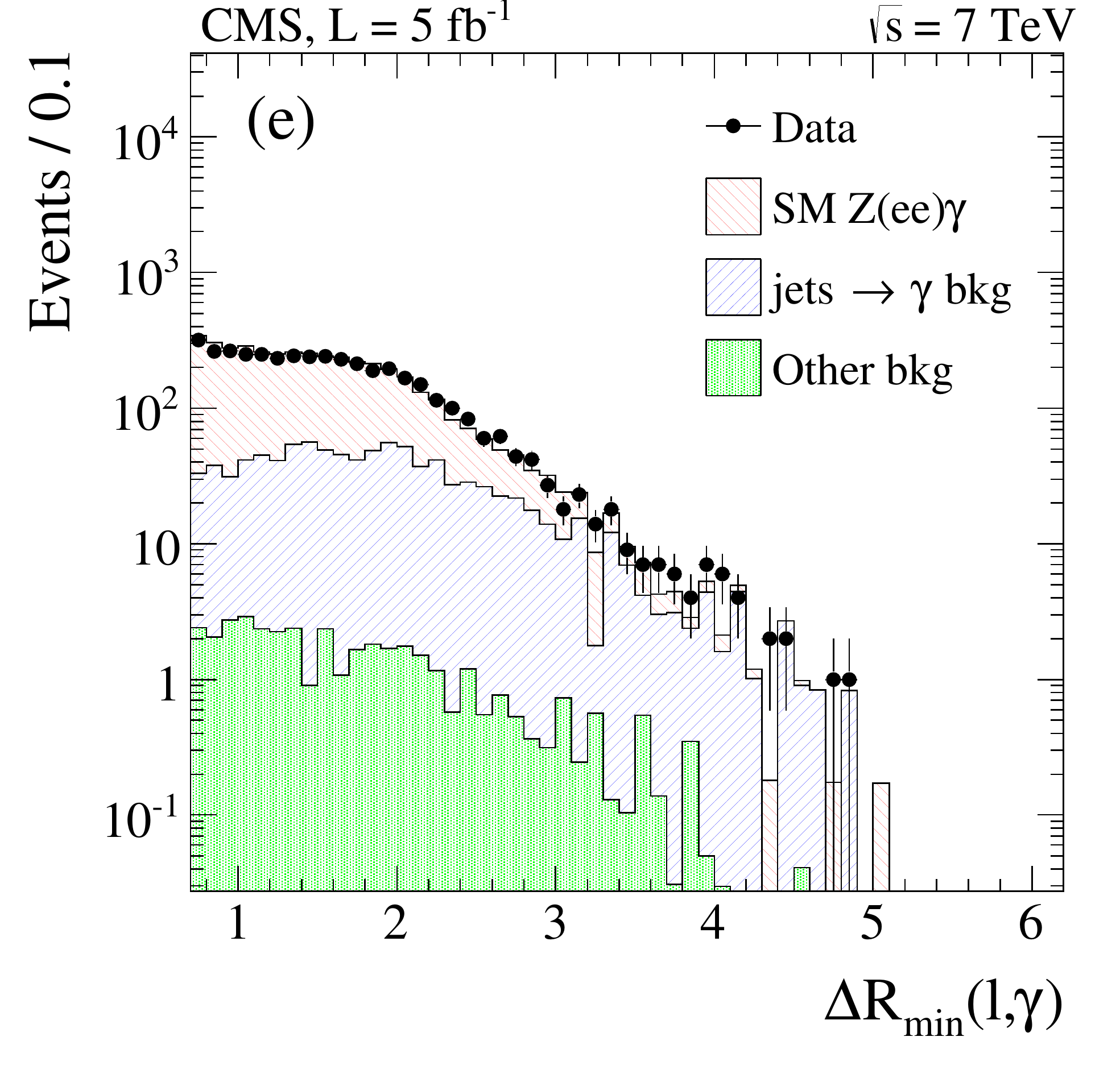}
    \includegraphics[width=0.42\textwidth]{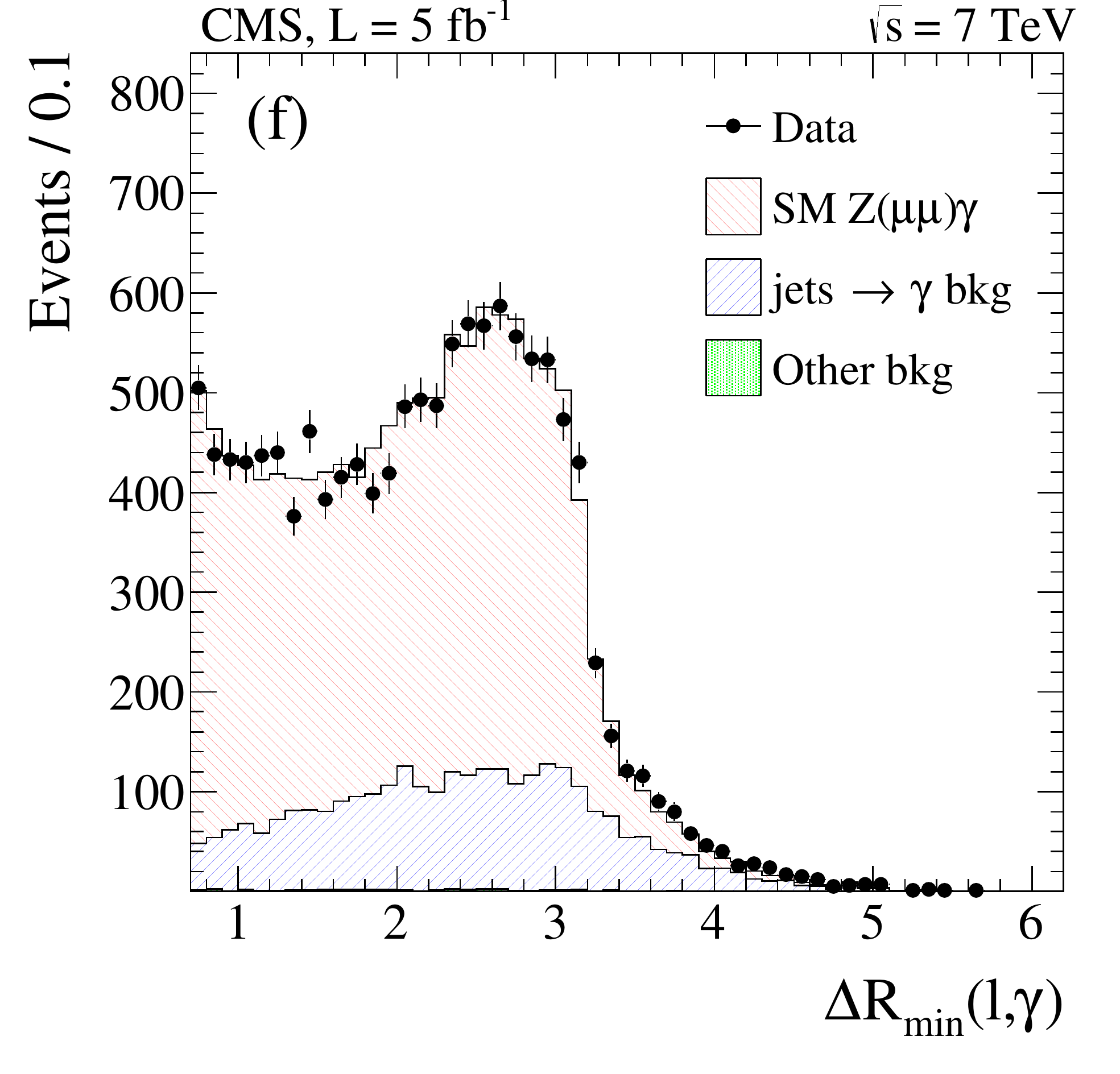}
    \end{center}
    \caption{
      Distributions in the dilepton invariant mass for $\cPZ\gamma$
      candidate events in data, with signal and background MC simulation
      contributions to (a)~$\cPZ\gamma\to \Pe\Pe\gamma$ and
      (b)~$\cPZ\gamma\to\mu\mu\gamma$ channels shown for comparison. That for
      the dilepton plus photon invariant mass is given in (c) and (d).
      The smallest separation in $R$ between any charged leptons and the
      photon is given in (e) for the electron channel, and that for muon
      channel is illustrated in (f).
    }
    \label{fig:app2}
\end{figure*}

\begin{figure*}[hbtp]
  \begin{center}
    \includegraphics[width=0.49\textwidth]{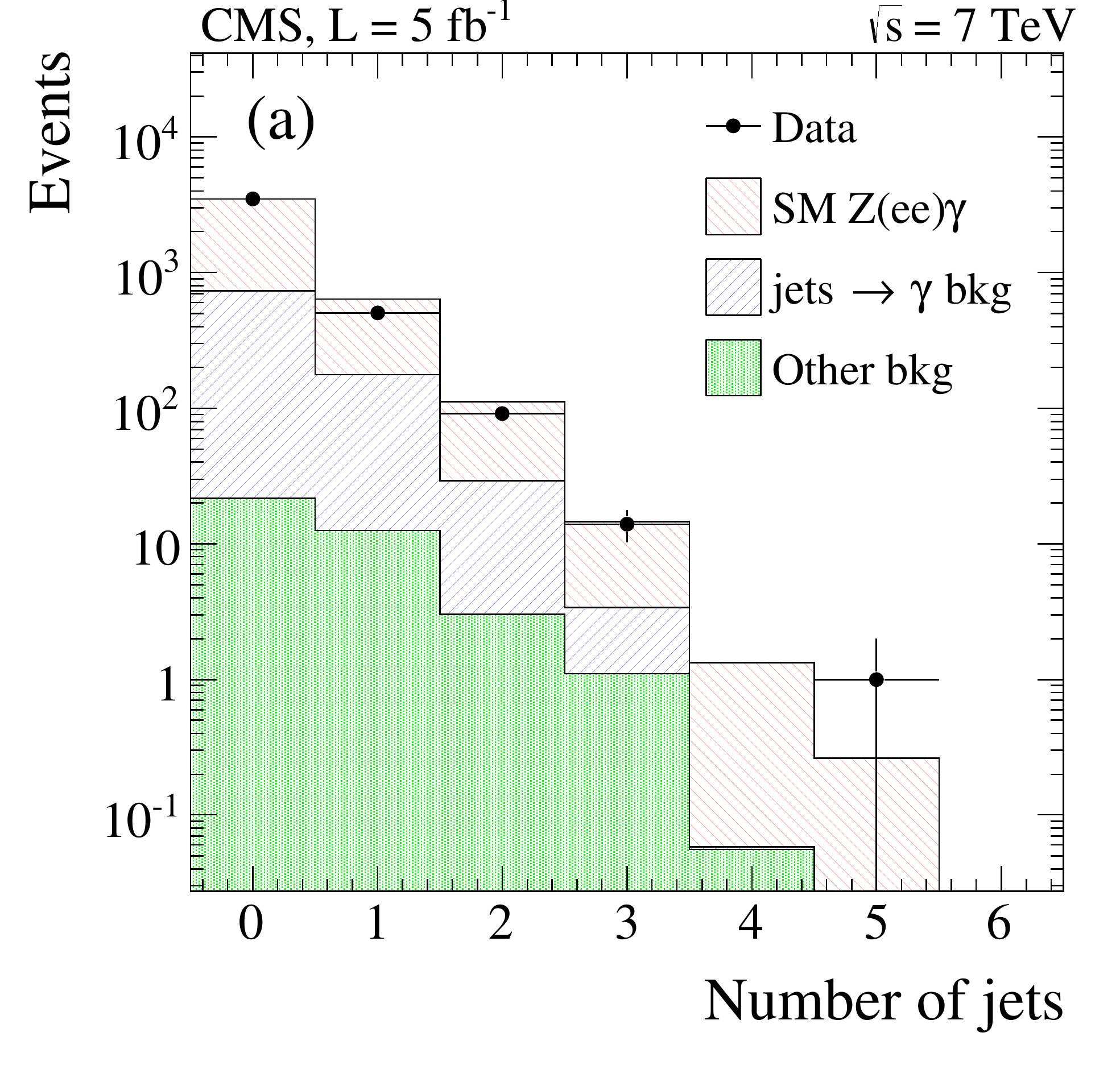}
    \includegraphics[width=0.49\textwidth]{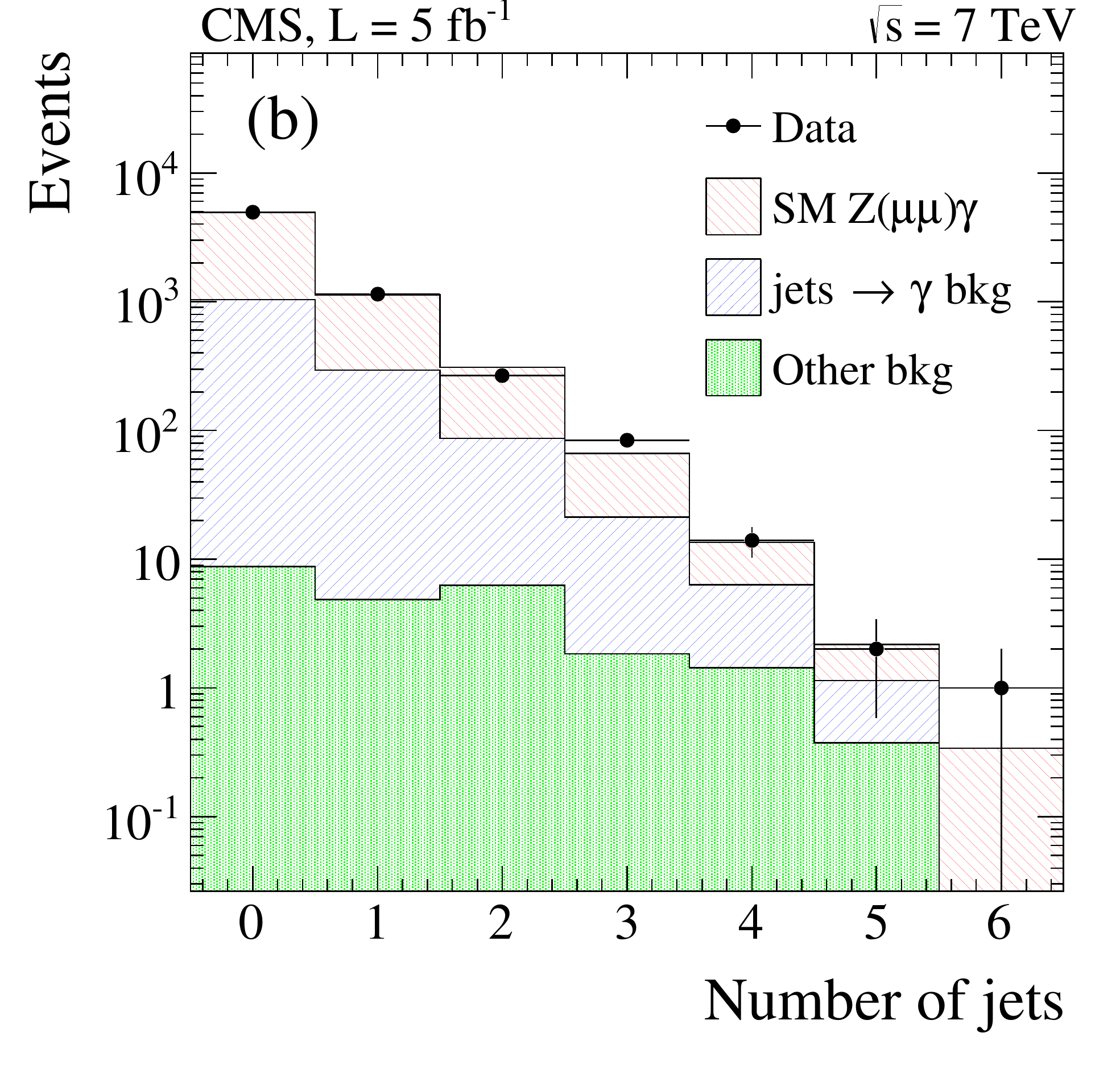}
    \end{center}
    \caption{
      Distributions in the number of jets with $\pt > 30\GeV$
      for $\cPZ\gamma$ candidate events in data, with signal and background
      MC simulation contributions to (a)~$\cPZ\gamma\to \Pe\Pe\gamma$ and
      (b)~$\cPZ\gamma\to\mu\mu\gamma$ channels shown for comparison.
    }
    \label{fig:app3}
\end{figure*}

The various kinematic distributions of the selected
$\PW\gamma$ and $\cPZ\gamma$ candidate events in data overlaid with
the background predictions are shown in Fig.~\ref{fig:app1},~\ref{fig:app2},
and~\ref{fig:app3}.

\cleardoublepage \section{The CMS Collaboration \label{app:collab}}\begin{sloppypar}\hyphenpenalty=5000\widowpenalty=500\clubpenalty=5000\textbf{Yerevan Physics Institute,  Yerevan,  Armenia}\\*[0pt]
S.~Chatrchyan, V.~Khachatryan, A.M.~Sirunyan, A.~Tumasyan
\vskip\cmsinstskip
\textbf{Institut f\"{u}r Hochenergiephysik der OeAW,  Wien,  Austria}\\*[0pt]
W.~Adam, T.~Bergauer, M.~Dragicevic, J.~Er\"{o}, C.~Fabjan\cmsAuthorMark{1}, M.~Friedl, R.~Fr\"{u}hwirth\cmsAuthorMark{1}, V.M.~Ghete, N.~H\"{o}rmann, J.~Hrubec, M.~Jeitler\cmsAuthorMark{1}, W.~Kiesenhofer, V.~Kn\"{u}nz, M.~Krammer\cmsAuthorMark{1}, I.~Kr\"{a}tschmer, D.~Liko, I.~Mikulec, D.~Rabady\cmsAuthorMark{2}, B.~Rahbaran, C.~Rohringer, H.~Rohringer, R.~Sch\"{o}fbeck, J.~Strauss, A.~Taurok, W.~Treberer-Treberspurg, W.~Waltenberger, C.-E.~Wulz\cmsAuthorMark{1}
% ...
\vskip\cmsinstskip
% ...
1:~~Also at Vienna University of Technology, Vienna, Austria\\
2:~~Also at CERN, European Organization for Nuclear Research, Geneva, Switzerland\\
% ...

\end{sloppypar}
\end{document}